\documentclass[a4paper,superscriptaddress,twocolumn,prd,showkeys,nofootinbib]{revtex4}
\usepackage[caption=false]{subfig}

\usepackage{graphicx,amsmath,lipsum}

\usepackage{siunitx}
\usepackage{color}
\usepackage[utf8]{inputenc}

\usepackage{xcolor}
\usepackage{dcolumn}
\usepackage{bm}

\usepackage{aas_macros}

\usepackage{url}
\usepackage[caption=false]{subfig} 
\usepackage{booktabs}

\newcolumntype{H}{>{\setbox0=\hbox\bgroup}c<{\egroup}@{}}
\newcolumntype{Z}{>{\setbox0=\hbox\bgroup}c<{\egroup}@{\hspace*{-\tabcolsep}}}

\newcommand*\diff{\mathop{}\!\mathrm{d}}

\newcommand{\ergmpcyr}{\ensuremath{\mathrm{erg}\,\mathrm{Mpc}^{-3}\,\mathrm{yr}^{-1}}}

\newcommand{\gevcms}{\ensuremath{\mathrm{GeV}\,\mathrm{cm}^{-2}\,\mathrm{s}^{-1}}}

\newcommand{\nhone}{\ensuremath{N_H^{(1)}=5\times10^{25}~\mathrm{cm}^{-2}}}
\newcommand{\nhtwo}{\ensuremath{N_H^{(2)}=10^{26}~\mathrm{cm}^{-2}}}

\begin{document}

\title{Obscured $pp$-channel neutrino sources} 

\author{Matthias Vereecken}
\affiliation{IIHE/ELEM, Vrije Universiteit Brussel, Pleinlaan 2, 1050 Brussels, Belgium}
\affiliation{TENA, Vrije Universiteit Brussel, International Solvay institutes, Pleinlaan 2, 1050 Brussels, Belgium}

\author{Krijn D. de Vries}
\affiliation{IIHE/ELEM, Vrije Universiteit Brussel, Pleinlaan 2, 1050 Brussels, Belgium}

\date{\today}

\keywords{neutrino astronomy}

\begin{abstract}
  We explore the possibility that the astrophysical neutrinos are
  produced in $pp$-interactions with a gas cloud near the source
  acting as a beam dump, which is sufficiently dense to significantly
  attenuate the associated gamma-ray flux through pair-production on
  this gas. In this way, such sources could potentially supply the
  astrophysical neutrino flux whilst avoiding the existing constraints
  on the non-blazar contribution to the extragalactic gamma-ray
  background. After defining our model, we implement a Monte Carlo
  simulation and apply this to different scenarios. First, we
  investigate a set of active galaxies which exhibit signs of
  obscuration. We find that, currently, the expected neutrino flux
  from these objects in our model is below the existing exclusion limits, but
  can already constrain the amount of protons accelerated in such
  sources. Second, we investigate the diffuse neutrino flux generated
  by a population of obscured sources. We find that such a population
  can indeed alleviate the tension with the extragalactic background
  light. We discuss the possibility that ultra-luminous infrared
  galaxies represent such a source class.
\end{abstract}

\maketitle

\section{Introduction}
\label{sec:pp-neutrinos}

While the presence of a high-energy neutrino flux has been firmly
established in recent years by the IceCube Neutrino Observatory, the
sources of these neutrinos remain unknown. Recently, the first
neutrino source has been identified: the flaring blazar
TXS~0506+056. However, sources such as this one are unlikely to be
responsible for the bulk of the observed diffuse neutrino
flux~\cite{Murase:2018iyl,Hooper:2018wyk}, although there is no
consensus on this (see e.g.~\cite{Palladino:2018lov}). IceCube
analyses show no significant clustering in either the high-energy
starting events~\cite{Aartsen:2017mau}, which is a very pure sample at
high energy, or the most recent all-sky point source searches using
eight years of data~\cite{Aartsen:2018ywr}, which includes events at
lower energy, but has more contamination by background.


The results from the all-sky clustering search indicate that the total
diffuse flux is not dominated by a few individually powerful
sources. However, it could still be dominated by a single source
class, with blazars and gamma-ray bursts standing out as candidate
sources, since they are rare and powerful enough to be
well-detectable. Stacked searches for neutrinos from
blazars~\cite{Aartsen:2016lir,Aartsen:2017kru}, selected for their
bright (gamma-ray) emission, has limited the contribution of these
blazars to a few percent up to $\sim30\%$ of the diffuse neutrino
flux, depending on the energy range and spectral index of the emitted
neutrinos. Similarly, searches for neutrinos from gamma-ray burst in
their prompt phase~\cite{Aartsen:2017wea}, find results compatible
with background, limiting the total contribution from such events to
$\leq~1\%$.

Despite these null-results, it is possible to infer some general
properties of the neutrino source population by combining the
non-detection of such a population with the requirement that the
sources still need to supply the detected astrophysical
flux~\cite{Lipari:2008zf,2010PhRvD..81b3001S,Murase:2012df,Ahlers:2014ioa,Kowalski:2014zda,Murase:2016gly}.
Each individual neutrino source needs to have a low neutrino
luminosity in order to not have shown up in any point source
search. On the other hand, this means that the population needs to be
sufficiently numerous in order to supply the detected astrophysical
neutrino flux. These constraints already disfavour several important
source classes as the (dominant) source of the observed astrophysical
neutrinos, such as BL~Lacs and standard FSRQ
scenarios. 

In addition to direct searches, possible neutrino sources are also
strongly constrained by measurements of the (diffuse) gamma-ray flux,
since neutrinos and gamma rays are produced together in proton-photon
or proton-proton interactions. A priori, one can use the extragalactic
gamma-ray background (EGB) for this, since it is the total gamma-ray
flux from outside our galaxy. However, tighter constraints can be
achieved by subtracting known sources of gamma rays that do not
contribute to the neutrino flux. Most of the EGB flux is due to
blazars, whose spectra can typically be explained with leptonic
emission only and whose total neutrino emission is
constrained. Therefore, the gamma-ray
flux associated to neutrino emission can instead be compared with the
unresolved gamma-ray flux, the isotropic diffuse gamma-ray background
(IGRB). Alternatively, one can also compare with the estimated (i.e.\
model-dependent) non-blazar contribution to the EGB, which attempts to
also subtract unresolved blazars from the diffuse flux.

The comparison between the diffuse neutrino and gamma-ray fluxes gives
very important constraints in the case of CR reservoir models, which
feature $pp$-interactions.  In such models, cosmic rays are confined
in e.g.\ a starburst galaxy or galaxy cluster, such that the
interaction probability becomes appreciable after integrating over
the cosmic ray path length, while the produced gamma rays and neutrinos can
escape unhindered.  Earlier studies for $pp$-sources in
general~\cite{Murase:2013rfa} and for star-forming galaxies in
particular~\cite{Tamborra:2014xia} found that the neutrino flux
observed by IceCube was compatible with the observed gamma-ray
background for spectral indices $\alpha$ around $\sim 2.1$.  Later
studies, however, disfavour star-forming galaxies as the dominant
source of neutrinos~\cite{Bechtol:2015uqb}. Using stronger constraints
from the bound on the non-blazar contribution to the EGB, which is
about
28\%~\cite{TheFermi-LAT:2015ykq,Lisanti:2016jub,Ajello:2015mfa,Ackermann:2015tah,DiMauro:2014wha,Inoue:2014ona,Costamante:2013sva},
it is difficult to explain the neutrino flux without simultaneously
violating this bound on the gamma-ray flux. This is due to the
combination of two effects. First, the gamma-ray spectrum produced in
$pp$-interaction goes down to low energy (since the energy required to
for proton-proton interaction is almost trivially satisfied in the
centre-of-mass frame). Second, during propagation, high-energy gamma
rays initiate an electromagnetic cascade by interacting with the
cosmic microwave background and extragalactic background
light. Consequently, there is a build-up of lower-energy gamma rays in
the 1--100~GeV band observed by
Fermi-LAT~\cite{Atwood:2009ez,Atwood:2013rka}. As a result, the
diffuse gamma-ray flux in this band violates the existing
bounds. More recently, however, a study of hadronically powered
gamma-ray galaxies~\cite{Palladino:2018bqf}, such as starbursts and
ultra-luminous infrared galaxies, finds instead that a spectral index
$\alpha<2.12$ is still compatible with all observations, including the
most recent estimates of the non-blazar contribution to the EGB.

In $p\gamma$-source scenarios, there is a lower limit on the energy of
the produced gamma-rays from $\pi^0$-decay, such that their flux at
lower energies, i.e.\ in the Fermi band, is only generated through the
electromagnetic cascade. Therefore, the constraint from the above
argument is slightly weaker (although dependent on specific models for
the target radiation field). Still, given that the non-blazar
contribution to the EGB is already constrained below 28\% and bright
gamma-ray blazars detected by Fermi are disfavoured as the main source
of the neutrinos, a tension remains when considering the gamma-ray
flux produced by $p\gamma$-neutrino sources. This tension can be
resolved when considering photon-photon annihilation inside these
sources, caused by the same radiation field as the one responsible for
$p\gamma$-interactions. As shown in~\cite{Murase:2015xka},
$p\gamma$-neutrino sources bright in X-rays or MeV gamma-rays
(assuming these can escape the system unhindered) can be such sources.
As a result of the many constraints above, conventional sources like
gamma-ray bursts and active galactic nuclei bright in gamma rays are
unlikely to form the dominant contribution to the diffuse neutrino
flux. Instead, neutrino sources are likely dim in GeV gamma
rays~\cite{Murase:2015xka}.

In this context, we investigate a model of neutrino sources which are
obscured in gamma rays, as well as X-rays, by a dense gas cloud with
column density \nhone{} or \nhtwo{} close to the source. We envision a
scenario where this dense gas gets bombarded by accelerated cosmic
rays, acting as a beam dump. Such a scenario was first considered
in~\cite{Maggi:2016bbi}. We develop a Monte Carlo code to calculate
the predicted neutrino and gamma-ray fluxes from such a scenario.

We apply our model first to a selection of active galaxies which are
possibly obscured. We compare our prediction with current constraints
from IceCube and put constraints on the possible proton content of
these sources under our model. Afterwards, we calculate the diffuse
neutrino and gamma-ray fluxes from a population of obscured sources to
investigate whether such a population of sources can explain the
diffuse neutrino flux without exceeding the measured extragalactic
gamma-ray background in our model. We find that our model indeed
alleviates the tension. In this context, we also consider in more
detail ultra-luminous infrared galaxies as a promising source
category.

\section{Neutrinos from obscured $pp$-sources}

In this section, we introduce our model for obscured $pp$-sources,
discuss the properties of the gas cloud and the attenuation of gamma
rays near the source.

\subsection{Model definition}
\label{sec:setup}

\begin{figure}
  \centering
  \includegraphics[width=0.7\columnwidth]{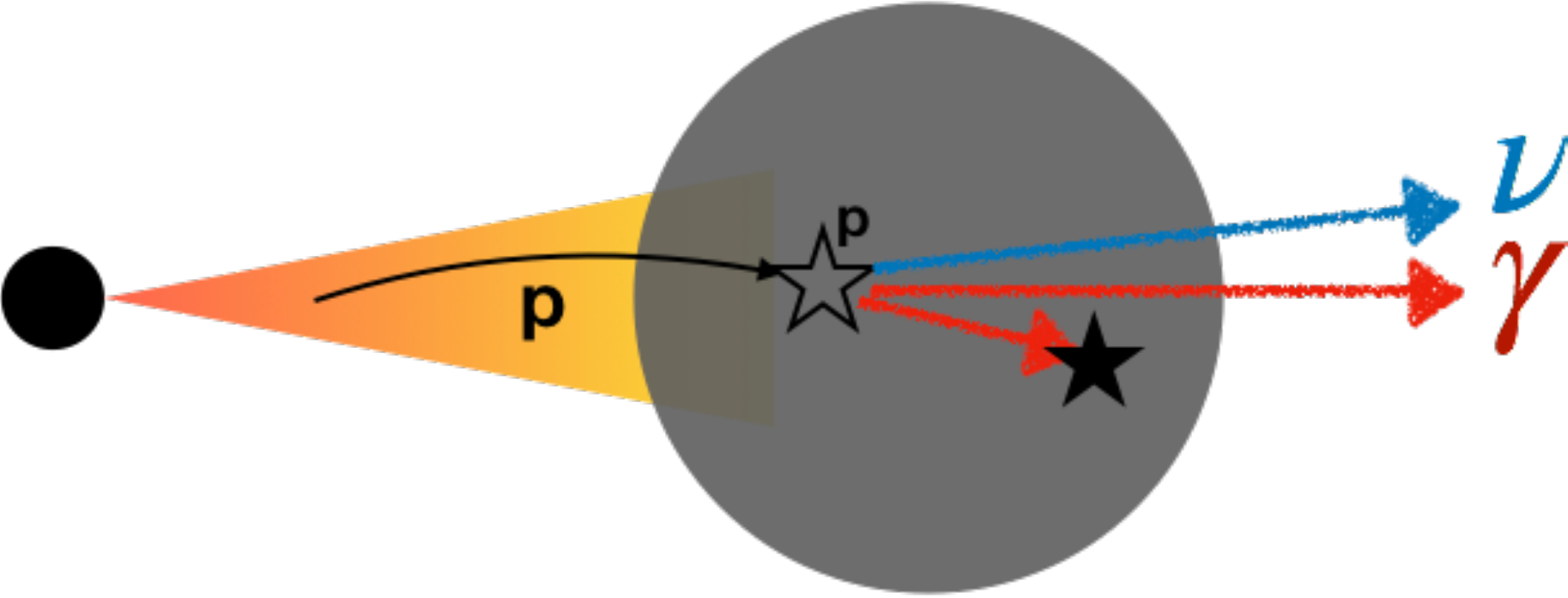}
  \caption[Our model for neutrino production on boscured
  sources]{Schematic representation of the model we consider for
    neutrino and gamma-ray production from $pp$-interactions of cosmic
    rays with a dense gas cloud near the cosmic-ray source acting as a
    beam dump. Due to the high integrated column density of the gas
    cloud, the produced gamma rays are also attenuated by the same
    cloud through Bethe-Heitler pair production. The relative size of
    the cloud and the source/outflow can vary. Note that while the
    figure features a jet, this is not a requirement for this
    mechanism to work and we do not initially assume its existence in
    our calculations.}
  \label{fig:setup}
\end{figure}

In our model, we consider the setup shown schematically in
Figure~\ref{fig:setup}. A source of accelerated cosmic rays is
obscured from our line of sight by a sufficiently large and dense
cloud of gas near to the source, which acts as a beam dump for
accelerated cosmic rays. The cosmic rays interact hadronically with
the gas in $pp$, $pA$, or $AA$-interactions, although we will only
consider $pp$-interactions. In these interactions, neutrinos and gamma
rays are produced. The gas cloud is sufficiently thick that the gamma
rays can interact again with the remaining part of the gas column
after their production, undergoing Bethe-Heitler pair
production~\cite{Bethe:1934za}. Therefore, the source can be hidden
(or at the very least obscured) in gamma rays.

The amount of matter present in the gas cloud is expressed by the
equivalent hydrogen column density $N_H$, defined as the line-of-sight
integral of the hydrogen density $n_H$
\begin{equation}
  \label{eq:NHdef}
  N_H = \int\diff l\, n_H(l),
\end{equation}
and denotes the amount of hydrogen atoms per cm$^2$.  The column
density required for the gamma-ray attenuation at GeV energies and
above to be significant can be estimated from the Bethe-Heitler pair
production cross section with matter. Using the approximate value
$\sigma_{\rm BH} \approx 20\mathrm{~mb} =
2\times10^{-26}\mathrm{~cm}^2$, we see that we require a column
density of approximately
\begin{equation}
  \label{eq:NHestimate}
  N_H = 5\times10^{25}\mathrm{~cm}^{-2}.
\end{equation}
While such a column density is rather high, we will motivate its use
in Section~\ref{sec:motiv-high-column}. In this work, we explore
the possibility that neutrinos are produced in highly obscured
sources, investigating two benchmark values of the column density:
$N_H^{(1)}=5\times10^{25}$~cm$^{-2}$, which is easily motivated, or
the higher value $N_H^{(2)}=10^{26}$~cm$^{-2}$.

As a consequence of this high column density, (nearly) all cosmic rays
will interact with the gas before traversing the cloud, since the
proton-proton cross section $\sigma_{pp}$ (starting at $\sim 30$~mb
and rising with energy, see e.g.\ the analytic expression
in~\cite{Kelner:2006tc}) is a few times higher than the Bethe-Heitler
pair production cross section $\sigma_\mathrm{BH}$. Therefore, such a
neutrino source would be a poor cosmic ray source, although similar
but unobscured sources would still make excellent cosmic ray
sources. The remaining gas column after interaction is sufficiently
thick for secondary protons to interact with the cloud again, giving a
slight boost to the total neutrino and gamma-ray flux, although this
contribution will turn out to be negligible.

While the sketch in Figure~\ref{fig:setup} shows an AGN-like scenario
featuring a jet, and we will apply the model to objects of this scale,
the mechanism is not restricted to this case and we do not a priori
assume the existence of a jet in our calculations. The only
requirement is that there exists a compact source of accelerated
cosmic rays obscured by a dense gas cloud near to the source. In
principle, such a model can feature both transient and continuous
neutrino production, although we restrict ourselves to continuous
emission and assume that the configuration is stable for a
sufficiently long time, where ``sufficiently long'' depends on the
specific case under consideration.

The beam dump model considered here should be contrasted with cosmic ray
reservoir models of neutrino production. There, cosmic rays have a
significant interaction probability with the target gas only by
integrating the gas density over the cosmic ray trajectory inside a
galaxy or cluster, since cosmic rays are confined in these
structures. Gamma rays, which are not confined, escape
unattenuated. Instead, in our model discussed here, the interaction
happens close to the source with a thin, but dense target (i.e.\ the
cosmic rays are not confined to this region),
such that the cosmic rays and gamma rays traverse
the same gas column. On the other hand, cosmic ray accelerator models
with a similar configuration (e.g.~\cite{Nellen:1992dw,Tjus:2014dna})
typically feature lower column densities, such that the gamma ray flux is either
unattenuated or attenuated only by invoking a strong radiation field.

Another model with more similarities to our scenario but of a
different scale is one of the microquasar SS433, where a supergiant
star feeds a $10$~M$_\odot$ black
hole~\cite{Reynoso:2007us,Reynoso:2008nk,Reynoso:2019vrp}. Inside the
jet, accelerated particles undergo $pp$-interactions with cold matter
and produce both neutrinos and gamma rays. In this model, attenuation
of gamma rays by both Bethe-Heitler pair production and photomeson
production are taken into account, on top of the more standard
$\gamma\gamma$-annihilation.

Typical neutrino models consider sources from which strong non-thermal
emission, usually X-ray or gamma ray, has been observed. Instead, our
model features objects with obscured gamma- and X-ray
emission. Therefore, strong sources can only be selected at radio or
infrared wavelengths, where the presence of an accelerated particle
population can also be inferred. The phenomenology of our model also
differs from that of the hidden sources in~\cite{Murase:2015xka},
where strong X-ray emission is required in order to attenuate the
gamma rays.

\subsection{Cloud dynamics}
\label{sec:cloud-dynamics}


As already mentioned, often our source will feature a jet which is
impacting the gas cloud. While we assume that this configuration is
stable for a sufficiently long time, either on the observing timescale
or a relevant period in the cosmological history of the source,
eventually such a jet will break up the configuration if it is
sufficiently strong at the location of the cloud. The exact physics
depends on the location of the cloud, its total mass and the strength
of the jet. A full study of this is outside the scope of this
work. However, similar studies exist in the literature, which we
briefly discuss here.

The interaction of a cloud of cold gas with a jet\footnote{Note that
  the jet and cloud have different properties: a jet is a Poynting
  flux (i.e.\ dominated by radiation) or extremely relativistic matter,
  while a gas cloud is dense and cold matter.} was studied
in~\cite{2010MNRAS.405..821P,2010A&A...522A..97A}. Due to the pressure
of the jet, some or all of the cloud material can be blown away and
swept up by the jet. Initially, the interaction will cause shocks in
both the jet and cloud material, which can serve as a potential site
for particle acceleration. As the material is swept up by the jet,
eventually it will spread out and move along with the jet, at which
point significant interactions between the cloud and jet
cease. Depending on the kinetic energy of the jet and the mass of the
cloud, the jet might be slowed down.


Several scenarios have been studied in the literature, often with a
cloud or a star interacting with an AGN jet. Due to the scale of the
cloud/star compared to the jet (i.e.\ typically much smaller than the
jet at entry), these scenarios feature transient emission and have
been invoked to explain gamma-ray flares in
blazars~\cite{Dar:1996qv,2010A&A...522A..97A}. These models differ
from the scenario we envision mostly due to the scale of the cloud,
since we focus on a case where a possible jet is obscured for an
extended amount of time\footnote{In this sense, our model has some
  similarities with low-luminosity gamma-ray burst models, where the
  jet is stopped by a cocoon of matter~\cite{Senno:2015tsn}.}.

Jet-cloud/star models have been used to explain the neutrinos from the
blazar TXS~0506+056.  The bright gamma-ray flare associated with a
high-energy neutrino has been modelled using clouds with
$N_H >10^{24}$~cm$^{-2}$ present in the broad-line region of the
blazar~\cite{Liu:2018utd}. The electromagnetic cascade initiated by
the $pp$-interactions (modelled using the analytical fits to the
kinematic distributions of final state particles) ionises the cloud,
which becomes optically thick for optical to X-ray photons while the
gamma rays and the neutrinos escape. The SED at lower frequencies can
then be explained with a second leptonic emission zone. This scenario
can explain the observed neutrino and electromagnetic emission with
moderate proton luminosity, in contrast with $p\gamma$-models which
have a low interaction rate in order to avoid efficient
$\gamma\gamma$-annihilation and therefore require high proton
luminosity.  The neutrino flare during the quiescent state of the
blazar has been explained with unbound layers from a tidally-disrupted
red giant\footnote{A red giant is necessary since only there the outer
  layers are sufficiently weakly bound to be blown away and interact
  with a significant portion of the
  jet~\cite{2012ApJ...755..170B}.}~\cite{Wang:2018zln}. In this case,
the target has a column density of $N_H=5\times10^{25}$~cm$^{-2}$, the
same as our first benchmark point.  Again, the cloud becomes ionised
and is optically thick to X-rays, but now the Comptonised radiation
field is also sufficiently thick to efficiently attenuate gamma rays
through $\gamma\gamma$-pair production.

There are additional models featuring jet-cloud/star interactions
built only to explain gamma-ray emission. A model of jets interacting
with a BLR cloud of lower density than those above was investigated
in~\cite{delPalacio:2019mql}. In~\cite{Khangulyan:2013xxa}, particle
acceleration is induced by a strong star wind and loss of mass of a
star in an FSRQ jet.
Another study investigates M87 TeV flares from jet-cloud interaction which accelerates particles
and leads to $pp$-interactions~\cite{2012ApJ...755..170B}.
Finally, in an alternative scenario for orphan flares, such flares are caused by
the interaction of accelerated particles with star radiation inside jet
blobs~\cite{Banasinski:2016kpe}.

\subsection{High column densities}
\label{sec:motiv-high-column}


The column densities required for the gamma-ray attenuation to be
significant are high compared to typical astrophysical environments.
The average column density of our own galaxy is $N_H^{\rm
  gal}\sim10^{20}$~cm$^{-2}$, driven mainly by the
interstellar medium and not the compact objects (stars) within. On the
other hand, Earth's atmosphere has a column density of
$10^3$~g\,cm$^{-2}$ or $N_H\sim10^{26}$~cm$^{-2}$, similar to the
column density we are considering.
Still, the latter is a compact object and thus not a good comparison
for the viability of our scenario,
although it is more
appropriate as a comparison for the jet-cloud/star scenario. In order
to have longer lasting neutrino production and obscuration, a more
extended gas cloud is necessary.

Though rare, astrophysical environments with a gas column of
the required magnitude do exist.  As already mentioned in the previous
section, models of AGN jets interacting with dense clouds possessing
column densities up to $5\times 10^{25}$~cm$^{-2}$ have been invoked
to explain the TXS~0506+056 neutrino and gamma-ray flares. More
generally, supermassive black hole in AGNs are surrounded by gas
and dust in the broad-line region and torus (for a review,
see~\cite{Bianchi:2012vn,2017NatAs...1..679R}). In case an AGN is
observed edge-on, the central engine is frequently hidden from our
view by the dusty torus. The column density of this torus varies and
an AGN is generally considered obscured when $N_H \geq 10^{22}$~cm$^{-2}$.
Below $N_H\sim10^{24}$~cm$^{-2}$, the photon attenuation cross section
in the $2$--$10$~keV
X-ray regime
has the correct magnitude to probe
this column density\footnote{More specifically, the main indicator is the
  photo-electric cut-off, induced by the sharply rising cross section
  of photo-electric absorption towards low energies.}.
If the column density is higher than the inverse of the Thompson cross
section, $N_H \geq 1.5\times10^{24}$~cm$^{-2}$, the AGN is called
Compton-thick~\cite{Comastri:2004nd}.
At these values, the density can be probed by hard X-rays ($E_\mathrm{X-ray}\geq 10$~keV), where
Compton scattering dominates.
However, for densities above $N_H = 10^{25}$~cm$^{-2}$
(heavily Compton-thick), the X-ray emission is suppressed even above
$10$~keV, since the
photons are down-scattered by Compton interactions and subsequently absorbed.
Other (indirect) methods need to be used to probe these obscuring
columns (e.g.\ through reflected X-rays).
In this way, Compton-thick AGNs have been found with column densities
exceeding
$N_H = 10^{25}$~cm$^{-2}$~\cite{Comastri:2004nd,Ricci:2016lbt}.
AGNs obscured by column densities significantly higher than
$10^{25}$~cm$^{-2}$ are suspiciously missing from surveys. From the above
discussion, it is clear that there is an observational bias against
such highly obscured objects.

The Chandra X-ray Observatory~\cite{2000SPIE.4012....2W} detects
X-rays between $0.1$-$10$~keV and has been used to detect obscured
AGNs. An analysis fitting AGN spectra in Chandra Deep Fields with physical
models, finds many highly obscured AGNs~\cite{Li:2019zqc}.
While their analysis corrects for observational bias, they are unable
to constrain the number of AGNs with $N_H\geq 10^{25}$~cm$^{-2}$,
since these sources are missed in their sample and therefore the
missing number of sources cannot be determined. As such, sources of
this kind may contribute to the heavily obscured AGN population. This
agrees with earlier results presented in~\cite{Risaliti:1999yd}. Other
analyses do find multiple objects with $N_H\geq 10^{25}$~cm$^{-2}$,
even up to $N_H\sim 10^{26}$~cm$^{-2}$ in Chandra
surveys~\cite{Brightman:2014pva}. Another analysis studies torus model
properties~\cite{2019arXiv190403694G} with an ultra-hard X-ray sample
($14$--$195$~keV) of Seyfert galaxies from
Swift/BAT~\cite{Krimm:2013lwa}, which can identify more strongly
obscured objects.  They find that even from such a selection, a
population of the most obscured objects is still missing, agreeing
with~\cite{Ricci:2016lbt,2016ApJ...825...85K,Mateos:2017rsi}.

An example of such an obscured source is NGC~4418, a luminous infrared
galaxy (LIRG). It has a core bright in IR along with the deepest known
silicate absorption, but it has not been detected in
X-rays~\cite{2015MNRAS.449.2598R,2019arXiv190403694G}. Its inferred
column density is $N_H > 10^{25}$~cm$^{-2}$, with a spectrum
consistent with AGNs as the main power source and showing similarities
with ARP 220~\cite{Sakamoto:2013py} (which will end up as a source in
the analysis of Section~\ref{sec:obsc-flat-spectr}). Another analysis
finds that an AGN is only allowed in case the column density exceeds
this same value~\cite{Costagliola:2013tea}.

While the objects above do have strong obscuration, this does not yet
mean that the obscuring material is bombarded by cosmic rays or blocks
a jet. However, there exist several models of tilted tori that envision such
a scenario~\cite{Bianchi:2012vn,2010arXiv1002.1759L}. We show that
such configurations provide interesting sites for neutrino production
through the model proposed in this paper.

It follows that, while extreme, column densities
$N_H > 10^{25}$~cm$^{-2}$ in astrophysical environments are very
likely to occur. Therefore, our lower benchmark value
$N_H^{(1)}=5\times10^{25}$~cm$^{-2}$ for obscured neutrino sources is
compatible with the conventional view of various astrophysical
objects. On the other hand, values much higher than this have not been
observed, such that we consider our second benchmark
$N_H^{(2)}=10^{26}$~cm$^{-2}$ as a more extreme, but still realistic,
case.

\subsection{Photon attenuation}
\label{sec:photon-atten}

In this section, we discuss the attenuation of photons at the source,
in particular X-rays and gamma rays, since photons at these energies
are sensitive to the presence of surrounding matter and their
appearance is linked to high-energy particle acceleration and
interactions. Since attenuation can be both due to matter and
radiation, we discuss these separately.



\subsubsection{Attenuation by matter}
\label{sec:attenuation-matter}

The interaction of X-rays and gamma rays with matter, which is the
main attenuation channel in our model, happens through different
processes. The photo-electric effect is important for soft X-rays
traversing an unionised medium. For hard X-rays propagating through
any medium, Compton scattering is important.  Finally, for gamma rays,
Compton scattering dominates at MeV energies, while at GeV energies
pair production on either the electrons or the nuclei is dominant. The
strength of these interactions depends on the energy of the photons
and the target composition. For GeV gamma rays, which will be
considered in this work\footnote{In order to properly model X-rays
  propagating through a medium with Compton scattering, multiple
  scattering needs to be taken into account, in particular when the
  densities are high. This requires the use of specialised codes, such
  as \texttt{XSPEC}~\cite{1996ASPC..101...17A}.}, the only relevant
process is pair production, either on the nuclei or on electrons.

The target gas is expected to have a composition similar to the
interstellar medium, which is dominated by
hydrogen~\cite{Anders:1989zg}. The hydrogen equivalent column density
reported by astronomers, derived from spectral modelling, takes this
composition into account.  The total amount of matter in the column is
then given by summing over all elements, with a density equal to $N_H$
times the abundance $a_Z$ of that element relative to hydrogen
\begin{equation}
  \label{eq:11}
  N = \sum_Z a_Z N_H,
\end{equation}
see e.g.~\cite{2015ApJ...800...29G}, and the absorption models included in
\texttt{XSPEC}~\cite{1996ASPC..101...17A}.
However, given the dominance of hydrogen, we assume that the target is
a pure proton gas when considering the interactions with accelerated
particles\footnote{For neutrino and gamma-ray production, the
  additional mass (increase by a factor of $\sim 1.4$) due to the
  heavier elements can be included by rescaling $N_H$, see for
  example~\cite{Mannheim:1994sv,Tjus:2014dna}. However, for the
  purpose of our work, we do not explicitly take into account this
  small numerical factor.}.
For modelling the X-ray attenuation, this introduces a rather large
error, since both photo-electric absorption and Compton scattering are
strongly dependent on the target element. For our model, however,
where we are mainly interested in the gamma-ray attenuation, this is a
reasonable assumption, which we explicitly checked using attenuation
cross sections retrieved from the \texttt{XCOM} database~\cite{XCOM}.


The gamma-ray attenuation is dominated by Bethe-Heitler pair production~\cite{Bethe:1934za},
the expression for which can be given as a series expansion in~\cite{1992ApJ...400..181C} (see also
the review~\cite{Ruffini:2015oha} and an earlier treatment
in~\cite{1990ApJ...362...38B}). The threshold for pair production is
$E^{\rm thres.}_{\rm BH} = 2 m_e c^2$.
For the full pair production cross section $\sigma_{\rm BH}(E_\gamma)$ (on proton \emph{and}
electron), we use this expression multiplied by $2$. Since the cross
section rises only logarithmically with the photon energy, the cross
section is approximately constant with a value of
$\sigma_{\rm BH}\approx 20$~mb. This validates our estimation of the required
column density $N_H$ in Eq.~\eqref{eq:NHestimate}.

For high-energy gamma rays in matter, the interaction cross section
with matter starts to decrease due to the
``Landau-Pomeranchuk-Migdal'' (LPM)
effect~\cite{Landau:1953um,Landau:1953gr,Migdal:1956tc,haar2013collected,Klein:1998du}
(see also the PDG review~\cite{Tanabashi:2018oca}). This effect is due
to destructive interference between amplitudes from different, nearby,
scattering centres. However, for an astrophysical gas cloud, the
density is too low for this effect to be relevant, even if the
integrated column density is high.

Another possible source of gamma-ray attenuation would be through
photohadronic $p\gamma$-interactions. However, the cross section for
this process is much lower than for pair production\footnote{This is
  the reason that $p\gamma$-interaction models require strong
  radiation fields (leading to gamma-ray attenuation) and much larger
  proton luminosities than $pp$-interaction models.} and we will
therefore neglect it (see also the related discussion in
Section~\ref{sec:addit-modell-assumpt}).


\subsubsection{Attenuation by a radiation field}
\label{sec:atten-radi-field}

Gamma rays can also interact with a radiation field at the source and undergo pair
production if the centre of mass energy exceeds $2m_e$, attenuating
the gamma-ray flux. The optical depth to pair production depends on
the number density of photons (i.e.\ the energy density of the
radiation field). We can ignore this attenuation channel if the
radiation fields at the
cloud are weak enough for $\gamma\gamma$-pair production. In
case of a strongly radiating source (as will usually be the case),
this means that the cloud can not be too close to the source.
Otherwise, pair production on radiation will further decrease the
gamma-ray flux.

Even if there is not a sufficiently strong radiation field present at
first, it can be generated by the particle cascade initiated in the
$pp$-interactions. Depending on the density and location of the gas
cloud, the gamma rays and $e^\pm$-pairs produced in pion decay can
initiate an electromagnetic cascade through repeated creation of
photons through synchrotron radiation, bremsstrahlung, and
inverse-Compton scattering and the creation of $e^{\pm}$-pairs through
pair production. As a result, the cloud can become fully ionised and
the photon field can become Comptonised. The cloud is then optically
thick to X-rays (due to the free electrons) and the generated photon
field can act as a target for gamma-ray pair production (see
e.g.~\cite{Wang:2018zln}), decreasing the gamma-ray flux.

Since $\gamma\gamma$-attenuation introduces a stronger model
dependency, we do not take this additional attenuation channel into
account. In that sense, the amount of gamma-ray obscuration we obtain
in our model is conservative.



\subsection{Model assumptions}
\label{sec:addit-modell-assumpt}

In addition to intrinsic features of our model, we also make
some further assumptions in order to reduce the parameter space. The
cosmic-ray composition is considered to be pure proton, which is a
reasonable approximation in the energy range relevant for IceCube.
As already mentioned in Section~\ref{sec:photon-atten}, the target gas
is approximated as pure hydrogen.
Therefore, we can model the interactions as pure
$pp$-collisions, without complications from nuclear effects.

We do not include additional sources of neutrinos or gamma rays, i.e.\
$p\gamma$-interactions and leptonic processes, since these are highly
model dependent. Typically, $p\gamma$-interaction models require
higher proton luminosities than $pp$-models to obtain sizeable
fluxes\footnote{This is due to their much lower cross section, which
  might not be compensated completely by the high target density of a
  radiation field}. Therefore, if $pp$-interactions are present, they
can be expected to be dominant compared to
$p\gamma$-interactions. Additional gamma rays can be produced in
leptonic processes, such as through inverse-Compton scattering. Both
$p\gamma$-interactions and leptonic processes increase the total
gamma-ray flux. This increase is counterbalanced by attenuation
through $\gamma\gamma$-pair production on the intense radiation fields
present inside the cloud (which we do not take into account), reducing
the total gamma-ray flux. In addition, if the latter process is not
dominant, it is safe to ignore photomeson production, since its
interaction rate is roughly three orders of magnitude smaller than
$\gamma\gamma$-annihilation for the same photon
field~\cite{2012ApJ...755..147D,Murase:2015xka,Liu:2018utd}.

We do not take into account synchrotron losses of the muons, pions and kaons.
However, for the scenario we envision, these losses should be negligible: while
the obscuring gas should be close to the source, it can not be too close,
otherwise we need to take into account the effect of strong radiation fields on
the gas. Therefore, if the gas is sufficiently far removed from the source, one
can expect the synchrotron losses due to the magnetic fields to be negligible.
Indeed, following the approach of~\cite{Hummer:2010ai}, we can estimate above
which critical energy these losses become dominant (i.e.\ the timescale
associated to synchrotron losses is shorter than the decay time), for an AGN jet
scenario with the gas cloud at parsec scales. We find that this
energy can easily exceed $10^6$~GeV even for the magnetic fields associated to
jets, using the magnetic fields found in~\cite{2009MNRAS.400...26O} or higher;
if no (strong) jet is present at the location of the cloud, magnetic fields are
expected to be even weaker at the parsec scale.

Unless otherwise stated, the protons are assumed to follow an
$E^{-2}$-spectrum, consistent with Fermi acceleration. However, this
immediately implies that the predicted neutrino spectrum also follows
an $E^{-2}$-spectrum, while IceCube observes a softer
spectrum~\cite{Aartsen:2017mau}. Therefore, the computation here can
only serve to explain the observed neutrino flux above $\sim100$~TeV,
while there must be a second, softer, component below this energy that
we do not model.

With these assumptions and the choice of two benchmark values of
$N_H$, the only free parameters left are the normalisation of the
injected proton flux (used in Section~\ref{sec:obsc-flat-spectr}) or
the resulting neutrino or gamma-ray flux (used in
Section~\ref{sec:diffuse-flux}) and the energy range in which the
protons are injected. The maximum energy of the protons will be fixed
at $10^8$~GeV, which is sufficient to explain the neutrinos observed
by IceCube without violating the limits at the highest energy.  The
minimum proton energy is determined by the acceleration mechanism. In
the case of shock acceleration, one typically has
$E_p^{\rm min}\sim\Gamma m_p c^2$~\cite{Waxman:1997ti}, with $\Gamma$
the Lorentz factor of the shock, which is the minimum energy with
which the particles can efficiently participate in the acceleration
process. Sometimes, also the value $E_p^{\rm min}\sim\Gamma^2 m_p c^2$
is used~\cite{Murase:2015xka}. Since we will consider AGNs as the
central engine in the following, with
$\Gamma\sim10-30$~\cite{1992ApJ...387..449P},
we take $E_p^{\rm min}=10^2$~GeV.
For an $E^{-2}$-spectrum, the final result is not very sensitive to
the exact value of the minimum energy (as long as it is around the
same order of magnitude), since the luminosity in this case is divided
equal per decade of energy. More concretely, we have
\begin{equation}
  L \propto \int_{E_{\rm min}}^{E_{\rm max}}\diff E\, E E^{-2} =
  \ln\left(\frac{E_{\rm max}}{E_{\rm min}}\right),
\end{equation}
resulting in only a logarithmic dependence on the minimum
energy. Nevertheless, varying the minimum energy with more than an
order of magnitude has a significant effect on the total normalisation
of the flux.  More importantly, if the spectral index deviates from
$2$, the luminosity quickly becomes very sensitive to the minimum
energy, see also the discussion in~\cite{Merten:2017mzg} and
Section~\ref{sec:normalising-spectrum}. Since cosmic ray experiments
are only sensitive to the maximum energy of extragalactic cosmic rays
(below the knee galactic cosmic rays dominate), this is an important
source of uncertainty. However, the choice we make above is
theoretically well motivated.

\section{Calculating the $\nu$ and $\gamma$-ray flux}
\label{sec:neutrino-production}

In this section, we describe the method for calculating the neutrino
and gamma-ray flux for our model of neutrino production in obscured
sources, before applying it to specific scenarios in the following
sections. We perform the calculation in two ways. For the first one,
we use analytical fits of the neutrino and gamma-ray spectrum from
$pp$-collisions. For the second one, we use a full Monte Carlo
simulation for the $pp$-interactions. For our final results, we use
the Monte Carlo simulation, since it includes more details, while the
analytical method serves only as a consistency check on our results.

The analytical calculation of neutrino and gamma-ray spectra from
cosmic ray interactions with a cloud of integrated column density
$N_H$ follows the method of~\cite{Kelner:2006tc}. These authors
provided analytical fits to the neutrino and gamma-ray spectra using
the meson spectra simulated with the Monte Carlo generators
\texttt{SIBYLL}~\cite{Fletcher:1994bd} and
\texttt{QGSJET}~\cite{KALMYKOV199717}, with their subsequent decay to photons,
neutrinos and electrons treated analytically. Note that using
these fits, we have no access to the secondary proton spectrum and
therefore can not implement the interactions of secondary protons. For
simplicity, and because this calculation serves only as a sanity
check, we implement the gamma-ray attenuation using the full column
density of the cloud, ignoring that the gamma rays are created
somewhere along the column and do not see the entire gas column.

These analytical fits are accurate and faster than performing
Monte Carlo simulations. Nevertheless, a full Monte Carlo simulation
was performed in order to model the neutrino production in more
detail. The two main reasons are that this allows us to use an updated
Monte Carlo generator for performing the $pp$-interactions and that in
this way we have access to secondary protons, which can interact again
with the gas column. We implemented our model with the Monte Carlo generator
\texttt{SIBYLL}~2.3~\cite{Ahn:2009wx}, which is an updated version of
the code used for the analytical fits above. In
particular, it includes the contribution from charmed meson
decay~\cite{Engel:2015dxa,Riehn:2015oba}, although
this does not noticeably impact our results\footnote{This is in
  contrast with the result for atmospheric neutrinos, where charmed
  meson decay produces a distinct, harder spectrum. However, the
  reason for this difference is that in atmospheric neutrino
  production there is a competition between the decay of the meson and
  its interaction with an air nucleus. The latter produces softer
  spectra, since it initiates a new cascade. Because charmed mesons
  have a shorter lifetime, their contribution produces a harder
  spectrum. In astrophysical scenarios, this competition is not
  present, as also mentioned in~\cite{Nellen:1992dw}.}.

In the simulation, protons are propagated through a matter column of
specified integrated density $N_H$ and allowed to interact using
standard Monte Carlo techniques\footnote{See e.g.\ the
  \texttt{GEANT4}~\cite{Agostinelli:2002hh} physics reference
  manual.}.  Initially, in order to build up sufficient statistics at
high energies, we inject a proton with an energy which is drawn from a
power law distribution following $E^{-1}$. Afterwards, events are
reweighted to the distribution under study ($\propto
E_p^{-2}$ unless otherwise stated).

The mean free path of a proton of energy $E_p$ under
$pp$-interactions propagating through a medium with density $n$ is
given by
\begin{equation}
  \label{eq:mfp}
  \lambda(E) = \frac{1}{n \sigma(E_p)},
\end{equation}
which is determined using the cross section tables calculated by
\texttt{SIBYLL}. The interaction point of a proton can be determined
by sampling from the distribution
\begin{equation}
  \label{eq:mpfdist}
  P(n_r < n_\lambda) = 1 - e^{-n_\lambda}.
\end{equation}
If this point is beyond the total depth of the gas column at the
respective proton energy $n_{\lambda,{\rm tot}}(E_p)= N_H
\sigma(E_p)$, the proton escapes and is saved in the final
output. Otherwise, a collision is performed using \texttt{SIBYLL} and
the final state particles $\nu_\alpha$, $\bar{\nu}_\alpha$, $e^\pm$
and $\gamma$ are saved, while secondary $p$, $n$ and their
antiparticles are allowed to interact again with the remaining
column\footnote{For simplicity, the cross section of interactions with
  protons is put equal to the proton-proton cross section for all
  these particles (i.e. also for $n$ and anti-$p$/$n$), which is a
  good approximation at high energies.}. These secondary interactions
have a minor effect. This is a simple consequence of the power law
proton spectrum: the $N$ secondary protons produced by a proton of
energy $E_p$ carry on average a fraction $x$ of the parent proton
energy and are dominated by the primary protons at the lower energy $x
E_p$, which are more numerous by a factor $\frac{\left(x
  E_p\right)^{-2}}{N E_p^{-2}}=\frac{1}{x^2N}$. Since the sum of all
secondary energies (including leptons and gamma rays) needs to total
$E_p$, we have $xN<1$ and $x<1$, so the primary protons are indeed
dominant.

The decay of pions and other mesons to neutrinos is performed by the
\texttt{SIBYLL} decay routines. In order to obtain a better accuracy,
these decay routines are often replaced by analytical calculations or
by interfacing the output to other codes like
\texttt{Pythia}~\cite{Sjostrand:2014zea}. However, these inaccuracies
are mainly important for air shower simulations, where the full
particle spectrum of individual events needs to be well modelled. For
our purposes, where we only care about the total neutrino spectrum,
the \texttt{SIBYLL} routines should suffice. Neutrons are considered
stable in this simulation (the additional neutrinos from their decay
outside of the source are at low energy, since most of the energy goes
towards to resulting proton).

The attenuation of photons, taking into account only the remaining gas
column, is taken into account by reweighting each photon
by\footnote{Note that by using this formula, we implicitly reduce the
  simulation to a one-dimensional one, ignoring the photon momentum in
  the direction perpendicular to the initial proton direction. While
  taking this into account would increase the gas column seen by the
  photon in the case of an infinite ``plane'' of gas, this correction
  is minor due to the beaming. Moreover, taking into account a more
  realistic geometry than a flat infinite plane would decrease the
  column slightly.}
\begin{equation}
  \label{eq:wphoton}
  w_\gamma(E_\gamma, E_p) = e ^{- N_H\, \sigma_{\rm BH}(E_\gamma)\left(1 - \frac{n_\lambda}{n_{\lambda,{\rm tot}}(E_p)}\right)},
\end{equation}
where $n_\lambda$ is the
number of mean free paths the proton travelled before interacting
(i.e. where the photon is produced) and
$n_{\lambda,\ \mathrm{tot}}(E_p)$ is the total number of proton mean
free paths of the cloud for the energy of the parent proton.

Finally, the overall normalisation
of the produced spectra is determined in different ways, depending on
the scenario under consideration.

\section{Obscured flat-spectrum radio AGNs}
\label{sec:obsc-flat-spectr}

In this section, our
model is applied to a set of AGNs selected for their possible
obscuration by matter. We first review the object selection~\cite{Maggi:2016bbi} which we
use. Next, we discuss our normalisation, based on the measured radio
flux from these AGN. Finally, we show the results and, using
existing limits on their neutrino emission from
IceCube~\cite{Aartsen:2017kru}, derive limits on the cosmic ray
content of these sources.

The original motivation for this object selection is the possibility
of having AGNs with tilted
tori~\cite{Bianchi:2012vn,2010arXiv1002.1759L}. In this case, the jet
can penetrate the dust torus, which has a considerable integrated
column density, and produce neutrinos efficiently while being obscured
in X-rays.  However, the exact origin of the gas cloud is not
important for the details of our calculation.  We assume the cloud is
stable on observation timescales (i.e.\ at least years). Due to the
strong electromagnetic radiation from the jet, the cloud becomes
ionised~\cite{Maggi:2016bbi}. At the same time, these radiation fields
also attenuate the gamma rays through $\gamma\gamma$ pair production,
which we do not model. The gamma ray flux predicted is therefore an
upper limit on the hadronic gamma rays. On the other hand, leptonic
processes might create additional gamma rays. However, for this
application, we are only interested in the neutrino emission from
these objects and do not model their complete SED.

\subsection{Selected objects}
\label{sec:selobj}

We investigate the obscured flat-spectrum radio AGN selected in~\cite{Maggi:2016bbi}, which
targets nearby sources which are candidate cosmic ray accelerators,
feature beamed emission (i.e.\ a jet, so that all emission is boosted
towards Earth) and which exhibit signs of obscuration by matter by
selecting those objects with a lower-than-expected X-ray luminosity,
relative to the radio luminosity. Whereas X-rays are hindered by gas,
radio waves can propagate through gas unimpeded. Moreover, radio
emission is usually explained by synchrotron emission from a
non-thermal population of electrons. Therefore, the radio emission
from these sources characterises the strength of the inner
engine. Under the assumption that the reduced X-ray emission relative
to the radio emission is (mainly) due to attenuation by matter, these
objects are potential strong neutrino sources through
$pp$-interactions (in addition to the $p\gamma$-interactions, which we
ignore).

The study in~\cite{Maggi:2016bbi} first considers an unbiased set of nearby
cosmic ray source candidates which exhibit signs of beamed emission
towards Earth. Assuming that their original set consists of generic
sources and that the observed spread in X-ray intensity relative to
the radio intensity is due to the presence of gas and dust, they
select the 25\% weakest sources as heavily obscured, retaining 15
objects.


In order to estimate the fraction of interacting protons from the
X-ray obscuration, we need to know the column density of the obscuring
matter. This can be determined from the observed X-ray flux if we know
which process is responsible for X-ray attenuation. At higher X-ray
energies, this is always Compton scattering (which we already
discussed in Section~\ref{sec:photon-atten}), but at $1.24$~keV
photo-electric absorption dominates if the electrons are still bound
in atoms. Due to the strong radiation of these sources, however, the
obscuring gas cloud is completely ionised for natural
geometries. Indeed, following the analysis in~\cite{Maggi:2016bbi},
for natural values of the cloud thickness and distance from the
central engine, i.e.\ between $1$ and $10$ pc, and for typical values
of AGN luminosity and ionisation, the cloud will be fully ionised due
to Compton scattering.  In this case, the column density leading to an
obscuration in X-rays of $\sim 90\%$ is
$N_H\sim10^{26}$~cm$^{-2}$. This also corresponds to the required
column density for significant neutrino production to occur, since
Compton scattering on hydrogen and proton-proton interactions have a
similar interaction cross section. Using the Compton scattering cross
section, the fraction of interacting protons derived
in~\cite{Maggi:2016bbi} for the objects in the selection above varies
between $0.80$ and $0.99$, which corresponds to column densities
$N_H\sim5\times10^{25}$--$10^{26}$~cm$^{-2}$, exactly those we
consider in our model.

An IceCube analysis was done for 14
objects~\cite{Aartsen:2017kru}. Two objects from the original
selection above were not analysed since they are located in the
southern sky where IceCube has a lower sensitivity. On the other hand,
NGC~3628 was added back in the IceCube analysis, since it was omitted
from the original selection for its lack of X-ray emission above
background, making it, however, an interesting target.
The IceCube analysis finds no significant signal and gives an upper limit on
the $E^2\Phi$-flux for each of these objects, assuming an $E^{-2}$-flux
between $1$~TeV and $1$~PeV. The final list of objects, their
classification and the limits on their neutrino emission are shown in
Table~\ref{tab:selectedobj}.

Finally, it is important to remark that since almost all objects in
this selection are a subset of blazars, it is unlikely that sources
from this class (i.e.\ ``obscured blazars'') are responsible for the
bulk of the neutrino flux. However, this selection is interesting for
two reasons. First, given the luminosity of these objects, they are
good targets to test whether the model under consideration occurs in
nature. Second, if the scenario is indeed applicable to these objects,
the neutrino flux thus produced allows us to directly probe the amount
of accelerated hadrons in these blazars.

\begin{table*}
\centering
\caption[Final objects in the obscured AGN selection]{Final objects in
  the obscured flat-spectrum radio AGN selection for which an IceCube analysis exists and the upper
  limit on their neutrino emission in units of
  $10^{-9}$~GeV\,cm$^{-2}$\,s$^{-1} $, from~\cite{Aartsen:2017kru}.
}
\label{tab:selectedobj}
{
  \renewcommand{\arraystretch}{1.2}
  \setlength\tabcolsep{.2cm}
\begin{tabular}{lrrrHHHHHHHHHHHHHHHZ}
\hline \hline
Name & RA ($^\circ$) & Dec ($^\circ$) & $E^2\Phi_\nu^{90\%}$  &
$E^2\Phi_\nu^{\rm (1),~No~sec.}$ & $E^2\Phi_\nu^{\rm (1),~No~sec.}$ & $E^2\Phi_\nu^{(1)}$
& $E^2\Phi_\nu^{(1)}$ &
$E^2\Phi_\nu^{\rm (2),~No~sec.}$ & $E^2\Phi_\nu^{\rm (2),~No~sec.}$ & $E^2\Phi_\nu^{(2)}$
& $E^2\Phi_\nu^{(2)}$ &
$f_e^{\rm (1),~No~sec.}$ & $f_e^{\rm (1),~No~sec.}$ & $f_e^{(1)}$
& $f_e^{(1)}$ &
$f_e^{\rm (2),~No~sec.}$ & $f_e^{\rm (2),~No~sec.}$ & $f_e^{(2)}$
& $f_e^{(2)}$ \\

\hline
PKS1717+177 & 259.80 & 17.75 & 0.754 & 0.038 & 0.052 & 0.047 & 0.062 & 0.042 & 0.055 & 0.052 & 0.070 & $5.0\times10^{-3}$ & $6.8\times10^{-3}$ & $6.1\times10^{-3}$ & $8.1\times10^{-3}$ & $5.5\times10^{-3}$ & $7.2\times10^{-3}$ & $6.9\times10^{-3}$ & $9.2\times10^{-3}$\\ 
CGCG186-048 & 176.84 & 35.02 & 0.856 & 0.020 & 0.027 & 0.024 & 0.032 & 0.022 & 0.029 & 0.027 & 0.037 & $2.3\times10^{-3}$ & $3.1\times10^{-3}$ & $2.8\times10^{-3}$ & $3.7\times10^{-3}$ & $2.5\times10^{-3}$ & $3.3\times10^{-3}$ & $3.1\times10^{-3}$ & $4.2\times10^{-3}$\\ 
RGBJ1534+372 & 233.70 & 37.27 & 0.899 & 0.001 & 0.001 & 0.001 & 0.002 & 0.001 & 0.002 & 0.002 & 0.002 & $1.2\times10^{-4}$ & $1.6\times10^{-4}$ & $1.5\times10^{-4}$ & $1.9\times10^{-4}$ & $1.3\times10^{-4}$ & $1.7\times10^{-4}$ & $1.6\times10^{-4}$ & $2.2\times10^{-4}$\\ 
NGC3628 & 170.07 & 13.59 & 0.719 & 0.024 & 0.033 & 0.030 & 0.039 & 0.026 & 0.035 & 0.033 & 0.045 & $3.3\times10^{-3}$ & $4.5\times10^{-3}$ & $4.1\times10^{-3}$ & $5.4\times10^{-3}$ & $3.6\times10^{-3}$ & $4.8\times10^{-3}$ & $4.6\times10^{-3}$ & $6.1\times10^{-3}$\\ 
SBS1200+608 & 180.76 & 60.52 & 1.090 & 0.008 & 0.010 & 0.009 & 0.012 & 0.008 & 0.011 & 0.010 & 0.014 & $6.9\times10^{-4}$ & $9.4\times10^{-4}$ & $8.5\times10^{-4}$ & $1.1\times10^{-3}$ & $7.5\times10^{-4}$ & $1.0\times10^{-3}$ & $9.5\times10^{-4}$ & $1.2\times10^{-3}$\\ 
GB6J1542+6129 & 235.74 & 61.50 & 1.070 & 0.006 & 0.008 & 0.007 & 0.009 & 0.006 & 0.008 & 0.008 & 0.011 & $5.4\times10^{-4}$ & $7.3\times10^{-4}$ & $6.6\times10^{-4}$ & $8.7\times10^{-4}$ & $5.9\times10^{-4}$ & $7.7\times10^{-4}$ & $7.4\times10^{-4}$ & $9.9\times10^{-4}$\\ 
4C+04.77 & 331.07 & 4.67 & 0.650 & 0.043 & 0.058 & 0.053 & 0.069 & 0.047 & 0.062 & 0.059 & 0.079 & $6.5\times10^{-3}$ & $8.9\times10^{-3}$ & $8.0\times10^{-3}$ & $1.0\times10^{-2}$ & $7.1\times10^{-3}$ & $9.4\times10^{-3}$ & $9.0\times10^{-3}$ & $1.2\times10^{-2}$\\ 
MRK0668 & 211.75 & 28.45 & 0.879 & 0.096 & 0.130 & 0.117 & 0.155 & 0.105 & 0.138 & 0.132 & 0.176 & $1.0\times10^{-2}$ & $1.4\times10^{-2}$ & $1.3\times10^{-2}$ & $1.7\times10^{-2}$ & $1.1\times10^{-2}$ & $1.5\times10^{-2}$ & $1.4\times10^{-2}$ & $2.0\times10^{-2}$\\ 
3C371 & 271.71 & 69.82 & 1.180 & 0.149 & 0.202 & 0.182 & 0.240 & 0.162 & 0.214 & 0.204 & 0.274 & $1.2\times10^{-2}$ & $1.7\times10^{-2}$ & $1.5\times10^{-2}$ & $2.0\times10^{-2}$ & $1.3\times10^{-2}$ & $1.8\times10^{-2}$ & $1.7\times10^{-2}$ & $2.3\times10^{-2}$\\ 
B21811+31 & 273.40 & 31.74 & 0.850 & 0.009 & 0.012 & 0.011 & 0.015 & 0.010 & 0.013 & 0.013 & 0.017 & $1.0\times10^{-3}$ & $1.4\times10^{-3}$ & $1.3\times10^{-3}$ & $1.7\times10^{-3}$ & $1.1\times10^{-3}$ & $1.5\times10^{-3}$ & $1.4\times10^{-3}$ & $1.9\times10^{-3}$\\ 
SBS0812+578 & 124.09 & 57.65 & 1.090 & 0.005 & 0.006 & 0.006 & 0.008 & 0.005 & 0.007 & 0.007 & 0.009 & $4.3\times10^{-4}$ & $5.9\times10^{-4}$ & $5.3\times10^{-4}$ & $7.0\times10^{-4}$ & $4.7\times10^{-4}$ & $6.2\times10^{-4}$ & $5.9\times10^{-4}$ & $8.0\times10^{-4}$\\ 
2MASXJ05581173+5328180 & 89.55 & 53.47 & 1.080 & 0.018 & 0.024 & 0.021 & 0.028 & 0.019 & 0.025 & 0.024 & 0.032 & $1.6\times10^{-3}$ & $2.1\times10^{-3}$ & $1.9\times10^{-3}$ & $2.6\times10^{-3}$ & $1.7\times10^{-3}$ & $2.3\times10^{-3}$ & $2.2\times10^{-3}$ & $2.9\times10^{-3}$\\ 
1H1720+117 & 261.27 & 11.87 & 0.695 & 0.005 & 0.006 & 0.006 & 0.007 & 0.005 & 0.006 & 0.006 & 0.008 & $6.4\times10^{-4}$ & $8.7\times10^{-4}$ & $7.9\times10^{-4}$ & $1.0\times10^{-3}$ & $7.0\times10^{-4}$ & $9.3\times10^{-4}$ & $8.8\times10^{-4}$ & $1.1\times10^{-3}$\\ 
ARP220 & 233.74 & 23.50 & 0.746 & 0.019 & 0.026 & 0.023 & 0.031 & 0.021 & 0.027 & 0.026 & 0.035 & $2.5\times10^{-3}$ & $3.4\times10^{-3}$ & $3.0\times10^{-3}$ & $4.0\times10^{-3}$ & $2.7\times10^{-3}$ & $3.6\times10^{-3}$ & $3.4\times10^{-3}$ & $4.6\times10^{-3}$\\ 

\hline \hline
\end{tabular}
\begin{tabular}{l}
  \hline \hline
  Classification \\
  \hline
  BL Lac \\
  BL Lac \\
  BL Lac \\
  Radio gal. \\
  BL Lac \\
  BL Lac \\
  BL Lac \\
  FSRQ \\
  BL Lac \\
  BL Lac \\
  BL Lac \\
  FSRQ \\
  BL Lac \\
  ULIRG \\
  \hline \hline
\end{tabular}
}
\end{table*}

\subsection{Normalising the neutrino flux}
\label{sec:normalising-spectrum}

In the following, we calculate the neutrino flux expected in our model
for the objects in the final IceCube analysis cited above. We fix the
column density to the
benchmark values $N_H^{(1,2)}$, compatible with the values derived for
the objects individually, and investigate the resulting neutrino production.

Given the choice of benchmark values of $N_H$ in our model, the only
parameter left to determine in order to predict the neutrino flux from
the set of objects selected above is the normalisation of the
flux. The first selection criterion for the object selection presented
in~\cite{Maggi:2016bbi} was strong radio emission. Moreover, the radio
flux is unattenuated by matter in between the source and the observer,
giving a direct view of the inner engine, a feature which was also
already exploited in this analysis. Therefore, it is natural to
normalise the expected neutrino flux based on the radio flux.

The radio emission from astrophysical objects is typically attributed
to synchrotron emission from accelerated electrons in the magnetic
field of the source. Therefore, it is expected that the radio and
electron luminosity are comparable in size. The exact relation between
the radio and electron luminosities was derived
in~\cite{Tjus:2014dna}. It is obtained by integrating the synchrotron
emission from electrons of energy $E_e$, or equivalently $\gamma_e$, over
the electron spectrum. Assume an electron spectrum
$\frac{\diff N_e}{\diff \gamma_e}\propto\gamma_e^{-2}$, a minimum energy
$\gamma_e^{\rm min}=\{1,~10\}$ (estimated from efficient cooling, i.e.\ strong radio
emission of the electrons) and maximum energy $\gamma_e^{\rm
  max}=10^9$--$10^{11}$ (corresponding to the assumption that protons are
co-accelerated up to energies $10^{18}$--$10^{21}$~eV). For these limits
of the electron energy, the authors of~\cite{Tjus:2014dna} find that
\begin{equation}
  \label{eq:rad-e-corr}
  \chi = \frac{L_e}{L_R}\approx 100,
\end{equation}
over a large range of magnetic field strengths. Deviations from this
value occur fastest for $\gamma_e^{\rm min}=10$, with significant changes
starting at $10$~G. Modelling the properties of bright Fermi blazars,
one finds that typical magnetic field strengths are between $0.1-2$~G
for BL~Lacs and $1$--$10$~G for
FSRQs~\cite{2010MNRAS.402..497G,beckmann2012active}. Since the objects
in our selection are assumed to be typical, apart from their
obscuration, we use $\chi=100$.

From the electron luminosity, we can obtain the proton luminosity,
since these two species are co-accelerated. This has been discussed in
detail by~\cite{Merten:2017mzg}, from which we will repeat the main
arguments here. Typically, one assumes that the number of accelerated
protons and electrons are equal\footnote{From charge balance, this is
  true for the total number of electrons and protons. If the only
  requirement for initiating the acceleration is sufficient energy,
  this is then also true for the number of particles above the energy
  threshold, since in a plasma the particle energies follow a
  Maxwell-Boltzmann distribution which is independent of the particle
  mass~\cite{Merten:2017mzg}.}, $N_e=N_p$ and that their spectral
indices are the same. It is then straightforward to calculate the
electron-proton luminosity ratio and this results in
$f_e=\frac{L_e}{L_p}\approx1/100$~\cite{schlickeiser2014cosmic,Merten:2017mzg},
which is true also for the differential luminosity (i.e.\ independent
of energy). This result is supported by observations from the galaxy:
when comparing the observed electrons with cosmic rays up to the knee,
one obtains a luminosity ratio $1/100$. On the other hand, for
extragalactic sources, this value is estimated to be closer to $1/10$,
obtained by comparing the observed radio luminosity with that of
cosmic rays above the ankle. Since obtaining a more precise value
requires extrapolating the extragalactic cosmic ray flux to energies
below the ankle, with an unknown energy spectrum and minimum energy,
this value is quite uncertain. Moreover, this extrapolation is
dependent on the exact spectral index of extragalactic cosmic rays,
which is still uncertain due to the degeneracy with composition and
maximum energy.  In case of a spectral index deviating from $2$, the
luminosity integral becomes very sensitive on the minimum energy,
making a strong constraint on $f_e$ difficult. However, for sure
$f_e\ll 1$.
From particle-in-cell simulations, it is found that the assumption of
equal spectral indices for protons and electrons might not be true. In
this case the luminosity ratio becomes energy dependent, further
complicating the conversion from electron to proton luminosity. In the
following, we will assume a fixed ratio
\begin{equation}
  f_e = \frac{L_e}{L_p}=\frac{1}{10},
\end{equation}
which is conservative (i.e.\ relatively few protons).

Summarising, in order to determine the expected neutrino flux from the
objects in the selection above, we integrate the observed radio flux
from each object individually and convert this to a proton luminosity
\begin{equation}
  \label{eq:Lpconversion}
  L_p = \frac{\chi \cdot L_R}{f_e}.
\end{equation}
A proton population with this total luminosity is then allowed to
interact with a gas cloud of column density $N_H=N_H^{(1,2)}$. Note
that it is not needed to convert the observed flux to luminosity at
the source, since the same factor $d_L^2$ appears when propagating the
obtained neutrino flux at the source back to Earth.

\subsection{Results}
\label{sec:obscuredagn-results}

In this section, we present the neutrino flux predicted by our model
for the objects in the obscured flat-spectrum AGN selection. For the
simulation, we generate $5\times 10^4$ events, with the proton energy
between $100$~GeV and $10^8$~GeV and a spectral index of 2. The column
density is set to either of the two benchmark values: \nhone{} or
\nhtwo{}. As argued in Section~\ref{sec:photon-atten}, we include only
gamma-ray attenuation at the source by matter, not radiation fields.
While propagating to Earth, the gamma-ray flux is also attenuated by
interaction with the EBL and CMB (this is discussed in more detail in
Section~\ref{sec:diffuse-gamma-ray}). This attenuation, but not the
full EM cascade to lower energy gamma rays, is included in the final
gamma-ray flux\footnote{This is the usual approach. For a point
  source, including the full cascade would require more detailed
  modelling than the simple approach of
  Section~\ref{sec:diffuse-gamma-ray}. Since the predicted gamma-ray
  flux turns out to be very low compared to the observed flux, the
  additional modelling is not important.}, using the optical depth
$\tau(E_\gamma, z)$
for gamma rays at the redshift of the source from~\cite{Inoue:2012bk}
\begin{equation}
  \label{eq:gammaattebl}
 E^2_\gamma \Phi_\gamma(E_\gamma) =  e^{-\tau(E_\gamma, z)}E^2_\gamma \Phi_\gamma^0(E_\gamma).
\end{equation}



\begin{figure}
  \centering
  \includegraphics[width=\columnwidth]{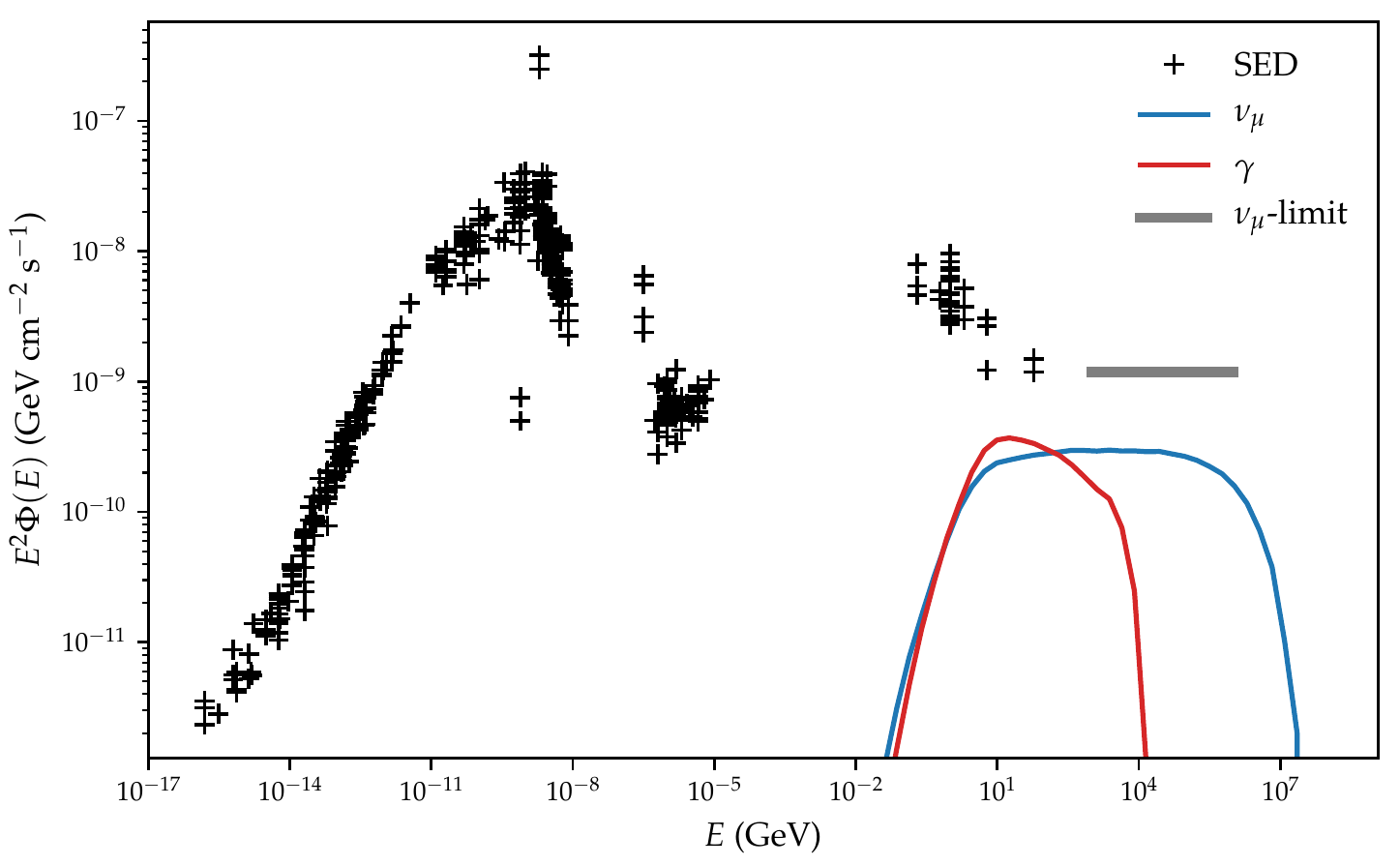}
  \caption[Hybrid SED for 3C371 predicted in our obscured neutrino
  source model]{Hybrid SED for 3C371, one of the objects with a
    predicted muon neutrino flux closest to the current upper
    limit~\cite{Aartsen:2017kru}, showing the measured electromagnetic
    data together with the predicted muon neutrino flux and gamma-ray
    flux in our obscured neutrino source model. Here we assumed
    \nhtwo{} and $E_p \in [10^2,\ 10^8]$~GeV. Electromagnetic spectrum
    data
    from~{\cite{2002babs.conf...63G,2003MNRAS.341....1M,2007ApJS..171...61H,1970ApJS...20....1D,1981A&AS...45..367K,2007AJ....133.1947N,1998AJ....115.1693C,1990IRASF.C......0M,1994yCat.2125....0J,2011A&A...536A...7P,1996ApJS..103..427G,1992ApJS...79..331W,2014A&A...571A..28P,2015arXiv150702058P,2009ApJS..180..283W,2010AJ....140.1868W,2011MNRAS.411.2770B,2013A&A...551A.142D,0067-0049-210-1-8,1992ApJS...80..257E,1999A&A...349..389V,2016A&A...588A.103B,2008A&A...480..611S,2010ApJS..188..405A,2012ApJS..199...31N,2015ApJS..218...23A,0004-637X-779-1-27}}
    retrieved using the \texttt{SSDC SED Builder}~\cite{sedbuilder}.}
  \label{fig:obj}
\end{figure}

The hybrid spectral energy distribution (SED) for the object with the
highest expected neutrino emission, 3C371, is shown in
Figure~\ref{fig:obj}, with \nhtwo{}. The figure includes the measured
photon SED across all wavelengths, the predicted gamma-ray emission
from $pp$-interactions, as well as the predicted muon neutrinos
flux\footnote{The muon neutrino flux is $1/3$ of the total neutrino
  flux, assuming full mixing between the neutrino flavours. This is
  the reason why the neutrino flux is below the gamma ray flux.} and
the IceCube limit on the muon neutrino flux from this object. The SEDs
of the other objects are shown in Appendix~\ref{sec:sed-all-objects}.
The predicted neutrino flux for 3C371 is well below the limit from
IceCube, such that our model is not ruled out and could only be
observable with next-generation cosmic neutrino detectors, such as
IceCube~Gen2 (with an estimated improvement of the point source
sensitivity with a factor of about five over
IceCube~\cite{Aartsen:2019swn}). The gamma-ray flux from
$pp$-interactions in this model is well below the observed flux from
this object, leaving the model also unconstrained here.

\begin{figure*}
  \centering
  \includegraphics[width=.7\linewidth]{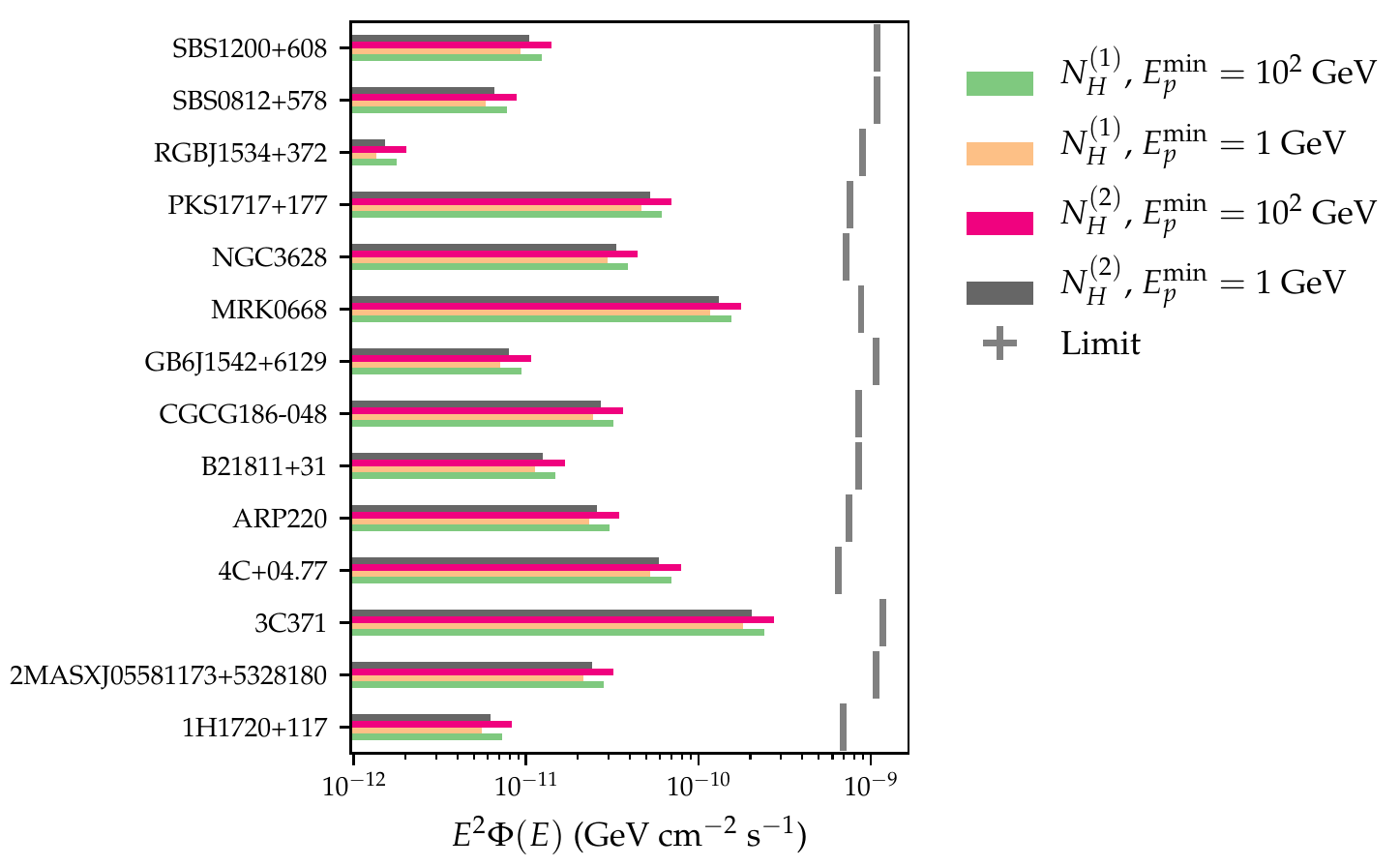}
  \caption[Summary of obscured neutrino sources for a set of selected
  objects]{Summary of the predicted muon neutrino flux in our obscured
    neutrino source model and the IceCube upper limits
    (from~\cite{Aartsen:2017kru}) for the objects in the obscured
    flat-spectrum radio AGN selection.}
  \label{fig:obscuredagnresults}
\end{figure*}

The same conclusions are also true for the other objects in the
selection. Their calculated neutrino fluxes can be found in
Table~\ref{tab:obsc-agn-res}, for both \nhone{} and \nhtwo{}. The same
results are shown in Figure~\ref{fig:obscuredagnresults}, which also
includes the flux in case the minimum proton energy is lowered to
$1$~GeV. For all objects, the predicted neutrino flux using natural
choices for the values of the parameters $\chi$ and $f_e$ is below the
limit placed by IceCube. Given the expected sensitivity of
IceCube~Gen2, however, some of these objects could be observable in
the near future in this model. The gamma-ray flux is in each case well
below the observed value, putting no constraint on the model. This
also immediately implies that, for this class of objects, there is no
constraint from the EGB, since their contribution is irrelevant
compared to the blazar contribution already present. On the other
hand, this also suggests that, even if this scenario were
applicable to all blazars (which is certainly not true), the neutrino
flux would not be high enough to explain the diffuse flux observed by
IceCube.

\begin{table*}
\centering
\caption[Summary of the predicted muon neutrino flux in our obscured
neutrino source model for the selected set of objects]{Summary of the
  predicted muon neutrino flux compared to the upper limits on their
  muon neutrino flux (from~\cite{Aartsen:2017kru}), for the objects in
  the obscured flat-spectrum radio AGN selection in units of
  $10^{-9}$~\gevcms, as well as the corresponding limits on
  $f_e$. Protons are injected within the energy range
  $E_p \in [10^2,\ 10^8]$~GeV and the considered column densities are
  \nhone{} and \nhtwo{}.  }
\label{tab:obsc-agn-res}
{
  \renewcommand{\arraystretch}{1.2}
  \setlength\tabcolsep{.2cm}
\begin{tabular}{lHHrHHHrHHHrHHHrHHHr}
\hline \hline
Name & RA ($^\circ$) & Dec ($^\circ$) & $E^2\Phi_\nu^{90\%}$  &
$E^2\Phi_\nu^{\rm (1),~No~sec.}$ & $E^2\Phi_\nu^{\rm (1),~No~sec.}$ & $E^2\Phi_\nu^{(1)}$
& $E^2\Phi_\nu^{(1)}$ &
$E^2\Phi_\nu^{\rm (2),~No~sec.}$ & $E^2\Phi_\nu^{\rm (2),~No~sec.}$ & $E^2\Phi_\nu^{(2)}$
& $E^2\Phi_\nu^{(2)}$ &
$f_e^{\rm (1),~No~sec.}$ & $f_e^{\rm (1),~No~sec.}$ & $f_e^{(1)}$
& $f_e^{(1)}$ &
$f_e^{\rm (2),~No~sec.}$ & $f_e^{\rm (2),~No~sec.}$ & $f_e^{(2)}$
& $f_e^{(2)}$ \\

\hline

\hline \hline
\end{tabular}
}
\end{table*}

\begin{figure*}
  \centering
  \includegraphics[width=.7\linewidth]{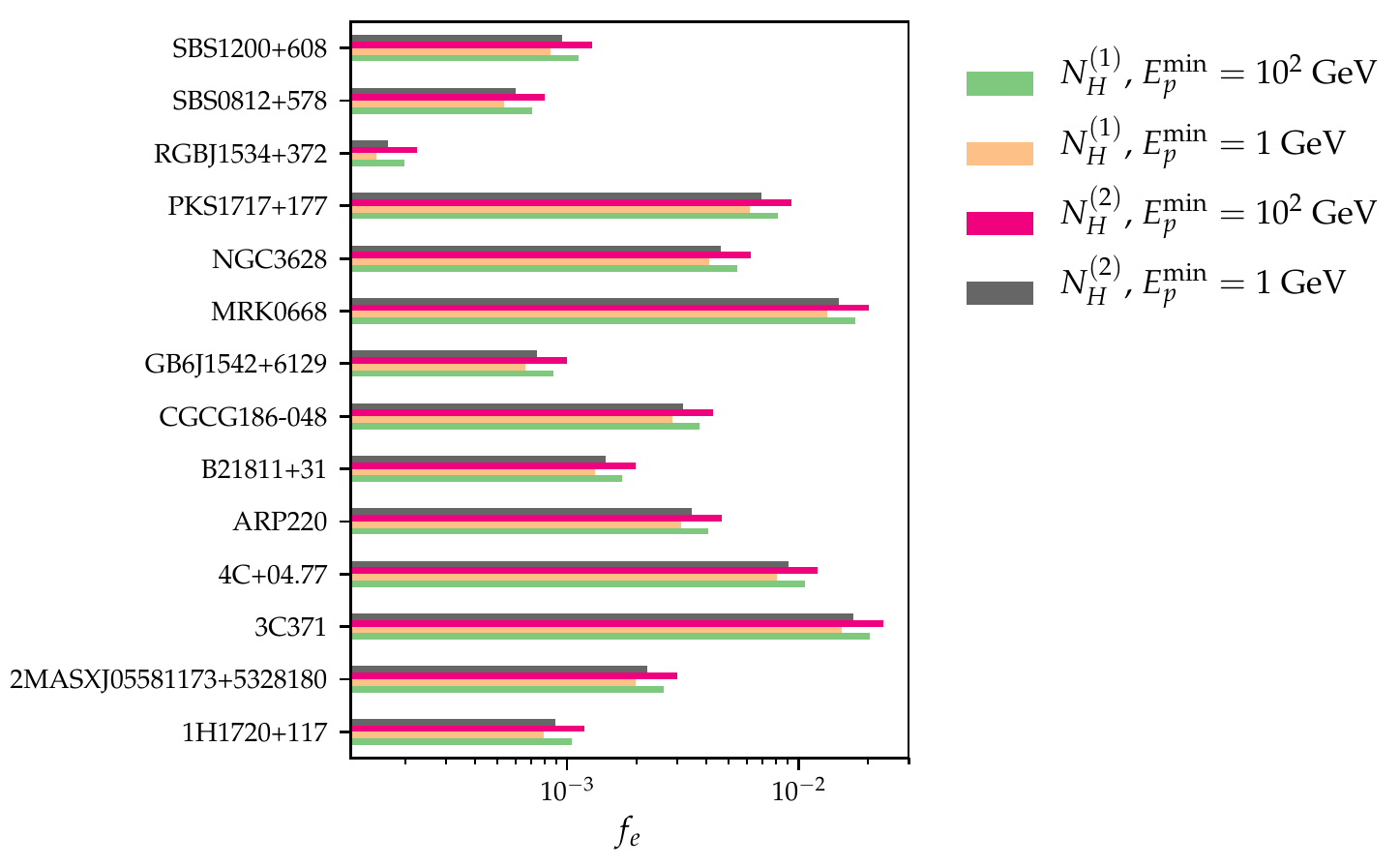}
  \caption[Summary of the lower limits on $f_e$]{Summary of
    the lower limits on $f_e$ for the objects in the
    obscured flat-spectrum radio AGN selection.  }
  \label{fig:feresults}
\end{figure*}

The above results already allow us to put meaningful constraints on
the only parameter which is not fixed by our model\footnote{Since we
  consider the value of $\chi$ to be quite robust.}: the
electron-proton luminosity ratio, for which we took
$f_e=1/10$. Translating the IceCube limit to a lower bound on $f_e$
(i.e.\ maximum amount of accelerated protons) in our model, we find a
bound of $f_e \approx 0.001 - 0.02$. This is also shown in
Table~\ref{tab:obsc-agn-res} and Figure~\ref{fig:feresults}.  Since
the galactic value of the electron-proton luminosity ratio is
$f_e\approx1/100$, these constraints on $f_e$ within our model are
already quite strong. The reason that these bounds are so strong, is
that for the column densities considered here, the full proton
population is depleted to produce neutrinos, as opposed to typical
scenarios where only part of the proton flux interacts. The advantage
of this was already discussed: even if such blazars are not the
dominant source of astrophysical neutrinos, identifying a few obscured
blazars gives independent constraints on the amount of accelerated
protons in blazars.


Finally, we can also investigate an alternative calculation, where
instead the predicted gamma-ray and neutrino flux are normalised to
the observed gamma-ray flux as opposed to the radio flux. In the case
of 3C371, this would lead to a neutrino emission comparable to the
IceCube limit. However, in this case more detailed SED modelling is
needed in order to explain the complete emission from this
object. Moreover, not all objects in our selection have sufficient
data to perform this analysis. Therefore, we refrain from doing this.

\section{Diffuse flux from a generic obscured population}
\label{sec:diffuse-flux}

In this section, we calculate the diffuse neutrino and gamma-ray flux from a
generic population of obscured sources. In particular, we investigate
whether sources obscured by a gas with column density $N_H^{(1,2)}$
can be responsible for the astrophysical neutrino flux measured by
IceCube, without violating the bound on the non-blazar contribution to
the EGB.

For this calculation, we will not be limiting ourselves to sources similar to
those in the selection of Section~\ref{sec:obsc-flat-spectr}. Instead,
we consider a generic population of obscured sources characterised
only by their evolution with redshift, while their total energy budget
will be fitted to reproduce the observed neutrino flux.
From generic arguments~\cite{Waxman:1998yy}, we know that a source
class with a power similar to the UHECR sources can reproduce the flux
observed by IceCube\footnote{Although, as we will see, we will end up
  requiring a maximum proton energy of $E_p \sim 10^{8}$~GeV in order
  to fit the IceCube flux and its upper limits at the higher
  energies. Therefore, these sources would not supply the UHECRs, but
  possess a similar energy budget.}. The only requirement on the
source configuration is that throughout the history of the universe, a
sufficiently large, possibly changing, population of obscured sources
exists. In our calculations, we do not assume the presence of a jet in
such sources.

\subsection{Populations}
\label{sec:populations}

We consider the possibilities of redshift evolution
following star formation, no evolution and the case of ULIRGs.  One
could also consider other scenarios, such as treating explicitly
BL~Lac and FSRQ evolution separately (see
e.g.~\cite{Murase:2016gly}).

\subsubsection{Star formation rate and flat evolution}
\label{sec:sfrandflat}

First, we consider a redshift evolution following the star formation
rate. This is a natural choice, since it follows the evolution of
galaxies throughout the cosmic history. This is particularly relevant
for GRBs and supernovae, since these events occur more frequently
during star formation when many large stars with short lifetimes are
formed. It is also relevant for AGN- and blazar-like scenarios, since
during intense star formation there could also be efficient accretion
around supermassive black holes in the centres of galaxies. Even if
the true rate deviates from the star formation rate, it is still a
good first approximation for the amount of activity in galaxies as a
function of redshift.

The star formation rate is measured using different observables, such
as the supernova rate, luminosity densities, limits on the diffuse
neutrino background from supernovae and GRBs. The star formation rate
is well-known up to redshifts of $z\approx1$. At higher redshifts,
measurements deviate, although the knowledge is improving. In the
following, we will use the star formation history derived
in~\cite{Yuksel:2008cu,Hopkins:2006bw}. Its form is
\begin{equation}
  \label{eq:SFR}
  \mathcal{H}_{\rm SFR} \propto(1+z)^{n_i},
\end{equation}
with
\begin{equation}
  \label{eq:SFRindices}
  n_i =
  \begin{cases}
    3.4 \quad &z<1 \\
    -0.3 \quad &1<z<4 \\
    -3.5 \quad &z > 4.
  \end{cases}
\end{equation}
The normalisation is such that $\mathcal{H}_{\rm SFR}(z=0)=1$ and the
function is continuous.

As an alternative to cosmic star formation, we consider also the
simplest case of no evolution
\begin{equation}
  \label{eq:Hnoevo}
  \mathcal{H}_{\rm flat}=1.
\end{equation}

\subsubsection{Ultra-luminous infrared galaxies}
\label{sec:ultra-lumin-infr}


In addition to the above generic scenarios, we will also consider in
more detail the possibility that ultra-luminous infrared galaxies
(ULIRG) could be responsible for the astrophysical neutrino flux
detected by IceCube through our model. ULIRGs showed up already in the
object selection used in Section~\ref{sec:selobj}, where ARP~220 was a
candidate object in the final selection. ARP~220 is the most
well-known and best studied ULIRG and is an object with a very high IR
luminosity formed by the merger of two galaxies, with a very high
column density of at least $10^{25}$~cm$^{-2}$ due to gas and
dust. Its nucleus is possibly powered by an AGN, which must however be
obscured and Compton-thick.~\cite{2014ApJ...789L..36W}.

More generally, ULIRGs are defined as galaxies with extreme infrared
luminosities $L_{\rm IR}>10^{12}~L_\odot$ (see e.g.\ the
review~\cite{Lonsdale:2006sr}). They are the mergers of gas rich
galaxies, with the central regions harbouring huge amounts of gas and
dust. The emission is caused by starburst activity (i.e.\ high star
formation rate), and possibly also AGN activity, triggered by the
merger~\cite{2010MNRAS.403..274C}.
It is believed that submillimetre galaxies are the high-redshift
counterparts of local ULIRGs.

The abundance and importance of AGN activity compared to starburst
activity in ULIRGs is still not completely clear. Surveys indicate that
ULIRGs contain radio cores which are due to AGN
activity~\cite{Nagar:2003an}. In studies of local
ULIRGs~\cite{Farrah:2003ka}, it was found that all of them require
starburst activity to explain their emission, while only half require
an AGN. Moreover, in $90\%$ of the cases, the starburst activity provides over
half of the IR luminosity, with an average fractional luminosity of
$82\%$. The AGN contribution does not increase with luminosity.
Other studies have found that over half of the ULIRGs contain an AGN,
with the fraction increasing with total IR luminosity~\cite{2009ApJ...704..789H}.
In~\cite{2010ApJ...719..425F}, it was found that only few ULIRGs are
dominated by AGNs (5\% at $z\sim1$ and 12\% at $z\sim2$) although, at a given
luminosity, the fraction of AGN activity is lower in high-$z$ ULIRGs.
In a $5-8~\mu$m analysis~\cite{doi:10.1111/j.1365-2966.2009.15357.x} of
local ULIRGs, signatures of AGN activity were found in $\sim70\%$ of the
sample. While most of the luminosity is due to the starburst activity,
$\sim23\%$ is due to the AGN, increasing with luminosity. More general,
part of the emission of star-forming galaxies at high redshifts can be
explained by AGN activity~\cite{2018A&A...617A.131S}.  So, while AGNs
are in general definitely not the dominant component of ULIRGs
emission, their contribution is not negligible. Moreover, high
obscuration of the central core may lead to underestimation of the AGN
power~\cite{Lonsdale:2006sr}.

The exact interplay between the central AGN and the starburst activity
is still unclear. Several scenarios are still possible: one could
evolve from the other, trigger the other or their coexistance might be
coincidental.  Typically, the AGN and ULIRG activity is unified in an
evolutionary scenario. Early on in the merger, starburst activity is
high when there is still plenty of gas. This merger might also relate
to the growth of a supermassive black hole and AGN activity at the
core~\cite{2002MNRAS.332..529K}. This was confirmed by observations of
stellar
kinematics~\cite{0004-637X-638-2-745,0004-637X-651-2-835}. Later, when
the gas is concentrated in a compact region in the centre, a
starburst-ULIRG phase occurs.
During the late merger state, there is a high accretion rate at the
centre, giving rise to an obscured QSO\footnote{Quasi-stellar object
  or quasar}/AGN-powered ULIRGs, due to the huge amount of
gas and dust driven to the centre. Afterwards, the galaxy enters its
most luminous phase with an optically-visible QSO which drives out the
remaining material~\cite{1538-4357-625-2-L71,Lonsdale:2006sr}. In this
sense, neutrino emission from ULIRGs would have an interesting
interplay with galaxy formation and evolution.
%
%
%
%
%
%

Estimates of the space densities of ULIRGs have been made by several
groups. In~\cite{Hopkins:2005fb}, they found at $z=0.15$ that
$n=3\times10^{-7}$~Mpc$^{-3}$ for $L_{\rm IR}=1.6\times10^{12}~L_\odot$ and
$n=9\times10^{-8}$~Mpc$^{-3}$ for $L_{\rm IR}=2.5\times10^{12}~L_\odot$. This
decreases with a factor $1.5$ to lower redshift ($z=0.04$), while it
increase to $\Phi(L>10^{11}~L_\odot)=1-3\times10^{-2}$~Mpc$^{-3}$ at $z\sim1$--$3$.
A full luminosity function was derived in~\cite{Chapman:2004fu},
although their analysis is only sensitive to $L_{\rm IR}>10^{12.3}~L_\odot$. In the same
redshift range $z\sim1$--$3$, they find $n>6\times10^{-6}$~Mpc$^{-3}$.
This leads to a redshift evolution
\begin{equation}
  \label{eq:HULIRG}
  \mathcal{H}_{\rm ULIRG} \propto
  \begin{cases}
    (1+z)^4 \quad &z <= 1 \\
    \mathrm{const.} \quad & 1 < z < 4,
  \end{cases}
\end{equation}
which we will use in the following, although other studies found more
extreme evolution up to $\propto(1+z)^7$ for
$z=0-1.5$~\cite{2004ApJ...603L..69C}.

We will consider the central AGN, obscured by gas and dust driven to
the centre by the merger, as a potential target for the obscured
$pp$-neutrino production mechanism. ULIRGs have been considered before
as the source of astrophysical
neutrinos~\cite{He:2013cqa,Palladino:2018bqf}, but in the context of
cosmic-ray reservoir models, through confined cosmic rays interacting
with gas in the galaxy. Instead, we consider only neutrino production
in a compact region near the core. This region can have a very high
column density, despite the overal surface gas density of starburst
being only about $1$~g\,cm$^{-2}$, or $N_H\sim10^{23}$~cm$^{-2}$, much
lower than the densities required by our mechanism. However, this
overal surface gas density is only valid for the overall gas density,
which is derived using the Kennicutt-Schmidt
law~\cite{1959ApJ...129..243S,1989ApJ...344..685K,1998ApJ...498..541K}
relating the star-formation rate with the gas density, and does not
exclude the existence of local, compact objects or regions with higher
densities. Finally, in our accelerator model, the cut-off of the
proton and neutrino spectra can be at higher energy than in reservoir
models, since there is no confinement criterion, although we will keep
the cut-off at $10^8$~GeV in order to not overshoot the IceCube flux
at high energies.

Now, we can estimate whether ULIRGs can provide the required
luminosity to supply the diffuse neutrino flux. Using the more
conservative value $n\approx5\times10^{-7}$~Mpc$^{-3}$ (integrating
over luminosity) of the number densities quoted above, along with
the minimum ULIRG luminosity $L_{\rm IR}=10^{12}~L_\odot$, we estimate
their local energy generation rate in the IR
\begin{equation}
  \label{eq:lir_ulirg}
  \mathcal{Q}_{\rm IR}= n(z=0)\cdot L_{\rm IR},
\end{equation}
Afterwards, this IR luminosity is converted to radio
luminosity using the radio-IR relation for
ULIRGs~\cite{2010ApJ...714L.190S}. Using the thermal infrared luminosity $L_{\rm TIR}\equiv
L(8-1000~\mu\mathrm{m})$, the TIR/radio flux ratio is defined as
\begin{equation}
  \label{eq:TIRradio}
  q_{\rm TIR} = \log\left( \frac{L_{\rm TIR}}{3.75\times10^{12}~\mathrm{W}}
  \right) - \log\left(\frac{L_{1.4~\mathrm{GHz}}}{\mathrm{W\,Hz}^{-1}}\right).
\end{equation}
On average, this ratio has the value $\left< q_{\rm TIR} \right>=2.6$,
with no evolution in redshift. After obtaining $L_{1.4~\mathrm{GHz}}$
from this relation, we estimate the total radio luminosity as
$L_R=1.4~\mathrm{GHz} \times L_{1.4~\mathrm{GHz}}$. Using the relations in
Section~\ref{sec:normalising-spectrum}, the radio luminosity can be
converted into the proton luminosity (or energy generation rate)
$\mathcal{Q}_p = \frac{\chi\cdot \mathcal{Q}_R}{f_e}$. We find
\begin{equation}
  \label{eq:Qulirg}
  \mathcal{Q}^{\rm ULIRG}_p \approx 10^{43.4}~\mathrm{erg}\,\mathrm{Mpc}^{-3}\,\mathrm{yr}^{-1}.
\end{equation}
This luminosity is slightly below the value estimated in the
Waxman-Bahcall calculation~\cite{Waxman:1998yy}, such that ULIRGs do
not initially seem capable of explaining the full astrophysical
neutrino flux. Moreover, in this calculation it was assumed that all
of the ULIRG luminosity is related to AGN activity, while in reality
the AGN contribution to the luminosity is at least one order of
magnitude lower (integrated over all ULIRGs). On the other hand, we
assumed that all ULIRGs have exactly $L_{\rm IR}=10^{12}~L_\odot$, while
many have higher luminosity. In addition, we used the standard value
for the parameter $f_e$, while it can easily be lower by at least an order
of magnitude, increasing $\mathcal{Q}_p$ by the same factor.

\subsection{Diffuse neutrino flux}
\label{sec:diff-neutr-flux}

In order to obtain the total diffuse flux of neutrinos from all
sources in the observable universe, we need to perform an integral
over cosmic history or, equivalently, over cosmological
distance. Following the method
in~\cite{Hogg:1999ad,weinberg2008cosmology,Ahlers:2014ioa}, the
diffuse flux is given by
\begin{align}
  E_\nu^2\Phi_\nu^{\rm diffuse}(E_\nu) ={}& \frac{c}{4\pi}
                                         \frac{1}{H_0}
                                         \int\frac{\mathcal{H}(z)}{(1+z)^2E(z)}
                                         \nonumber \\
          & \times \left.\varepsilon_\nu
            \mathcal{Q}_{\varepsilon_\nu}(\varepsilon_\nu)\right|_{\varepsilon_\nu=(1+z)E_\nu}\,\diff z,
  \label{eq:diffnufinal}
\end{align}
under the assumption that the redshift evolution can be factorised out
of the luminosity function.  The factor $\mathcal{Q}_{E_\nu}(E_\nu)$
is defined as $\mathcal{Q}_{E_\nu}(E_\nu)=E_\nu \Phi_\nu(E_\nu)$, such
that $\int\diff E_\nu\, E_\nu \Phi_\nu(E_\nu)=\mathcal{Q}_\nu$, with
$\mathcal{Q}_\nu$ the total injected neutrino luminosity per comoving
volume. Its form is obtained from our numerical simulation, where the
normalisation of $\mathcal{Q}_{E_\nu}$ is free to be determined either
by fitting the final diffuse flux $\Phi_\nu^{\rm diffuse}$ or by
fixing the total injected proton luminosity.

The $z$-integral becomes energy-independent for the case of a power
law spectrum\footnote{I.e.\ this is no longer true once we include a
  cut-off (e.g.\ an exponential) in the spectrum. This agrees with the
  intuition that the contribution from the end of the spectrum changes
  with redshift. However, as long as we calculate the flux at an
  energy sufficiently far from the cut-off, the calculation for a
  power law is valid. For our own result, we use the full integral,
  which does not suffer from this subtlety.}
$\mathcal{Q}_{E_\nu}\propto E_\nu\cdot E_\nu^{-\gamma}$ and the previous assumption that
the redshift evolution and $\mathcal{Q}_{E_\nu}(E_\nu)$ are independent. We then get
\begin{equation}
  \xi_z = \int\frac{\diff z}{E(z)}
  \mathcal{H}(z) (1+z)^{-\gamma},
\end{equation}
which can be solved numerically. The simplified formula for
the final flux then becomes the standard result~\cite{Waxman:1998yy}
\begin{equation}
  E_\nu^2\Phi_\nu^{\rm diffuse}(E_\nu) = \frac{c}{4\pi} \xi_z \frac{1}{H_0} E_\nu \mathcal{Q}_{E_\nu}(E_\nu).
\end{equation}
For the case of $\gamma=2$, we find $\xi_z=2.4$ for a redshift evolution
following star formation $\mathcal{H}(z)=\mathcal{H}_{\rm SFR}(z)$
(Eqs.~\eqref{eq:SFR}~and~\eqref{eq:SFRindices}). For no
evolution we have $\xi_z=0.53$ and for the case of ULIRGs, we find
$\xi_z=3.6$.

\subsection{Diffuse gamma-ray flux}
\label{sec:diffuse-gamma-ray}

In order to obtain the diffuse gamma-ray background from the neutrino
sources considered here, an additional effect needs to be taken into
account. During propagation, gamma rays can interact with the
extragalactic background light (EBL) and the cosmic microwave
background (CMB), producing an
$e^+e^-$-pair. The gamma-ray flux from redshift $z$ is thus cut
off above the energy where the optical depth in the EBL becomes equal
to $1$. In turn, these electrons can up-scatter EBL and CMB photons
back to gamma-ray energies through inverse-Compton scattering or by emitting
synchrotron radiation, initiating an electromagnetic cascade. In this
way photons of energies above $\sim100$~GeV are reprocessed and
accumulate at energies at and below $100$~GeV, significantly
increasing the flux at these energies.

\begin{sloppypar}
We follow the procedure outlined in~\cite{Murase:2012df} in order to
calculate the gamma ray spectrum after cascading in the EBL
analytically. As described in~\cite{Berezinsky:1975zz,Coppi:1996ze},
after the EM cascade has sufficiently developed, it attains a
universal form given by (using $G_{E_\gamma} = E_\gamma \Phi_\gamma$ in order to denote
the cascade)
\begin{equation}
  \label{eq:G_E}
  G_{E_\gamma} \propto
  \begin{cases}
    \left(\frac{E_\gamma}{E_\gamma^{\rm br}}\right)^{-1/2} \quad &(E_\gamma < E_\gamma^{\rm
      br}) \\
    \left(\frac{E_\gamma}{E_\gamma^{\rm br}}\right)^{1-\beta} \quad &(E_\gamma^{\rm br}< E_\gamma < E_\gamma^{\rm cut}),
  \end{cases}
\end{equation}
normalised to $\int\diff E_\gamma \, G_{E_\gamma} = 1$. Typically,
$\beta \approx 2$. The cut-off energy $E_\gamma^{\rm cut}$ is the
energy where suppression due to pair production occurs. It can be
obtained from the requirement\footnote{As opposed to the formula
  in~\cite{Coppi:1996ze}, where an extra factor (1+z) is included in
  the energy.}  $\tau(E_\gamma^{\rm cut}, z)=1$, for which we use the
optical depth tables provided in~\cite{Inoue:2012bk}.
The break energy is given by\footnote{This break energy is due to the
  lowest energy at which electrons are created that can up-scatter photons
  from the CMB through inverse-Compton scattering. From the energy
  loss rate of an electron through IC scattering and the number of
  photons scattered per unit time, one finds the average energy of
  scattered photons as $\epsilon_\gamma \approx \frac{4}{3}\gamma_e^2\epsilon_{\rm CMB}$, with $\gamma_e$
the Lorentz factor of the electron. This roughly corresponds to the
handwaving argument that the photons gains two Lorentz factors of
energy: one from transforming to the electron frame (where scattering
is easy) and one from transforming back.}
$E_\gamma^{\rm br} \approx \frac{4}{3}\left(\frac{E'^{\rm cut}_\gamma}{2 m_e
    c^2}\right)^2 \varepsilon_{CMB} \approx 0.034~\mathrm{GeV} \left(\frac{E^{\rm
      cut}_\gamma}{0.1~\mathrm{TeV}}\right)^2\left(\frac{1+z}{2}\right)^2$,
where $\varepsilon_{CMB}$ is the typical CMB energy.
Above the cut-off energy $E_\gamma^{\rm cut}$, but below
$\min \left[\frac{E^{\rm max}_\gamma}{2},
  \frac{4}{3}\left(\frac{E'^{max}_\gamma}{2 m_e c^2}\right)^2
  \varepsilon_{CMB}\right]$, the cascade is not sufficiently developped and its
exact form depends on the details of the injection. With far away
sources, for gamma rays scattering in the Thompson regime
one can assume a simple exponential cut-off $e^{-\tau_{\gamma\gamma}}$ for
$d>\lambda_{\gamma\gamma}$ with
$\tau_{\gamma\gamma}(E_\gamma, z)= d(z)/\lambda_{\gamma\gamma}(E_\gamma)$. On the other hand, in the
Klein-Nishina regime pairs are continuously supplied, giving a shape
$\frac{1-e^{-\tau_{\gamma\gamma}}}{\tau_{\gamma\gamma}}$ as long as the pair injection length
($\lambda_{\rm BH}$) is longer than $d$. For our purposes, we use the
exponential form, although the exact details do not significantly
influence our conclusion. For more details with a full numerical
calculation of the cascade,
see~\cite{Murase:2011yw,2012ApJ...749...63M,Murase:2012xs}.
\end{sloppypar}

Finally, the full diffuse spectrum is obtained by integrating the
cascaded spectrum from each redshift and weighting with the injected
luminosity from each redshift, giving
\begin{equation}
  \label{eq:GEdiffuse}
  E_\gamma^2\Phi_\gamma (E_\gamma) = \frac{c}{4\pi} \frac{1}{H_0}\int\frac{\mathcal{H}(z)\,\diff z}{(1+z)^2E(z)}
   E_\gamma G_{E_\gamma} \mathcal{Q}_\gamma.
\end{equation}
The factor $\mathcal{Q}_\gamma$ is the total integrated luminosity in gamma
rays per comoving volume injected by the sources (after attenuation by
the gas column) and is obtained from our simulation.

From the integrand, we see that the dominant injection is from sources
at $z\sim1$, so that we can estimate the diffuse gamma-ray flux
as~\cite{Murase:2012df}
\begin{equation}
  \label{eq:gam-diff-estimate}
  E_\gamma^2 \Phi_\gamma (E_\gamma) \approx \frac{c}{4\pi}\frac{1}{H_0} \xi_z \left. E_\gamma
    G_{E_\gamma}\right|_{z=1} \mathcal{Q}_\gamma,
\end{equation}
with $\left. E_\gamma G_{E_\gamma}\right|_{z=1}\approx0.1$ between the break and the
cut-off. Since the cut-off energy at $z=1$ is about $100$~GeV, there
will be an accumulation\footnote{This accumulation does indeed rise
  above the hypothetical flux in the absence of attenuation, since the
  non-attenuated spectrum can be substituted in
  Eq.~\eqref{eq:gam-diff-estimate} by replacing the factor $E_\gamma G_{E_\gamma}$ by
  $1/\ln \left(\frac{E^\mathrm{max}_\gamma}{E^\mathrm{min}_\gamma} \right)$. With an injected
  energy range between $E^\mathrm{max}_p/20$ and $\ll1$~GeV from pion decays,
  this is smaller than $0.1$.} of gamma rays at this energy, which is
then also typically the point most strongly constrained by the Fermi
measurements.

In order to get the full spectrum, however, this simple approximation
is not sufficient and we need to do the full integration of
Eq.~\eqref{eq:GEdiffuse} numerically.
As a consequence of this complicated integration over redshift, the
gamma-flux is sensitive to the details of the redshift evolution, in
contrast to the neutrino flux where there is a degeneracy between the
evolution (through $\xi_z$) and the normalisation.

We also verified the resulting spectrum ourselves with a numerical
simulation using \texttt{CRPropa 3}~\cite{Batista:2016yrx} and its
DINT module, which follows the description in~\cite{Lee:1996fp}. We
find that the resulting spectrum has the same form.

\subsection{Results}
\label{sec:diffuse-results}

\begin{figure*}
  \centering
  \includegraphics[width=.85\linewidth]{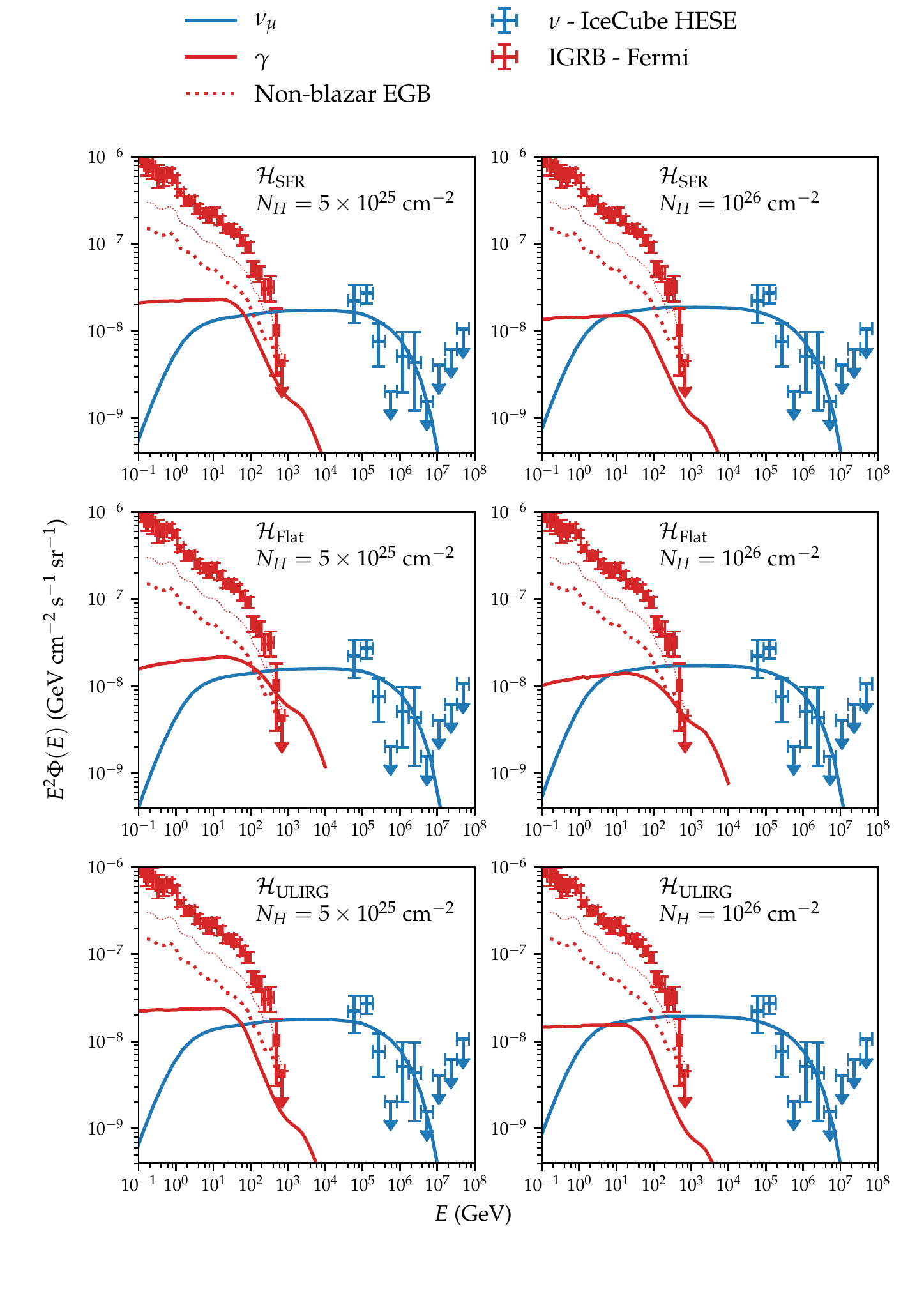}
  \caption[Diffuse neutrino and gamma-ray flux predicted in our
  obscured neutrino source model]{Results for the diffuse neutrino and gamma-ray flux, for
    different evolutions and column densities for the obscured
    $pp$-neutrino scenario, fitted to the IceCube single-flavour
    neutrino flux (HESE)~\cite{Aartsen:2017mau}, for a spectral index
    of 2 and proton energy between 100~GeV and $10^8$~GeV. The non-blazar contribution to the
    EGB shows both the best fit value ($14\%$ of the EGB measured by
    Fermi~\cite{Ackermann:2014usa}) and the weakest upper limit
    ($28\%$).}
  \label{fig:diff}
\end{figure*}

\begin{figure*}
  \centering
  \includegraphics[width=.85\linewidth]{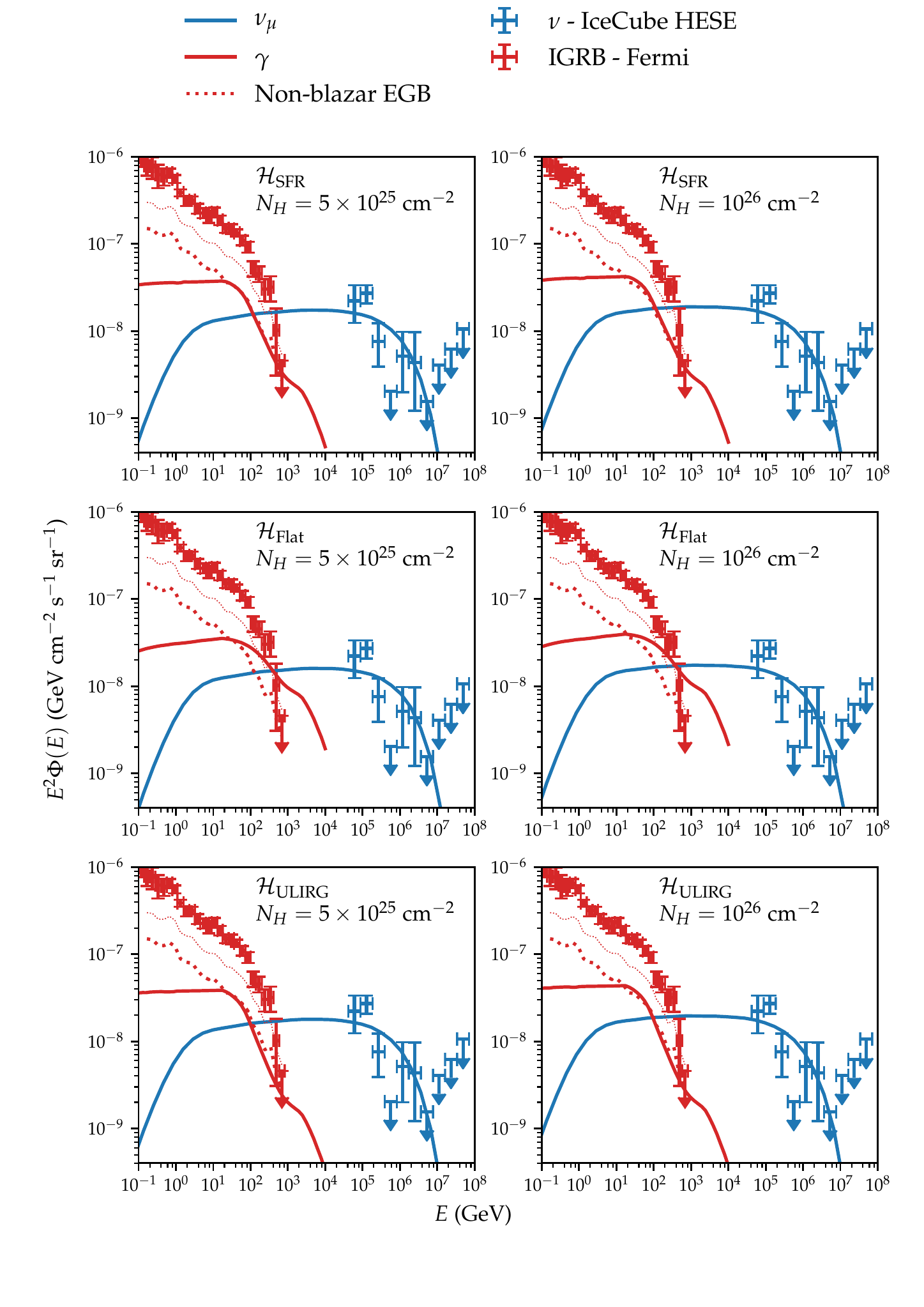}
  \caption[Diffuse neutrino and gamma-ray flux predicted in our
  obscured neutrino source model, without gamma-ray attenuation]{Same
    as Figure~\ref{fig:diff}, now without gamma-ray attenuation at the
    source.}
  \label{fig:diff_noatt}
\end{figure*}

The diffuse flux for the different redshift evolutions above and the
assumed column densities \nhone{} and \nhtwo{} are calculated using
our Monte Carlo simulation, generating $5\times 10^{4}$ events with
proton energy between 100~GeV and $10^8$~GeV. The injected proton
energy budget is then normalised by fitting the resulting neutrino
spectrum (taking into account a flavour factor $1/3$ after
oscillation) to the single-flavour neutrino flux observed by
IceCube~\cite{Aartsen:2017mau}. The results of this calculation are
shown in Figure~\ref{fig:diff}, where the gamma ray flux can be
compared to the bound on the non-blazar contribution to the EGB (both
the best fit ($14\%$) and its weakest upper limit
($28\%$)~\cite{TheFermi-LAT:2015ykq,Lisanti:2016jub,Ajello:2015mfa,Ackermann:2015tah,DiMauro:2014wha,Inoue:2014ona,Costamante:2013sva}).
These results can be compared with Figure~\ref{fig:diff_noatt}, which
shows the result for the same situation in case there is no
attenuation of the gamma rays. This comparison seems a bit artificial
(although it serves only to show the effect of obscuration in our
model), since these large column densities automatically imply that
attenuation is present. Physically, it might be more relevant to
compare with the case $N_H=10^{24}$~cm$^{-2}$ if one is interested in
the effect of the higher density. This change lowers both the
attenuation of gamma rays and the fraction of protons that
interact\footnote{It also is more sensitive to the energy dependence
  of the proton-proton cross section, such that the neutrino spectrum
  is flatter than that of the original protons.}, the latter of which
needs to be compensated for by increasing the injected proton
luminosity to fit the IceCube flux. In the end, only the neutrino
luminosity normalisation and its correlation with the gamma ray
luminosity are then important. On the other hand, in cosmic ray
reservoir models, the protons effectively cross a high density target,
while the gamma rays escape immediately. This then corresponds to the
high density case shown here with gamma-ray attenuation turned off.

We find that in the case of a redshift evolution following star
formation $\mathcal{H}_{\rm SFR}$ or the ULIRG evolution
$\mathcal{H}_{\rm ULIRG}$, the bounds from the non-blazar contribution
to the EGB can be satisfied, as opposed to the case in which there is
no attenuation, although in the case \nhone{} the improvement is
small. In the case of a flat evolution $\mathcal{H}_{\rm Flat}$, in
which the contribution from low-$z$ sources is more important, this
bound can not be satisfied for \nhone{} and only marginally for
\nhtwo{}.

The fitted proton energy budgets for all the cases are shown in
Table~\ref{tab:plumi}. For the case of ULIRGs, these can be compared
with the optimistic estimate from
Section~\ref{sec:ultra-lumin-infr}. We find that even this estimate,
which did not take into account that most of the ULIRG luminosity is
due to starburst activity instead of a central engine, is about one
order of magnitude too low. In order to explain the diffuse neutrino
flux with objects like ULIRGs, other object classes are therefore
necessary. The slightly less luminous LIRG\footnote{Galaxies with
  luminosities $L>10^{11}~L_\odot$, which are more numerous than
  e.g.\ starburst galaxies.} are candidates for this. On the other
hand, the estimate also depended on the electron-proton luminosity
ratio $f_e$. Lowering this ratio can already increase the contribution
from ULIRGs significantly.

\begin{figure*}
  \centering
  \includegraphics[width=0.85\linewidth]{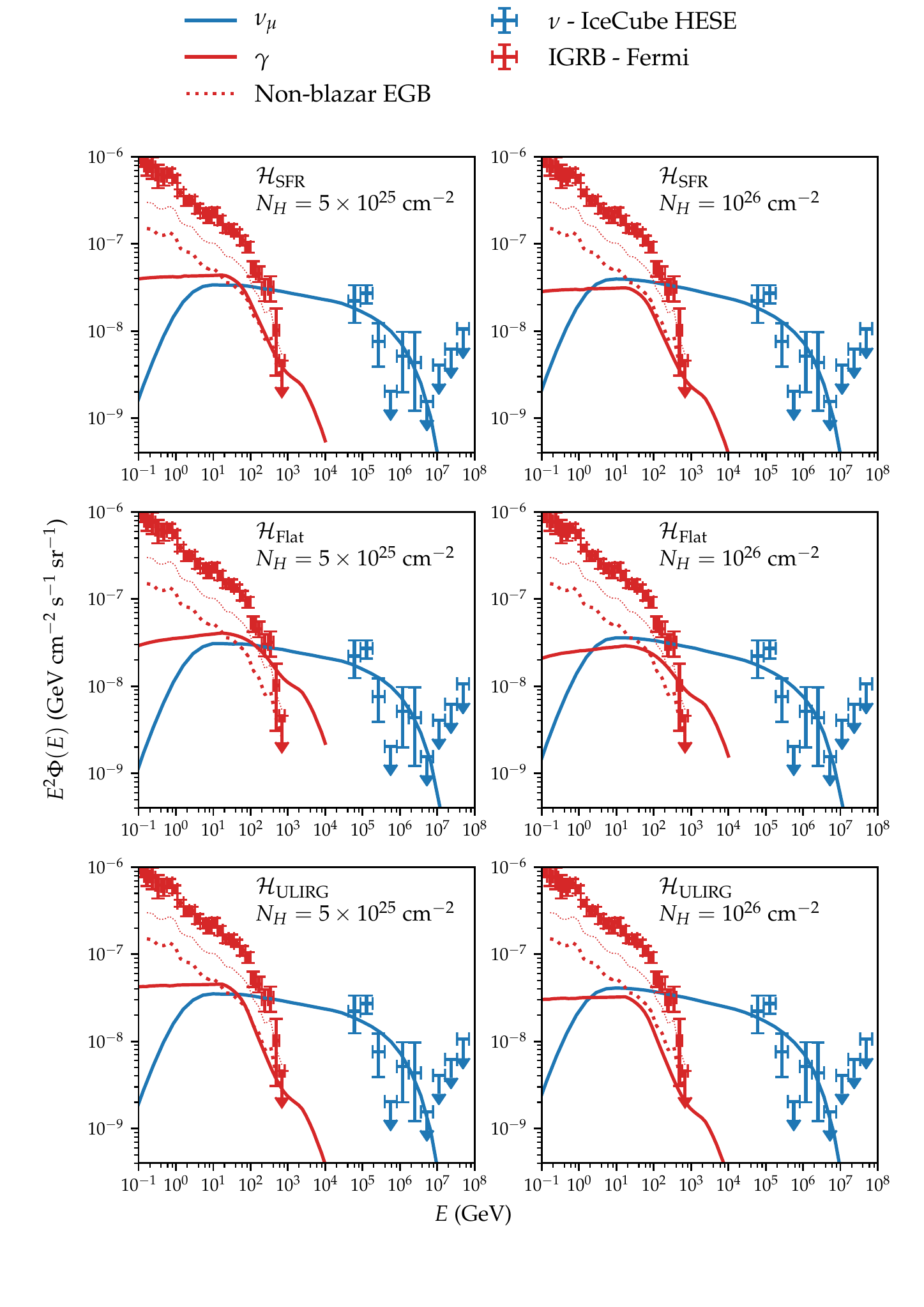}
  \caption[Diffuse neutrino and gamma-ray flux predicted in our
  obscured neutrino source model, for a spectral index of 2.1]{Same as
    Figure~\ref{fig:diff}, now with proton spectral index $\gamma=2.1$.}
  \label{fig:diff_2.1}
\end{figure*}

Finally, we show in Figure~\ref{fig:diff_2.1} results of the same
calculation for a proton index $\gamma=2.1$. In this case, only the
highest density \nhtwo{} can marginally satisfy the constraints from
the non-blazar contribution to the EGB for $\mathcal{H}_{\rm SFR}$ and
$\mathcal{H}_{\rm ULIRG}$.

\begin{table}
  \centering
  \caption[Required proton luminosities $\mathcal{Q}_p$ from the fit
  to the observed IceCube single-flavour neutrino flux]{Required
    proton luminosities $\mathcal{Q}_p$ from the fit to the observed
    IceCube single-flavour neutrino flux, in units of
    $10^{45}$~\ergmpcyr.}
  \label{tab:plumi}
  \renewcommand{\arraystretch}{1.2}
  \begin{tabular}{lrrr}
    \hline\hline
    \textbf{Evolution} & $\mathcal{H}_{\rm SFR}$ & $\mathcal{H}_{\rm
                                                   Flat}$ & $\mathcal{H}_{\rm ULIRG}$\\
    \hline
    \textbf{Column density}& & & \\
    \nhone{} & 2.30 & 9.71 & 1.60\\
    \nhtwo{} & 2.17 & 9.18 & 1.51 \\
    \hline \hline
  \end{tabular}
\end{table}

\section{Conclusion}
\label{sec:obscured-conclusion}

In this work, we investigated the possibility that AGN obscured by
matter emit high energy neutrinos created in $pp$-interactions with a
gas of a sufficiently high column density which acts as a beam
dump. In thic case, neutrino production can be efficient, whilst the
produced gamma rays can at the same time be attenuated through pair
production by the same gas cloud. Here, we took as benchmark values
for the column density \nhone{} and \nhtwo{}.

First, we calculated the neutrino spectra emitted from a set of active
galaxies whose electromagnetic spectrum can be explained with
obscuration by matter. The selection, performed
in~\cite{Maggi:2016bbi}, searches for strong radio-emitting galaxies
with a lower-than-expected X-ray flux. The resulting objects are
mainly blazars, along with one radio galaxy and one ULIRG. The
candidate class of obscured blazars, being a subset of blazars, is
unlikely to be responsible for the bulk of the IceCube flux, since
their number density is too low. Still, they are interesting targets
to test the viability of our model. Moreover, neutrino emission from
such objects would provide a simple measurement of the accelerated
proton content of blazars. We found that the predicted neutrino
emission from these objects is below the limit set by IceCube, leaving
the model unconstrained. For several of these objects, the scenario
can be tested in the future upgrade of IceCube, IceCube-Gen2, while
for many of them the predicted neutrino flux is too low to be detected
in the near future. On the other hand, current limits already allow to
constrain the amount of accelerated protons in blazars, with values
$f_e > 0.001$--$0.02$.

Second, we investigated the diffuse neutrino and gamma-ray flux from
an unspecified population of neutrino sources operating under our
model. In particular, we tested whether obscured $pp$-neutrino sources
can be the source of the IceCube flux without violating the bounds on
the non-blazar contribution to the EGB. In the case of a redshift
evolution following star formation or for ULIRGs, this is indeed
possible for both \nhone{} and \nhtwo{}, although in the former case
the gamma-ray flux is close to the EGB bound. In the case of no evolution with
redshift, the EGB bound is violated. This conclusion is valid for a neutrino
spectrum $\propto E_\nu^{-2}$, which has trouble explaining the low energy
events recorded by IceCube (although this can be solved with a second
distinct population). In the case of steeper spectra, the constraints
are more tight. Even for the thickest gas clouds considered here, the
gamma-ray flux is just barely below the EGB limit.

We also discussed in more detail whether a specific source class of
special interest for our model, ULIRGs, could be responsible for the
IceCube flux. Since ULIRGs occur when two galaxies merge, neutrino
emission from such objects could have an interesting interplay with
galaxy formation and evolution. While the predicted neutrino and
gamma-ray flux are compatible with observation\footnote{The gamma-ray
  flux might even be more attenuated by interactions at high energy
  with the IR field present in these galaxies, see
  e.g.~\cite{Murase:2015xka}.}, a simple, optimistic, estimate of
their the proton luminosity $\mathcal{Q}_p$ of the complete population
undershoots the required value. However, this conclusion depends on
the electron-proton luminosity ratio $f_e$, for which we assumed a
conservative value $1/10$. Taking lower values boosts the amount of
accelerated protons, increasing the provided luminosity. Another
possibility is that also the slightly less luminous LIRG contribute
through the same scenario.  In this sense, a separate analysis of the
highly obscured LIRG NGC~4418 would be interesting.

Finally, the same scenario could also be applicable to other
objects. One intriguing, though speculative, possibility is that AGNs
in the early universe can produce neutrinos through our model
when they first turn on. At this time, a lot of gas and dust is still
surrounding the centres of galaxies, potentially providing an ideal
target. While these early AGNs might not explain the full IceCube
neutrino flux (even more so because then the connection with the
observed total gamma-ray energy budget is less obvious), they could
make up part of it. This high-$z$ flux would constitute an unresolvable
component of the total neutrino flux. Moreover, gamma-rays from such a high
redshift would be cascaded down to lower energy than from
$z=1$-sources, such that the Fermi bounds from such a population are
much less stringent. Moreover, like ULIRGs, such a scenario would tie
neutrino production to galaxy formation and evolution.

\begin{acknowledgments}
  This work was performed while MV was aspirant FWO Vlaanderen. KDdV
  was supported in by the Flemish Foundation for Scientific Research
  FWO-12L3715N, and the European Research Council under the EU-ropean
  Unions Horizon 2020 research and innovation programme (grant
  agreement No 805486).
\end{acknowledgments}

\appendix

\section{SED of all the objects}
\label{sec:sed-all-objects}



\begin{figure*}
  \centering
  \includegraphics[width=.85\linewidth]{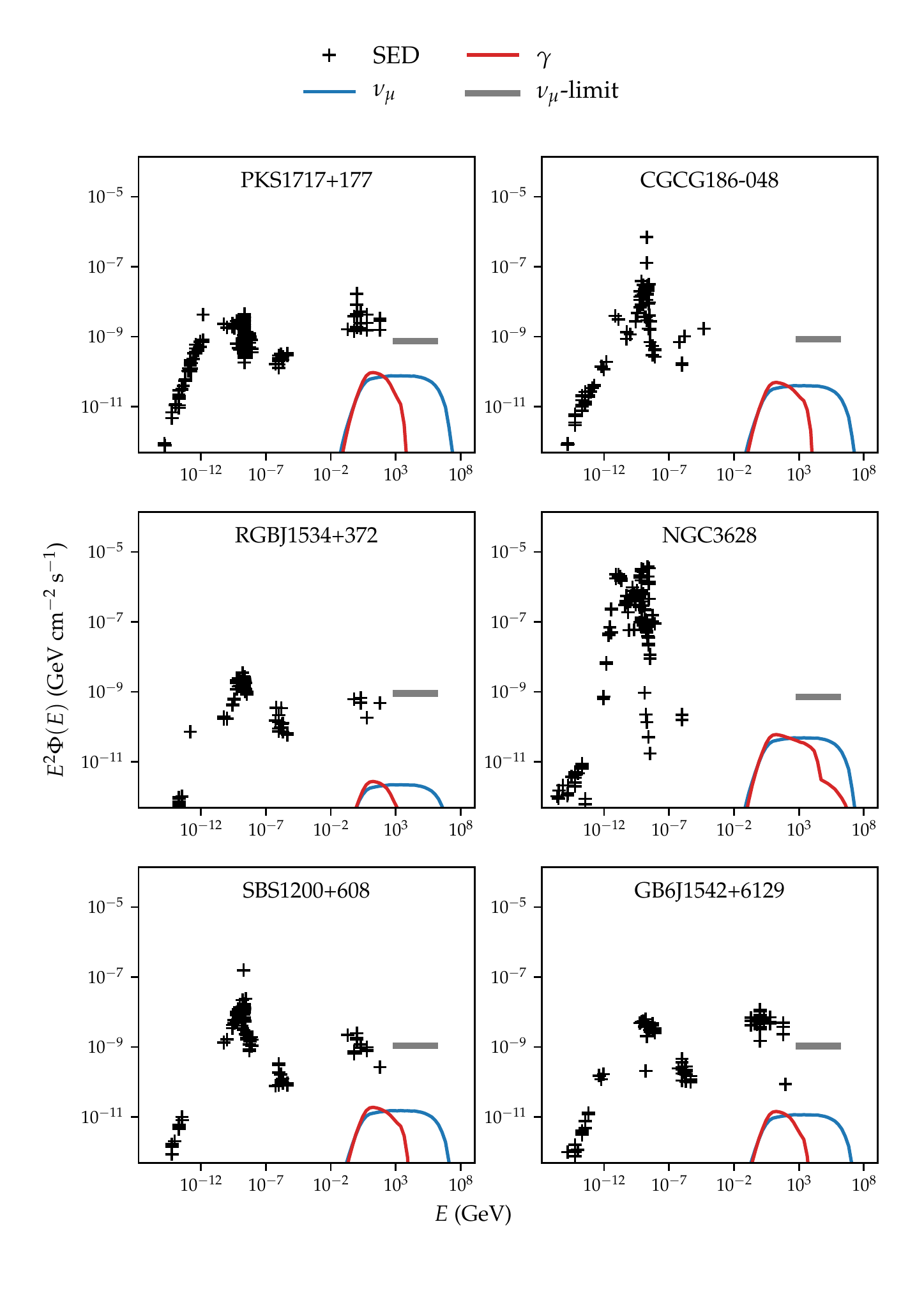}
  \caption[Hybrid SED predicted for other objects in our
  selection]{Hybrid SED for all the objects in the selection besides
    3C371, similar to Figure~\ref{fig:obj} for 3C371. Limit on the
    muon neutrino flux from~\cite{Aartsen:2017kru}. Electromagnetic
    data (citations in Table~\ref{tab:datacitations}) retrieved using
    the \texttt{SSDC SED Builder}~\cite{sedbuilder}.}
  \label{fig:otherobj}
\end{figure*}

\begin{figure*}
  \centering
  \includegraphics[width=.85\linewidth]{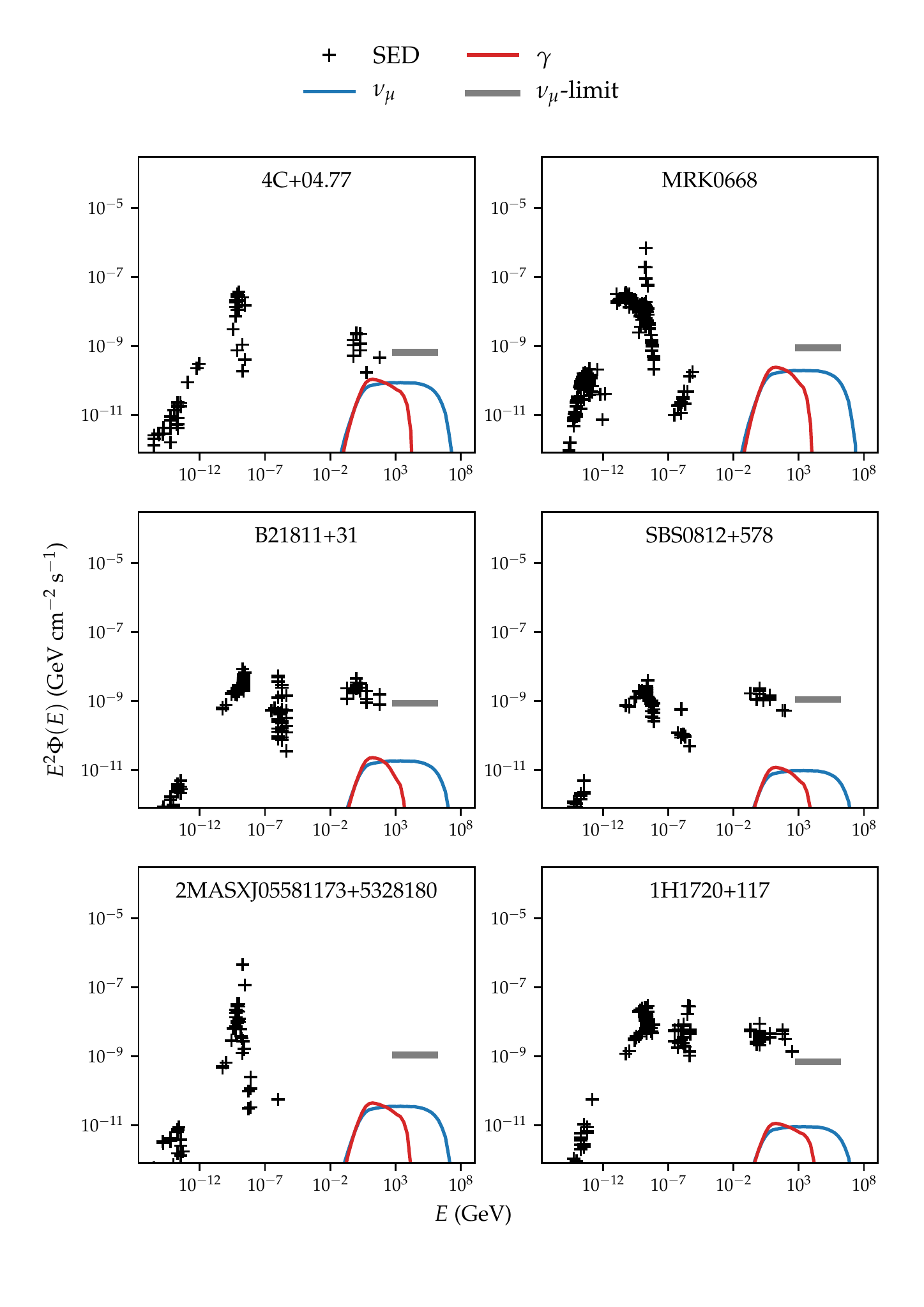}
  \caption[Hybrid SED predicted for other objects in our selection, continued]{Hybrid SED for all the other objects in the selection,
    continued.}
  \label{fig:otherobj2}
\end{figure*}

\begin{figure}
  \centering
  %
  \includegraphics[width=\columnwidth]{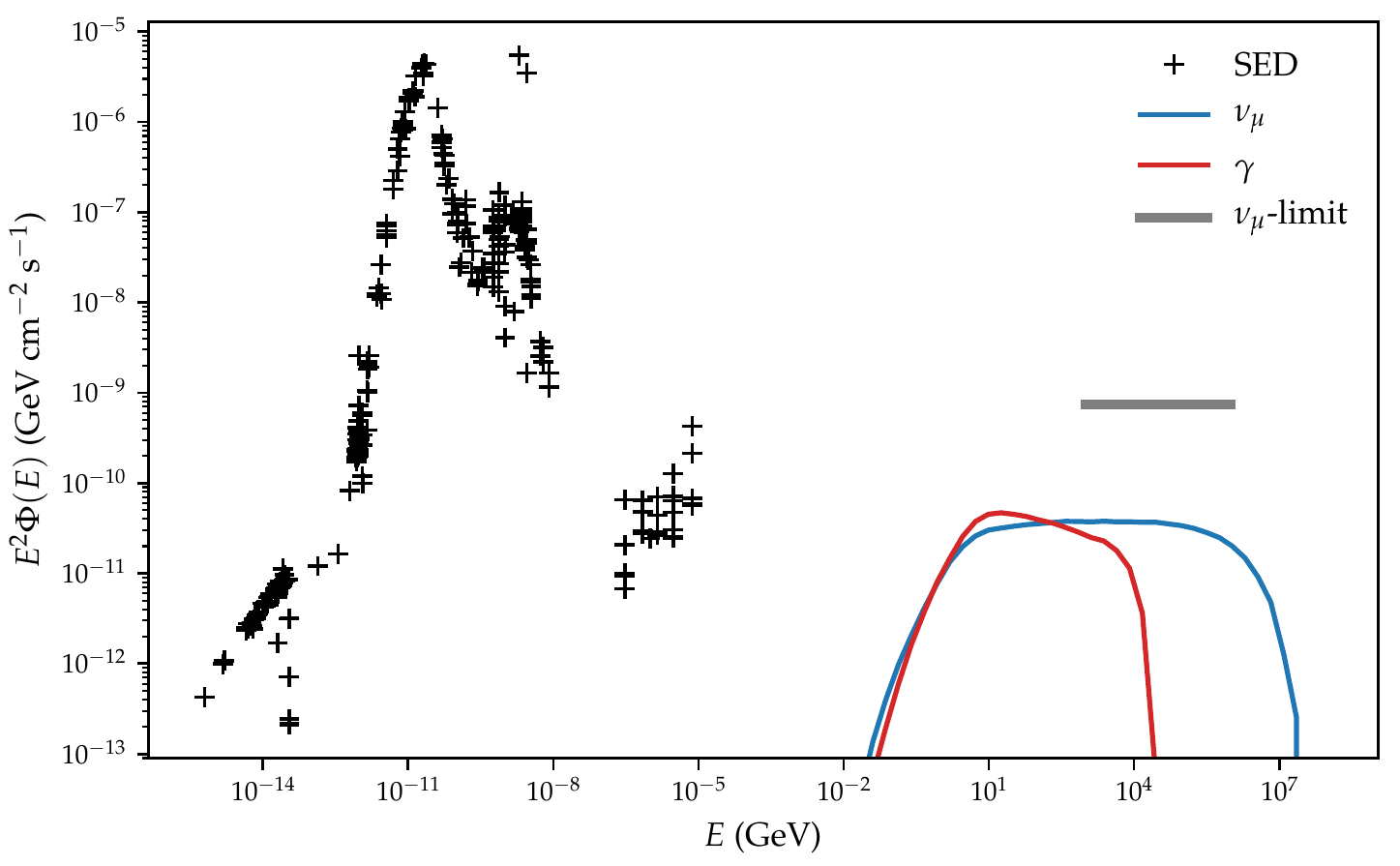}
  \caption[Hybrid SED predicted for ARP220.]{Hybrid SED for ARP~220. Note the different scale.}
  \label{fig:otherobj3}
\end{figure}

Figures~\ref{fig:otherobj},~\ref{fig:otherobj2}~and~\ref{fig:otherobj3}
show the hybrid SED for all other objects in the obscured
flat-spectrum radio AGN selection, as shown for 3C371 in
Figure~\ref{fig:obj}, for \nhtwo{}.

\section{Object citations}
\label{sec:objcit}

In table~\ref{tab:datacitations} we list the publications used for the
datapoints used in Figures~\ref{fig:obj}, \ref{fig:otherobj},
\ref{fig:otherobj2}, and \ref{fig:otherobj3}, retrieved using SED Builder~\cite{sedbuilder}.

\begin{table*}
  \centering
  \caption[Citations for the electromagnetic data used in the obscured
  AGN analysis]{Citations for the electromagnetic data used in
    Figures~\ref{fig:otherobj},~\ref{fig:otherobj2}~and~\ref{fig:otherobj3}.}
  \label{tab:datacitations}
  \small
  \begin{tabular}{ll}
   \hline\hline
    Object & Citations \\
    \hline
PKS1717+177 & {\cite{2003MNRAS.341....1M,2007ApJS..171...61H,1970ApJS...20....1D,2007MNRAS.376..371J,2007AJ....133.1947N,1998AJ....115.1693C,1990PKS...C......0W,2011A&A...536A...7P,1996ApJS..103..427G,1992ApJS...79..331W,2014A&A...571A..28P,2015arXiv150702058P,2010AJ....140.1868W,2011MNRAS.411.2770B,2013A&A...551A.142D,0067-0049-210-1-8,1999A&A...349..389V,2016A&A...588A.103B,2010ApJS..188..405A,2012ApJS..199...31N,2015ApJS..218...23A,0004-637X-779-1-27}} \\
CGCG186-048 & {\cite{2003MNRAS.341....1M,2007ApJS..171...61H,1970ApJS...20....1D,1997ApJ...475..479W,2007MNRAS.376..371J,1998AJ....115.1693C,1996ApJS..103..427G,1992ApJS...79..331W,2010AJ....140.1868W,2011MNRAS.411.2770B,1999A&A...349..389V,2016A&A...588A.103B,2008A&A...480..611S}} \\
RGBJ1534+372 & {\cite{2003MNRAS.341....1M,1997ApJ...475..479W,2007MNRAS.376..371J,2007AJ....133.1947N,1998AJ....115.1693C,1996ApJS..103..427G,2010AJ....140.1868W,2013A&A...551A.142D,0067-0049-210-1-8,1999A&A...349..389V,2016A&A...588A.103B,2008A&A...480..611S,2012ApJS..199...31N,2015ApJS..218...23A,0004-637X-779-1-27}} \\
NGC3628 & {\cite{2003MNRAS.341....1M,2007ApJS..171...61H,1997ApJ...475..479W,1998AJ....115.1693C,1990IRASF.C......0M,1994yCat.2125....0J,2011A&A...536A...7P,1996ApJS..103..427G,1992ApJS...79..331W,2014A&A...571A..28P,2015arXiv150702058P}} \\
SBS1200+608 & {\cite{2003MNRAS.341....1M,2007ApJS..171...61H,1997ApJ...475..479W,2007MNRAS.376..371J,1998AJ....115.1693C,1996ApJS..103..427G,1992ApJS...79..331W,2010AJ....140.1868W,2011MNRAS.411.2770B,2013A&A...551A.142D,0067-0049-210-1-8,1999A&A...349..389V,2016A&A...588A.103B,2010ApJS..188..405A,2012ApJS..199...31N,2015ApJS..218...23A,0004-637X-779-1-27}} \\
GB6J1542+6129 & {\cite{2003MNRAS.341....1M,2007ApJS..171...61H,1997ApJ...475..479W,1998AJ....115.1693C,1996ApJS..103..427G,1992ApJS...79..331W,2015arXiv150702058P,2013A&A...551A.142D,0067-0049-210-1-8,1999A&A...349..389V,2016A&A...588A.103B,2010ApJS..188..405A,2012ApJS..199...31N,2015ApJS..218...23A,0004-637X-779-1-27}} \\
4C+04.77 & {\cite{1970ApJS...20....1D,2014A&A...571A..28P,2010ApJS..188..405A,2012ApJS..199...31N,2015ApJS..218...23A}} \\
MRK0668 & {\cite{2003MNRAS.341....1M,2007ApJS..171...61H,1970ApJS...20....1D,1997ApJ...475..479W,2007MNRAS.376..371J,1981A&AS...45..367K,1998AJ....115.1693C,1990IRASF.C......0M,1994yCat.2125....0J,1996ApJS..103..427G,1992ApJS...79..331W,2010AJ....140.1868W,2011MNRAS.411.2770B,2013A&A...551A.142D,0067-0049-210-1-8,2015arXiv150407051R}} \\
3C371 & {\cite{2002babs.conf...63G,2003MNRAS.341....1M,2007ApJS..171...61H,1970ApJS...20....1D,1981A&AS...45..367K,2007AJ....133.1947N,1998AJ....115.1693C,1990IRASF.C......0M,1994yCat.2125....0J,2011A&A...536A...7P,1996ApJS..103..427G,1992ApJS...79..331W,2014A&A...571A..28P,2015arXiv150702058P,2009ApJS..180..283W,2010AJ....140.1868W,2011MNRAS.411.2770B,2013A&A...551A.142D,0067-0049-210-1-8,1992ApJS...80..257E,1999A&A...349..389V,2016A&A...588A.103B,2008A&A...480..611S,2010ApJS..188..405A,2012ApJS..199...31N,2015ApJS..218...23A,0004-637X-779-1-27}} \\
B21811+31 & {\cite{2003MNRAS.341....1M,2007ApJS..171...61H,1970ApJS...20....1D,2007MNRAS.376..371J,2007AJ....133.1947N,1998AJ....115.1693C,1996ApJS..103..427G,1992ApJS...79..331W,2010AJ....140.1868W,2013A&A...551A.142D,0067-0049-210-1-8,1999A&A...349..389V,2016A&A...588A.103B,2010ApJS..188..405A,2012ApJS..199...31N,2015ApJS..218...23A,0004-637X-779-1-27}} \\
SBS0812+578 & {\cite{2003MNRAS.341....1M,1997ApJ...475..479W,1998AJ....115.1693C,1996ApJS..103..427G,2010AJ....140.1868W,2011MNRAS.411.2770B,2013A&A...551A.142D,0067-0049-210-1-8,1999A&A...349..389V,2016A&A...588A.103B,2010ApJS..188..405A,2012ApJS..199...31N,2015ApJS..218...23A,0004-637X-779-1-27}} \\
2MASXJ05581173+5328180 & {\cite{2003MNRAS.341....1M,2007ApJS..171...61H,2007MNRAS.376..371J,2007AJ....133.1947N,1998AJ....115.1693C,1996ApJS..103..427G,1992ApJS...79..331W,2010AJ....140.1868W,2011MNRAS.411.2770B}} \\
1H1720+117 & {\cite{2003MNRAS.341....1M,2007ApJS..171...61H,2007MNRAS.376..371J,2007AJ....133.1947N,1998AJ....115.1693C,1996ApJS..103..427G,2010AJ....140.1868W,2011MNRAS.411.2770B,2013A&A...551A.142D,0067-0049-210-1-8,1999A&A...349..389V,2016A&A...588A.103B,1978ApJS...38..357F,2007A&A...472..705V,2008A&A...480..611S,2010ApJS..188..405A,2012ApJS..199...31N,2015ApJS..218...23A,0004-637X-779-1-27}} \\
ARP220 & {\cite{2003MNRAS.341....1M,2007ApJS..171...61H,1997ApJ...475..479W,2007MNRAS.376..371J,1998AJ....115.1693C,1990IRASF.C......0M,1994yCat.2125....0J,2011A&A...536A...7P,1996ApJS..103..427G,1992ApJS...79..331W,2014A&A...571A..28P,2015arXiv150702058P,2010AJ....140.1868W,2015arXiv150407051R}} \\

   \hline\hline
  \end{tabular}
\end{table*}

\bibliography{biblio,agnmerged}

\begin{thebibliography}{162}
\providecommand{\natexlab}[1]{#1}
\providecommand{\url}[1]{\texttt{#1}}
\expandafter\ifx\csname urlstyle\endcsname\relax
  \providecommand{\doi}[1]{doi: #1}\else
  \providecommand{\doi}{doi: \begingroup \urlstyle{rm}\Url}\fi

\bibitem[Murase et~al.(2018)Murase, Oikonomou, and Petropoulou]{Murase:2018iyl}
Kohta Murase, Foteini Oikonomou, and Maria Petropoulou.
\newblock {Blazar Flares as an Origin of High-Energy Cosmic Neutrinos?}
\newblock \emph{Astrophys. J.}, 865\penalty0 (2):\penalty0 124, 2018.
\newblock \doi{10.3847/1538-4357/aada00}.

\bibitem[Hooper et~al.(2019)Hooper, Linden, and Vieregg]{Hooper:2018wyk}
Dan Hooper, Tim Linden, and Abby Vieregg.
\newblock {Active Galactic Nuclei and the Origin of IceCube's Diffuse Neutrino
  Flux}.
\newblock \emph{JCAP}, 1902:\penalty0 012, 2019.
\newblock \doi{10.1088/1475-7516/2019/02/012}.

\bibitem[Palladino et~al.(2019)Palladino, Rodrigues, Gao, and
  Winter]{Palladino:2018lov}
Andrea Palladino, Xavier Rodrigues, Shan Gao, and Walter Winter.
\newblock {Interpretation of the diffuse astrophysical neutrino flux in terms
  of the blazar sequence}.
\newblock \emph{Astrophys. J.}, 871\penalty0 (1):\penalty0 41, 2019.
\newblock \doi{10.3847/1538-4357/aaf507}.

\bibitem[Aartsen et~al.(2017{\natexlab{a}})]{Aartsen:2017mau}
M.~G. Aartsen et~al.
\newblock {The IceCube Neutrino Observatory - Contributions to ICRC 2017 Part
  II: Properties of the Atmospheric and Astrophysical Neutrino Flux}.
\newblock 2017{\natexlab{a}}.

\bibitem[Aartsen et~al.(2019{\natexlab{a}})]{Aartsen:2018ywr}
M.~G. Aartsen et~al.
\newblock {Search for steady point-like sources in the astrophysical muon
  neutrino flux with 8 years of IceCube data}.
\newblock \emph{Eur. Phys. J.}, C79\penalty0 (3):\penalty0 234,
  2019{\natexlab{a}}.
\newblock \doi{10.1140/epjc/s10052-019-6680-0}.

\bibitem[Aartsen et~al.(2017{\natexlab{b}})]{Aartsen:2016lir}
M.~G. Aartsen et~al.
\newblock {The contribution of Fermi-2LAC blazars to the diffuse TeV-PeV
  neutrino flux}.
\newblock \emph{Astrophys. J.}, 835\penalty0 (1):\penalty0 45,
  2017{\natexlab{b}}.
\newblock \doi{10.3847/1538-4357/835/1/45}.

\bibitem[Aartsen et~al.(2017{\natexlab{c}})]{Aartsen:2017kru}
M.~G. Aartsen et~al.
\newblock {The IceCube Neutrino Observatory - Contributions to ICRC 2017 Part
  I: Searches for the Sources of Astrophysical Neutrinos}.
\newblock 2017{\natexlab{c}}.

\bibitem[Aartsen et~al.(2017{\natexlab{d}})]{Aartsen:2017wea}
M.~G. Aartsen et~al.
\newblock {Extending the search for muon neutrinos coincident with gamma-ray
  bursts in IceCube data}.
\newblock \emph{Astrophys. J.}, 843\penalty0 (2):\penalty0 112,
  2017{\natexlab{d}}.
\newblock \doi{10.3847/1538-4357/aa7569}.

\bibitem[Lipari(2008)]{Lipari:2008zf}
Paolo Lipari.
\newblock {Proton and Neutrino Extragalactic Astronomy}.
\newblock \emph{Phys. Rev.}, D78:\penalty0 083011, 2008.
\newblock \doi{10.1103/PhysRevD.78.083011}.

\bibitem[{Silvestri} and {Barwick}(2010)]{2010PhRvD..81b3001S}
A.~{Silvestri} and S.~W. {Barwick}.
\newblock {Constraints on extragalactic point source flux from diffuse neutrino
  limits}.
\newblock \emph{\prd}, 81\penalty0 (2):\penalty0 023001, January 2010.
\newblock \doi{10.1103/PhysRevD.81.023001}.

\bibitem[Murase et~al.(2012)Murase, Beacom, and Takami]{Murase:2012df}
Kohta Murase, John~F. Beacom, and Hajime Takami.
\newblock {Gamma-Ray and Neutrino Backgrounds as Probes of the High-Energy
  Universe: Hints of Cascades, General Constraints, and Implications for TeV
  Searches}.
\newblock \emph{JCAP}, 1208:\penalty0 030, 2012.
\newblock \doi{10.1088/1475-7516/2012/08/030}.

\bibitem[Ahlers and Halzen(2014)]{Ahlers:2014ioa}
Markus Ahlers and Francis Halzen.
\newblock {Pinpointing Extragalactic Neutrino Sources in Light of Recent
  IceCube Observations}.
\newblock \emph{Phys. Rev.}, D90\penalty0 (4):\penalty0 043005, 2014.
\newblock \doi{10.1103/PhysRevD.90.043005}.

\bibitem[Kowalski(2015)]{Kowalski:2014zda}
Marek Kowalski.
\newblock {Status of High-Energy Neutrino Astronomy}.
\newblock \emph{J. Phys. Conf. Ser.}, 632\penalty0 (1):\penalty0 012039, 2015.
\newblock \doi{10.1088/1742-6596/632/1/012039}.

\bibitem[Murase and Waxman(2016)]{Murase:2016gly}
Kohta Murase and Eli Waxman.
\newblock {Constraining High-Energy Cosmic Neutrino Sources: Implications and
  Prospects}.
\newblock \emph{Phys. Rev.}, D94\penalty0 (10):\penalty0 103006, 2016.
\newblock \doi{10.1103/PhysRevD.94.103006}.

\bibitem[Murase et~al.(2013)Murase, Ahlers, and Lacki]{Murase:2013rfa}
Kohta Murase, Markus Ahlers, and Brian~C. Lacki.
\newblock {Testing the Hadronuclear Origin of PeV Neutrinos Observed with
  IceCube}.
\newblock \emph{Phys. Rev.}, D88\penalty0 (12):\penalty0 121301, 2013.
\newblock \doi{10.1103/PhysRevD.88.121301}.

\bibitem[Tamborra et~al.(2014)Tamborra, Ando, and Murase]{Tamborra:2014xia}
Irene Tamborra, Shin'ichiro Ando, and Kohta Murase.
\newblock {Star-forming galaxies as the origin of diffuse high-energy
  backgrounds: Gamma-ray and neutrino connections, and implications for
  starburst history}.
\newblock \emph{JCAP}, 1409:\penalty0 043, 2014.
\newblock \doi{10.1088/1475-7516/2014/09/043}.

\bibitem[Bechtol et~al.(2017)Bechtol, Ahlers, Di~Mauro, Ajello, and
  Vandenbroucke]{Bechtol:2015uqb}
Keith Bechtol, Markus Ahlers, Mattia Di~Mauro, Marco Ajello, and Justin
  Vandenbroucke.
\newblock {Evidence against star-forming galaxies as the dominant source of
  IceCube neutrinos}.
\newblock \emph{Astrophys. J.}, 836\penalty0 (1):\penalty0 47, 2017.
\newblock \doi{10.3847/1538-4357/836/1/47}.

\bibitem[Ackermann et~al.(2016)]{TheFermi-LAT:2015ykq}
M.~Ackermann et~al.
\newblock {Resolving the Extragalactic $\gamma$-Ray Background above 50 GeV
  with the Fermi Large Area Telescope}.
\newblock \emph{Phys. Rev. Lett.}, 116\penalty0 (15):\penalty0 151105, 2016.
\newblock \doi{10.1103/PhysRevLett.116.151105}.

\bibitem[Lisanti et~al.(2016)Lisanti, Mishra-Sharma, Necib, and
  Safdi]{Lisanti:2016jub}
Mariangela Lisanti, Siddharth Mishra-Sharma, Lina Necib, and Benjamin~R. Safdi.
\newblock {Deciphering Contributions to the Extragalactic Gamma-Ray Background
  from 2 GeV to 2 TeV}.
\newblock \emph{Astrophys. J.}, 832\penalty0 (2):\penalty0 117, 2016.
\newblock \doi{10.3847/0004-637X/832/2/117}.

\bibitem[Ajello et~al.(2015)]{Ajello:2015mfa}
M.~Ajello et~al.
\newblock {The Origin of the Extragalactic Gamma-Ray Background and
  Implications for Dark-Matter Annihilation}.
\newblock \emph{Astrophys. J.}, 800\penalty0 (2):\penalty0 L27, 2015.
\newblock \doi{10.1088/2041-8205/800/2/L27}.

\bibitem[Ackermann et~al.(2015{\natexlab{a}})]{Ackermann:2015tah}
M.~Ackermann et~al.
\newblock {Limits on Dark Matter Annihilation Signals from the Fermi LAT 4-year
  Measurement of the Isotropic Gamma-Ray Background}.
\newblock \emph{JCAP}, 1509\penalty0 (09):\penalty0 008, 2015{\natexlab{a}}.
\newblock \doi{10.1088/1475-7516/2015/09/008}.

\bibitem[Di~Mauro et~al.(2014)Di~Mauro, Cuoco, Donato, and
  Siegal-Gaskins]{DiMauro:2014wha}
Mattia Di~Mauro, Alessandro Cuoco, Fiorenza Donato, and Jennifer~M.
  Siegal-Gaskins.
\newblock {Fermi-LAT $/gamma$-ray anisotropy and intensity explained by
  unresolved Radio-Loud Active Galactic Nuclei}.
\newblock \emph{JCAP}, 1411\penalty0 (11):\penalty0 021, 2014.
\newblock \doi{10.1088/1475-7516/2014/11/021}.

\bibitem[Inoue(2014)]{Inoue:2014ona}
Yoshiyuki Inoue.
\newblock {Cosmic Gamma-ray Background Radiation}.
\newblock In \emph{{5th International Fermi Symposium Nagoya, Japan, October
  20-24, 2014}}, 2014.
\newblock URL
  \url{https://inspirehep.net/record/1334100/files/arXiv:1412.3886.pdf}.

\bibitem[Costamante(2013)]{Costamante:2013sva}
Luigi Costamante.
\newblock {Gamma-rays from Blazars and the Extragalactic Background Light}.
\newblock \emph{Int. J. Mod. Phys.}, D22\penalty0 (13):\penalty0 1330025, 2013.
\newblock \doi{10.1142/S0218271813300255}.

\bibitem[Atwood et~al.(2009)]{Atwood:2009ez}
W.~B. Atwood et~al.
\newblock {The Large Area Telescope on the Fermi Gamma-ray Space Telescope
  Mission}.
\newblock \emph{Astrophys. J.}, 697:\penalty0 1071--1102, 2009.
\newblock \doi{10.1088/0004-637X/697/2/1071}.

\bibitem[Atwood et~al.(2013)]{Atwood:2013rka}
W.~Atwood et~al.
\newblock {Pass 8: Toward the Full Realization of the Fermi-LAT Scientific
  Potential}.
\newblock 2013.
\newblock URL
  \url{https://inspirehep.net/record/1223837/files/arXiv:1303.3514.pdf}.

\bibitem[Palladino et~al.(2018)Palladino, Fedynitch, Rasmussen, and
  Taylor]{Palladino:2018bqf}
Andrea Palladino, Anatoli Fedynitch, Rasmus~W. Rasmussen, and Andrew~M. Taylor.
\newblock {IceCube Neutrinos from Hadronically Powered Gamma-Ray Galaxies}.
\newblock 2018.

\bibitem[Murase et~al.(2016)Murase, Guetta, and Ahlers]{Murase:2015xka}
Kohta Murase, Dafne Guetta, and Markus Ahlers.
\newblock {Hidden Cosmic-Ray Accelerators as an Origin of TeV-PeV Cosmic
  Neutrinos}.
\newblock \emph{Phys. Rev. Lett.}, 116\penalty0 (7):\penalty0 071101, 2016.
\newblock \doi{10.1103/PhysRevLett.116.071101}.

\bibitem[Maggi et~al.(2016)Maggi, Buitink, Correa, de~Vries, Gentile, Tavares,
  Scholten, van Eijndhoven, Vereecken, and Winchen]{Maggi:2016bbi}
G.~Maggi, S.~Buitink, P.~Correa, K.~D. de~Vries, G.~Gentile, J.~León Tavares,
  O.~Scholten, N.~van Eijndhoven, M.~Vereecken, and T.~Winchen.
\newblock {Obscured flat spectrum radio active galactic nuclei as sources of
  high-energy neutrinos}.
\newblock \emph{Phys. Rev.}, D94\penalty0 (10):\penalty0 103007, 2016.
\newblock \doi{10.1103/PhysRevD.94.103007}.

\bibitem[Bethe and Heitler(1934)]{Bethe:1934za}
H.~Bethe and W.~Heitler.
\newblock {On the Stopping of fast particles and on the creation of positive
  electrons}.
\newblock \emph{Proc. Roy. Soc. Lond.}, A146:\penalty0 83--112, 1934.
\newblock \doi{10.1098/rspa.1934.0140}.

\bibitem[Kelner et~al.(2006)Kelner, Aharonian, and Bugayov]{Kelner:2006tc}
S.~R. Kelner, Felex~A. Aharonian, and V.~V. Bugayov.
\newblock {Energy spectra of gamma-rays, electrons and neutrinos produced at
  proton-proton interactions in the very high energy regime}.
\newblock \emph{Phys. Rev.}, D74:\penalty0 034018, 2006.
\newblock \doi{10.1103/PhysRevD.74.034018, 10.1103/PhysRevD.79.039901}.
\newblock [Erratum: Phys. Rev.D79,039901(2009)].

\bibitem[Nellen et~al.(1993)Nellen, Mannheim, and Biermann]{Nellen:1992dw}
Lukas Nellen, Karl Mannheim, and Peter~L. Biermann.
\newblock {Neutrino production through hadronic cascades in AGN accretion
  disks}.
\newblock \emph{Phys. Rev.}, D47:\penalty0 5270--5274, 1993.
\newblock \doi{10.1103/PhysRevD.47.5270}.

\bibitem[Becker~Tjus et~al.(2014)Becker~Tjus, Eichmann, Halzen, Kheirandish,
  and Saba]{Tjus:2014dna}
J.~Becker~Tjus, B.~Eichmann, F.~Halzen, A.~Kheirandish, and S.~M. Saba.
\newblock {High-energy neutrinos from radio galaxies}.
\newblock \emph{Phys. Rev.}, D89\penalty0 (12):\penalty0 123005, 2014.
\newblock \doi{10.1103/PhysRevD.89.123005}.

\bibitem[Reynoso et~al.(2008{\natexlab{a}})Reynoso, Christiansen, and
  Romero]{Reynoso:2007us}
Matias~M. Reynoso, Hugo~R. Christiansen, and Gustavo~E. Romero.
\newblock {Gamma-ray absorption in the microquasar SS433}.
\newblock \emph{Astropart. Phys.}, 28:\penalty0 565--572, 2008{\natexlab{a}}.
\newblock \doi{10.1016/j.astropartphys.2007.10.005}.

\bibitem[Reynoso et~al.(2008{\natexlab{b}})Reynoso, Romero, and
  Christiansen]{Reynoso:2008nk}
Matias~M. Reynoso, Gustavo~E. Romero, and Hugo~R. Christiansen.
\newblock {Production of gamma rays and neutrinos in the dark jets of the
  microquasar SS433}.
\newblock \emph{Mon. Not. Roy. Astron. Soc.}, 387:\penalty0 1745--1754,
  2008{\natexlab{b}}.
\newblock \doi{10.1111/j.1365-2966.2008.13364.x}.

\bibitem[Reynoso and Carulli(2019)]{Reynoso:2019vrp}
Matías~M. Reynoso and Agustín~M. Carulli.
\newblock {On the possibilities of high-energy neutrino production in the jets
  of microquasar SS433 in light of new observational data}.
\newblock \emph{Astropart. Phys.}, 109:\penalty0 25--32, 2019.
\newblock \doi{10.1016/j.astropartphys.2019.02.003}.

\bibitem[{Pittard} et~al.(2010){Pittard}, {Hartquist}, and
  {Falle}]{2010MNRAS.405..821P}
J.~M. {Pittard}, T.~W. {Hartquist}, and S.~A.~E.~G. {Falle}.
\newblock {The turbulent destruction of clouds - II. Mach number dependence,
  mass-loss rates and tail formation}.
\newblock \emph{\mnras}, 405\penalty0 (2):\penalty0 821--838, Jun 2010.
\newblock \doi{10.1111/j.1365-2966.2010.16504.x}.

\bibitem[{Araudo} et~al.(2010){Araudo}, {Bosch-Ramon}, and
  {Romero}]{2010A&A...522A..97A}
A.~T. {Araudo}, V.~{Bosch-Ramon}, and G.~E. {Romero}.
\newblock {Gamma rays from cloud penetration at the base of AGN jets}.
\newblock \emph{\aap}, 522:\penalty0 A97, November 2010.
\newblock \doi{10.1051/0004-6361/201014660}.

\bibitem[Dar and Laor(1997)]{Dar:1996qv}
Arnon Dar and Ari Laor.
\newblock {Hadronic production of TeV gamma-ray flares from blazars}.
\newblock \emph{Astrophys. J.}, 478:\penalty0 L5--L8, 1997.
\newblock \doi{10.1086/310544}.

\bibitem[Senno et~al.(2016)Senno, Murase, and Meszaros]{Senno:2015tsn}
Nicholas Senno, Kohta Murase, and Peter Meszaros.
\newblock {Choked Jets and Low-Luminosity Gamma-Ray Bursts as Hidden Neutrino
  Sources}.
\newblock \emph{Phys. Rev.}, D93\penalty0 (8):\penalty0 083003, 2016.
\newblock \doi{10.1103/PhysRevD.93.083003}.

\bibitem[Liu et~al.(2018)Liu, Wang, Xue, Taylor, Wang, Li, and
  Yan]{Liu:2018utd}
Ruo-Yu Liu, Kai Wang, Rui Xue, Andrew~M. Taylor, Xiang-Yu Wang, Zhuo Li, and
  Huirong Yan.
\newblock {A hadronuclear interpretation of a high-energy neutrino event
  coincident with a blazar flare}.
\newblock 2018.

\bibitem[{Barkov} et~al.(2012){Barkov}, {Bosch-Ramon}, and
  {Aharonian}]{2012ApJ...755..170B}
Maxim~V. {Barkov}, Valentí {Bosch-Ramon}, and Felix~A. {Aharonian}.
\newblock {Interpretation of the Flares of M87 at TeV Energies in the Cloud-Jet
  Interaction Scenario}.
\newblock \emph{\apj}, 755\penalty0 (2):\penalty0 170, Aug 2012.
\newblock \doi{10.1088/0004-637X/755/2/170}.

\bibitem[Wang et~al.(2018)Wang, Liu, Li, Wang, and Dai]{Wang:2018zln}
Kai Wang, Ruo-Yu Liu, Zhuo Li, Xiang-Yu Wang, and Zi-Gao Dai.
\newblock {Jet-cloud/star interaction as an interpretation of neutrino outburst
  from the blazar TXS 0506+056}.
\newblock 2018.

\bibitem[del Palacio et~al.(2019)del Palacio, Bosch-Ramon, and
  Romero]{delPalacio:2019mql}
S.~del Palacio, V.~Bosch-Ramon, and G.~E. Romero.
\newblock {Gamma rays from jets interacting with BLR clouds in blazars}.
\newblock \emph{Astron. Astrophys.}, 623:\penalty0 A101, 2019.
\newblock \doi{10.1051/0004-6361/201834231}.

\bibitem[Khangulyan et~al.(2013)Khangulyan, Barkov, Bosch-Ramon, Aharonian, and
  Dorodnitsyn]{Khangulyan:2013xxa}
D.~V. Khangulyan, M.~V. Barkov, V.~Bosch-Ramon, F.~A. Aharonian, and A.~V.
  Dorodnitsyn.
\newblock {Star-Jet Interactions and Gamma-Ray Outbursts from 3C454.3}.
\newblock \emph{Astrophys. J.}, 774:\penalty0 113, 2013.
\newblock \doi{10.1088/0004-637X/774/2/113}.

\bibitem[Banasiński et~al.(2016)Banasiński, Bednarek, and
  Sitarek]{Banasinski:2016kpe}
P.~Banasiński, W.~Bednarek, and J.~Sitarek.
\newblock {Orphan $\gamma$-ray flares from relativistic blobs encountering
  luminous stars}.
\newblock \emph{Mon. Not. Roy. Astron. Soc.}, 463\penalty0 (1):\penalty0
  L26--L30, 2016.
\newblock \doi{10.1093/mnrasl/slw149}.

\bibitem[Bianchi et~al.(2012)Bianchi, Maiolino, and Risaliti]{Bianchi:2012vn}
Stefano Bianchi, Roberto Maiolino, and Guido Risaliti.
\newblock {AGN Obscuration and the Unified Model}.
\newblock \emph{Adv. Astron.}, 2012:\penalty0 782030, 2012.
\newblock \doi{10.1155/2012/782030}.

\bibitem[{Ramos Almeida} and {Ricci}(2017)]{2017NatAs...1..679R}
Cristina {Ramos Almeida} and Claudio {Ricci}.
\newblock {Nuclear obscuration in active galactic nuclei}.
\newblock \emph{Nature Astronomy}, 1:\penalty0 679--689, Oct 2017.
\newblock \doi{10.1038/s41550-017-0232-z}.

\bibitem[Comastri(2004)]{Comastri:2004nd}
Andrea Comastri.
\newblock {Compton thick AGN: The Dark side of the x-ray background}.
\newblock \emph{Astrophys. Space Sci. Libr.}, 308:\penalty0 245, 2004.
\newblock \doi{10.1007/978-1-4020-2471-9_8}.

\bibitem[Ricci et~al.(2015)Ricci, Ueda, Koss, Trakhtenbrot, Bauer, and
  Gandhi]{Ricci:2016lbt}
C.~Ricci, Y.~Ueda, M.~J. Koss, B.~Trakhtenbrot, F.~E. Bauer, and P.~Gandhi.
\newblock {Compton-thick Accretion in the local Universe}.
\newblock \emph{Astrophys. J.}, 815:\penalty0 L13, 2015.
\newblock \doi{10.1088/2041-8205/815/1/L13}.

\bibitem[{Weisskopf} et~al.(2000){Weisskopf}, {Tananbaum}, {Van Speybroeck},
  and {O'Dell}]{2000SPIE.4012....2W}
Martin~C. {Weisskopf}, Harvey~D. {Tananbaum}, Leon~P. {Van Speybroeck}, and
  Stephen~L. {O'Dell}.
\newblock {Chandra X-ray Observatory (CXO): overview}.
\newblock In Joachim~E. {Truemper} and Bernd {Aschenbach}, editors, \emph{X-Ray
  Optics, Instruments, and Missions III}, volume 4012 of \emph{Society of
  Photo-Optical Instrumentation Engineers (SPIE) Conference Series}, pages
  2--16, Jul 2000.
\newblock \doi{10.1117/12.391545}.

\bibitem[Li et~al.(2019)]{Li:2019zqc}
Junyao Li et~al.
\newblock {Piercing Through Highly Obscured and Compton-thick AGNs in the
  Chandra Deep Fields: I. X-ray Spectral and Long-term Variability Analyses}.
\newblock 2019.

\bibitem[Risaliti et~al.(1999)Risaliti, Maiolino, and Salvati]{Risaliti:1999yd}
G.~Risaliti, R.~Maiolino, and M.~Salvati.
\newblock {The Distribution of absorbing column densities among seyfert 2
  galaxies}.
\newblock \emph{Astrophys. J.}, 522:\penalty0 157--164, 1999.
\newblock \doi{10.1086/307623}.

\bibitem[Brightman et~al.(2014)Brightman, Nandra, Salvato, Hsu, Rangel, and
  Aird]{Brightman:2014pva}
Murray Brightman, Kirpal Nandra, Mara Salvato, Li-Ting Hsu, Cyprian Rangel, and
  James Aird.
\newblock {Compton thick active galactic nuclei in Chandra surveys}.
\newblock \emph{Mon. Not. Roy. Astron. Soc.}, 443\penalty0 (3):\penalty0
  1999--2017, 2014.
\newblock \doi{10.1093/mnras/stu1175}.

\bibitem[{García-Bernete} et~al.(2019){García-Bernete}, {Almeida},
  {Alonso-Herrero}, {Ward}, {Acosta-Pulido}, {Pereira-Santaella},
  {Hern{\'a}n-Caballero}, {Ramos}, {Gonz{\'a}lez-Mart{\'\i}n}, {Levenson},
  {Mateos}, {Carrera}, {Ricci}, {Roche}, {Marquez}, {Packham}, {Masegosa}, and
  {Fuller}]{2019arXiv190403694G}
I.~{García-Bernete}, C.~Ramos {Almeida}, A.~{Alonso-Herrero}, M.~J. {Ward},
  J.~A. {Acosta-Pulido}, M.~{Pereira-Santaella}, A.~{Hern{\'a}n-Caballero},
  A.~Asensio {Ramos}, O.~{Gonz{\'a}lez-Mart{\'\i}n}, N.~A. {Levenson},
  S.~{Mateos}, F.~J. {Carrera}, C.~{Ricci}, P.~{Roche}, I.~{Marquez},
  C.~{Packham}, J.~{Masegosa}, and L.~{Fuller}.
\newblock {Torus model properties of an ultra-hard X-ray selected sample of
  Seyfert galaxies}.
\newblock \emph{arXiv e-prints}, art. arXiv:1904.03694, Apr 2019.

\bibitem[Krimm et~al.(2013)]{Krimm:2013lwa}
Hans~A. Krimm et~al.
\newblock {The Swift/BAT Hard X-ray Transient Monitor}.
\newblock \emph{Astrophys. J. Suppl.}, 209:\penalty0 14, 2013.
\newblock \doi{10.1088/0067-0049/209/1/14}.

\bibitem[{Koss} et~al.(2016){Koss}, {Assef}, {Balokovi{\'c}}, {Stern},
  {Gandhi}, {Lamperti}, {Alexander}, {Ballantyne}, {Bauer}, {Berney}, {Brand
  t}, {Comastri}, {Gehrels}, {Harrison}, {Lansbury}, {Markwardt}, {Ricci},
  {Rivers}, {Schawinski}, {Trakhtenbrot}, {Treister}, and
  {Urry}]{2016ApJ...825...85K}
Michael~J. {Koss}, R.~{Assef}, M.~{Balokovi{\'c}}, D.~{Stern}, P.~{Gandhi},
  I.~{Lamperti}, D.~M. {Alexander}, D.~R. {Ballantyne}, F.~E. {Bauer},
  S.~{Berney}, W.~N. {Brand t}, A.~{Comastri}, N.~{Gehrels}, F.~A. {Harrison},
  G.~{Lansbury}, C.~{Markwardt}, C.~{Ricci}, E.~{Rivers}, K.~{Schawinski},
  B.~{Trakhtenbrot}, E.~{Treister}, and C.~Megan {Urry}.
\newblock {A New Population of Compton-thick AGNs Identified Using the Spectral
  Curvature above 10 keV}.
\newblock \emph{\apj}, 825\penalty0 (2):\penalty0 85, Jul 2016.
\newblock \doi{10.3847/0004-637X/825/2/85}.

\bibitem[Mateos et~al.(2017)Mateos, Carrera, Barcons, Alonso-Herrero,
  Hernán-Caballero, Page, Ramos~Almeida, Caccianiga, Miyaji, and
  Blain]{Mateos:2017rsi}
S.~Mateos, F.~J. Carrera, X.~Barcons, A.~Alonso-Herrero, A.~Hernán-Caballero,
  M.~Page, C.~Ramos~Almeida, A.~Caccianiga, T.~Miyaji, and A.~Blain.
\newblock {Survival of the obscuring torus in the most powerful active galactic
  nuclei}.
\newblock \emph{Astrophys. J.}, 841\penalty0 (2):\penalty0 L18, 2017.
\newblock \doi{10.3847/2041-8213/aa7268}.

\bibitem[{Roche} et~al.(2015){Roche}, {Alonso-Herrero}, and
  {Gonzalez-Martin}]{2015MNRAS.449.2598R}
P.~F. {Roche}, A.~{Alonso-Herrero}, and O.~{Gonzalez-Martin}.
\newblock {The silicate absorption profile in the interstellar medium towards
  the heavily obscured nucleus of NGC 4418}.
\newblock \emph{\mnras}, 449\penalty0 (3):\penalty0 2598--2603, May 2015.
\newblock \doi{10.1093/mnras/stv495}.

\bibitem[Sakamoto et~al.(2013)Sakamoto, Aalto, Costagliola, Martin, Ohyama,
  Wiedner, and Wilner]{Sakamoto:2013py}
Kazushi Sakamoto, Susanne Aalto, Francesco Costagliola, Sergio Martin, Youichi
  Ohyama, Martina~C. Wiedner, and David~J. Wilner.
\newblock {Submillimeter Interferometry of the Luminous Infrared Galaxy NGC
  4418: A Hidden Hot Nucleus with an Inflow and an Outflow}.
\newblock \emph{Astrophys. J.}, 764:\penalty0 42, 2013.
\newblock \doi{10.1088/0004-637X/764/1/42}.

\bibitem[Costagliola et~al.(2013)Costagliola, Aalto, Sakamoto, Martín,
  Beswick, Muller, and Klöckner]{Costagliola:2013tea}
F.~Costagliola, S.~Aalto, K.~Sakamoto, S.~Martín, R.~Beswick, S.~Muller, and
  H.~R. Klöckner.
\newblock {A high-resolution mm and cm study of the obscured LIRG NGC 4418 - A
  compact obscured nucleus fed by in-falling gas?}
\newblock \emph{Astron. Astrophys.}, 556:\penalty0 A66, 2013.
\newblock \doi{10.1051/0004-6361/201220634}.

\bibitem[{Lawrence} and {Elvis}(2010)]{2010arXiv1002.1759L}
A.~{Lawrence} and M.~{Elvis}.
\newblock {Misaligned Discs as Obscurers in Active Galaxies}.
\newblock \emph{ArXiv e-prints}, February 2010.

\bibitem[{Arnaud}(1996)]{1996ASPC..101...17A}
K.~A. {Arnaud}.
\newblock {XSPEC: The First Ten Years}.
\newblock In G.~H. {Jacoby} and J.~{Barnes}, editors, \emph{Astronomical Data
  Analysis Software and Systems V}, volume 101 of \emph{Astronomical Society of
  the Pacific Conference Series}, page~17, 1996.

\bibitem[Anders and Grevesse(1989)]{Anders:1989zg}
E.~Anders and N.~Grevesse.
\newblock {Abundances of the elements: Meteroritic and solar}.
\newblock \emph{Geochim. Cosmochim. Acta}, 53:\penalty0 197--214, 1989.
\newblock \doi{10.1016/0016-7037(89)90286-X}.

\bibitem[{Gatuzz} et~al.(2015){Gatuzz}, {Garc{\'{\i}}a}, {Kallman}, {Mendoza},
  and {Gorczyca}]{2015ApJ...800...29G}
E.~{Gatuzz}, J.~{Garc{\'{\i}}a}, T.~R. {Kallman}, C.~{Mendoza}, and T.~W.
  {Gorczyca}.
\newblock {ISMabs: A Comprehensive X-Ray Absorption Model for the Interstellar
  Medium}.
\newblock \emph{\apj}, 800:\penalty0 29, February 2015.
\newblock \doi{10.1088/0004-637X/800/1/29}.

\bibitem[Mannheim and Schlickeiser(1994)]{Mannheim:1994sv}
K.~Mannheim and R.~Schlickeiser.
\newblock {Interactions of Cosmic Ray Nuclei}.
\newblock \emph{Astron. Astrophys.}, 286:\penalty0 983--996, 1994.

\bibitem[Berger et~al.(2010)Berger, Hubbell, Seltzer, Chang, Coursey, Sukumar,
  Zucker, and Olsen]{XCOM}
M.J. Berger, J.H. Hubbell, S.M. Seltzer, J.~Chang, J.S. Coursey, R.~Sukumar,
  D.S. Zucker, and K~Olsen.
\newblock {XCOM: Photon Cross Section Database (version 1.5) [online]}, 2010.
\newblock URL \url{http://physics.nist.gov/xcom}.
\newblock National Institute of Standards and Technology, Gaithersburg, MD.

\bibitem[{Chodorowski} et~al.(1992){Chodorowski}, {Zdziarski}, and
  {Sikora}]{1992ApJ...400..181C}
M.~J. {Chodorowski}, A.~A. {Zdziarski}, and M.~{Sikora}.
\newblock {Reaction rate and energy-loss rate for photopair production by
  relativistic nuclei}.
\newblock \emph{\apj}, 400:\penalty0 181--185, November 1992.
\newblock \doi{10.1086/171984}.

\bibitem[Ruffini et~al.(2016)Ruffini, Vereshchagin, and Xue]{Ruffini:2015oha}
R.~Ruffini, G.~V. Vereshchagin, and S.~S. Xue.
\newblock {Cosmic absorption of ultra high energy particles}.
\newblock \emph{Astrophys. Space Sci.}, 361:\penalty0 82, 2016.
\newblock \doi{10.1007/s10509-016-2668-5}.

\bibitem[{Begelman} et~al.(1990){Begelman}, {Rudak}, and
  {Sikora}]{1990ApJ...362...38B}
Mitchell~C. {Begelman}, Bronislaw {Rudak}, and Marek {Sikora}.
\newblock {Consequences of Relativistic Proton Injection in Active Galactic
  Nuclei}.
\newblock \emph{\apj}, 362:\penalty0 38, Oct 1990.
\newblock \doi{10.1086/169241}.

\bibitem[Landau and Pomeranchuk(1953{\natexlab{a}})]{Landau:1953um}
L.~D. Landau and I.~Pomeranchuk.
\newblock {Limits of applicability of the theory of bremsstrahlung electrons
  and pair production at high-energies}.
\newblock \emph{Dokl. Akad. Nauk Ser. Fiz.}, 92:\penalty0 535--536,
  1953{\natexlab{a}}.

\bibitem[Landau and Pomeranchuk(1953{\natexlab{b}})]{Landau:1953gr}
L.~D. Landau and I.~Pomeranchuk.
\newblock {Electron cascade process at very high-energies}.
\newblock \emph{Dokl. Akad. Nauk Ser. Fiz.}, 92:\penalty0 735--738,
  1953{\natexlab{b}}.

\bibitem[Migdal(1956)]{Migdal:1956tc}
Arkady~B. Migdal.
\newblock {Bremsstrahlung and pair production in condensed media at
  high-energies}.
\newblock \emph{Phys. Rev.}, 103:\penalty0 1811--1820, 1956.
\newblock \doi{10.1103/PhysRev.103.1811}.

\bibitem[Haar(2013)]{haar2013collected}
D.T. Haar.
\newblock \emph{Collected Papers of L.D. Landau}.
\newblock Elsevier Science, 2013.
\newblock ISBN 9781483152707.
\newblock URL \url{https://books.google.be/books?id=epc4BQAAQBAJ}.

\bibitem[Klein(1999)]{Klein:1998du}
Spencer Klein.
\newblock {Suppression of Bremsstrahlung and pair production due to
  environmental factors}.
\newblock \emph{Rev. Mod. Phys.}, 71:\penalty0 1501--1538, 1999.
\newblock \doi{10.1103/RevModPhys.71.1501}.

\bibitem[Tanabashi et~al.(2018)]{Tanabashi:2018oca}
M.~Tanabashi et~al.
\newblock {Review of Particle Physics}.
\newblock \emph{Phys. Rev.}, D98\penalty0 (3):\penalty0 030001, 2018.
\newblock \doi{10.1103/PhysRevD.98.030001}.

\bibitem[{Dermer} et~al.(2012){Dermer}, {Murase}, and
  {Takami}]{2012ApJ...755..147D}
Charles~D. {Dermer}, Kohta {Murase}, and Hajime {Takami}.
\newblock {Variable Gamma-Ray Emission Induced by Ultra-high Energy Neutral
  Beams: Application to 4C +21.35}.
\newblock \emph{\apj}, 755\penalty0 (2):\penalty0 147, Aug 2012.
\newblock \doi{10.1088/0004-637X/755/2/147}.

\bibitem[Hummer et~al.(2010)Hummer, Maltoni, Winter, and Yaguna]{Hummer:2010ai}
S.~Hummer, M.~Maltoni, W.~Winter, and C.~Yaguna.
\newblock {Energy dependent neutrino flavor ratios from cosmic accelerators on
  the Hillas plot}.
\newblock \emph{Astropart. Phys.}, 34:\penalty0 205--224, 2010.
\newblock \doi{10.1016/j.astropartphys.2010.07.003}.

\bibitem[{O'Sullivan} and {Gabuzda}(2009)]{2009MNRAS.400...26O}
S.~P. {O'Sullivan} and D.~C. {Gabuzda}.
\newblock {Magnetic field strength and spectral distribution of six
  parsec-scale active galactic nuclei jets}.
\newblock \emph{\mnras}, 400\penalty0 (1):\penalty0 26--42, Nov 2009.
\newblock \doi{10.1111/j.1365-2966.2009.15428.x}.

\bibitem[Waxman and Bahcall(1997)]{Waxman:1997ti}
Eli Waxman and John~N. Bahcall.
\newblock {High-energy neutrinos from cosmological gamma-ray burst fireballs}.
\newblock \emph{Phys. Rev. Lett.}, 78:\penalty0 2292--2295, 1997.
\newblock \doi{10.1103/PhysRevLett.78.2292}.

\bibitem[{Padovani} and {Urry}(1992)]{1992ApJ...387..449P}
P.~{Padovani} and C.~M. {Urry}.
\newblock {Luminosity Functions, Relativistic Beaming, and Unified Theories of
  High-Luminosity Radio Sources}.
\newblock \emph{\apj}, 387:\penalty0 449, March 1992.
\newblock \doi{10.1086/171098}.

\bibitem[Merten et~al.(2017)Merten, Becker~Tjus, Eichmann, and
  Dettmar]{Merten:2017mzg}
Lukas Merten, Julia Becker~Tjus, Björn Eichmann, and Ralf-Jürgen Dettmar.
\newblock {On the non-thermal electron-to-proton ratio at cosmic ray
  acceleration sites}.
\newblock \emph{Astropart. Phys.}, 90:\penalty0 75--84, 2017.
\newblock \doi{10.1016/j.astropartphys.2017.02.007}.

\bibitem[Fletcher et~al.(1994)Fletcher, Gaisser, Lipari, and
  Stanev]{Fletcher:1994bd}
R.~S. Fletcher, T.~K. Gaisser, Paolo Lipari, and Todor Stanev.
\newblock {SIBYLL: An Event generator for simulation of high-energy cosmic ray
  cascades}.
\newblock \emph{Phys. Rev.}, D50:\penalty0 5710--5731, 1994.
\newblock \doi{10.1103/PhysRevD.50.5710}.

\bibitem[Kalmykov et~al.(1997)Kalmykov, Ostapchenko, and
  Pavlov]{KALMYKOV199717}
N.N. Kalmykov, S.S. Ostapchenko, and A.I. Pavlov.
\newblock Quark-gluon-string model and eas simulation problems at ultra-high
  energies.
\newblock \emph{Nuclear Physics B - Proceedings Supplements}, 52\penalty0
  (3):\penalty0 17 -- 28, 1997.
\newblock ISSN 0920-5632.
\newblock \doi{https://doi.org/10.1016/S0920-5632(96)00846-8}.
\newblock URL
  \url{http://www.sciencedirect.com/science/article/pii/S0920563296008468}.

\bibitem[Ahn et~al.(2009)Ahn, Engel, Gaisser, Lipari, and Stanev]{Ahn:2009wx}
Eun-Joo Ahn, Ralph Engel, Thomas~K. Gaisser, Paolo Lipari, and Todor Stanev.
\newblock {Cosmic ray interaction event generator SIBYLL 2.1}.
\newblock \emph{Phys. Rev.}, D80:\penalty0 094003, 2009.
\newblock \doi{10.1103/PhysRevD.80.094003}.

\bibitem[Riehn et~al.(2015)Riehn, Engel, Fedynitch, Gaisser, and
  Stanev]{Engel:2015dxa}
Felix Riehn, Ralph Engel, Anatoli Fedynitch, Thomas~K. Gaisser, and Todor
  Stanev.
\newblock {Charm production in SIBYLL}.
\newblock \emph{EPJ Web Conf.}, 99:\penalty0 12001, 2015.
\newblock \doi{10.1051/epjconf/20159912001}.

\bibitem[Riehn et~al.(2016)Riehn, Engel, Fedynitch, Gaisser, and
  Stanev]{Riehn:2015oba}
Felix Riehn, Ralph Engel, Anatoli Fedynitch, Thomas~K. Gaisser, and Todor
  Stanev.
\newblock {A new version of the event generator Sibyll}.
\newblock \emph{PoS}, ICRC2015:\penalty0 558, 2016.
\newblock \doi{10.22323/1.236.0558}.

\bibitem[Agostinelli et~al.(2003)]{Agostinelli:2002hh}
S.~Agostinelli et~al.
\newblock {GEANT4: A Simulation toolkit}.
\newblock \emph{Nucl. Instrum. Meth.}, A506:\penalty0 250--303, 2003.
\newblock \doi{10.1016/S0168-9002(03)01368-8}.

\bibitem[Sjöstrand et~al.(2015)Sjöstrand, Ask, Christiansen, Corke, Desai,
  Ilten, Mrenna, Prestel, Rasmussen, and Skands]{Sjostrand:2014zea}
Torbjörn Sjöstrand, Stefan Ask, Jesper~R. Christiansen, Richard Corke,
  Nishita Desai, Philip Ilten, Stephen Mrenna, Stefan Prestel, Christine~O.
  Rasmussen, and Peter~Z. Skands.
\newblock {An Introduction to PYTHIA 8.2}.
\newblock \emph{Comput. Phys. Commun.}, 191:\penalty0 159--177, 2015.
\newblock \doi{10.1016/j.cpc.2015.01.024}.

\bibitem[{Ghisellini} et~al.(2010){Ghisellini}, {Tavecchio}, {Foschini},
  {Ghirlanda}, {Maraschi}, and {Celotti}]{2010MNRAS.402..497G}
G.~{Ghisellini}, F.~{Tavecchio}, L.~{Foschini}, G.~{Ghirlanda}, L.~{Maraschi},
  and A.~{Celotti}.
\newblock {General physical properties of bright Fermi blazars}.
\newblock \emph{\mnras}, 402:\penalty0 497--518, February 2010.
\newblock \doi{10.1111/j.1365-2966.2009.15898.x}.

\bibitem[Beckmann and Shrader(2012)]{beckmann2012active}
V.~Beckmann and C.~Shrader.
\newblock \emph{Active Galactic Nuclei}.
\newblock Physics textbook. Wiley, 2012.
\newblock ISBN 9783527410910.
\newblock URL \url{https://books.google.be/books?id=c44yvicEX70C}.

\bibitem[Schlickeiser(2014)]{schlickeiser2014cosmic}
R.~Schlickeiser.
\newblock \emph{Cosmic Ray Astrophysics}.
\newblock Astronomy and Astrophysics Library. Springer Berlin Heidelberg, 2014.
\newblock ISBN 9783662048153.
\newblock URL \url{https://books.google.be/books?id=1K8PswEACAAJ}.

\bibitem[Inoue et~al.(2013)Inoue, Inoue, Kobayashi, Makiya, Niino, and
  Totani]{Inoue:2012bk}
Yoshiyuki Inoue, Susumu Inoue, Masakazu A.~R. Kobayashi, Ryu Makiya, Yuu Niino,
  and Tomonori Totani.
\newblock {Extragalactic Background Light from Hierarchical Galaxy Formation:
  Gamma-ray Attenuation up to the Epoch of Cosmic Reionization and the First
  Stars}.
\newblock \emph{Astrophys. J.}, 768:\penalty0 197, 2013.
\newblock \doi{10.1088/0004-637X/768/2/197}.

\bibitem[Aartsen et~al.(2019{\natexlab{b}})]{Aartsen:2019swn}
M.~G. Aartsen et~al.
\newblock {Neutrino astronomy with the next generation IceCube Neutrino
  Observatory}.
\newblock 2019{\natexlab{b}}.

\bibitem[{Giommi} et~al.(2002){Giommi}, {Capalbi}, {Fiocchi}, {Memola},
  {Perri}, {Piranomonte}, {Rebecchi}, and {Massaro}]{2002babs.conf...63G}
P.~{Giommi}, M.~{Capalbi}, M.~{Fiocchi}, E.~{Memola}, M.~{Perri},
  S.~{Piranomonte}, S.~{Rebecchi}, and E.~{Massaro}.
\newblock {A Catalog of 157 X-ray Spectra and 84 Spectral Energy Distributions
  of Blazars Observed with BeppoSAX}.
\newblock In P.~{Giommi}, E.~{Massaro}, and G.~{Palumbo}, editors, \emph{Blazar
  Astrophysics with BeppoSAX and Other Observatories}, page~63, 2002.

\bibitem[{Myers} et~al.(2003){Myers}, {Jackson}, {Browne}, {de Bruyn},
  {Pearson}, {Readhead}, {Wilkinson}, {Biggs}, {Blandford}, {Fassnacht},
  {Koopmans}, {Marlow}, {McKean}, {Norbury}, {Phillips}, {Rusin}, {Shepherd},
  and {Sykes}]{2003MNRAS.341....1M}
S.~T. {Myers}, N.~J. {Jackson}, I.~W.~A. {Browne}, A.~G. {de Bruyn}, T.~J.
  {Pearson}, A.~C.~S. {Readhead}, P.~N. {Wilkinson}, A.~D. {Biggs}, R.~D.
  {Blandford}, C.~D. {Fassnacht}, L.~V.~E. {Koopmans}, D.~R. {Marlow}, J.~P.
  {McKean}, M.~A. {Norbury}, P.~M. {Phillips}, D.~{Rusin}, M.~C. {Shepherd},
  and C.~M. {Sykes}.
\newblock {The Cosmic Lens All-Sky Survey - I. Source selection and
  observations}.
\newblock \emph{\mnras}, 341:\penalty0 1--12, May 2003.
\newblock \doi{10.1046/j.1365-8711.2003.06256.x}.

\bibitem[{Healey} et~al.(2007){Healey}, {Romani}, {Taylor}, {Sadler}, {Ricci},
  {Murphy}, {Ulvestad}, and {Winn}]{2007ApJS..171...61H}
S.~E. {Healey}, R.~W. {Romani}, G.~B. {Taylor}, E.~M. {Sadler}, R.~{Ricci},
  T.~{Murphy}, J.~S. {Ulvestad}, and J.~N. {Winn}.
\newblock {CRATES: An All-Sky Survey of Flat-Spectrum Radio Sources}.
\newblock \emph{\apjs}, 171:\penalty0 61--71, July 2007.
\newblock \doi{10.1086/513742}.

\bibitem[{Dixon}(1970)]{1970ApJS...20....1D}
R.~S. {Dixon}.
\newblock {A Master List of Radio Sources}.
\newblock \emph{\apjs}, 20:\penalty0 1--503, July 1970.
\newblock \doi{10.1086/190216}.

\bibitem[{Kuehr} et~al.(1981){Kuehr}, {Witzel}, {Pauliny-Toth}, and
  {Nauber}]{1981A&AS...45..367K}
H.~{Kuehr}, A.~{Witzel}, I.~I.~K. {Pauliny-Toth}, and U.~{Nauber}.
\newblock {A catalogue of extragalactic radio sources having flux densities
  greater than 1 Jy at 5 GHz}.
\newblock \emph{\aaps}, 45:\penalty0 367--430, September 1981.

\bibitem[{Nieppola} et~al.(2007){Nieppola}, {Tornikoski},
  {L{\"a}hteenm{\"a}ki}, {Valtaoja}, {Hakala}, {Hovatta}, {Kotiranta},
  {Nummila}, {Ojala}, {Parviainen}, {Ranta}, {Saloranta}, {Torniainen}, and
  {Tr{\"o}ller}]{2007AJ....133.1947N}
E.~{Nieppola}, M.~{Tornikoski}, A.~{L{\"a}hteenm{\"a}ki}, E.~{Valtaoja},
  T.~{Hakala}, T.~{Hovatta}, M.~{Kotiranta}, S.~{Nummila}, T.~{Ojala},
  M.~{Parviainen}, M.~{Ranta}, P.-M. {Saloranta}, I.~{Torniainen}, and
  M.~{Tr{\"o}ller}.
\newblock {37 GHz Observations of a Large Sample of BL Lacertae Objects}.
\newblock \emph{\aj}, 133:\penalty0 1947--1953, May 2007.
\newblock \doi{10.1086/512609}.

\bibitem[{Condon} et~al.(1998){Condon}, {Cotton}, {Greisen}, {Yin}, {Perley},
  {Taylor}, and {Broderick}]{1998AJ....115.1693C}
J.~J. {Condon}, W.~D. {Cotton}, E.~W. {Greisen}, Q.~F. {Yin}, R.~A. {Perley},
  G.~B. {Taylor}, and J.~J. {Broderick}.
\newblock {The NRAO VLA Sky Survey}.
\newblock \emph{\aj}, 115:\penalty0 1693--1716, May 1998.
\newblock \doi{10.1086/300337}.

\bibitem[{Moshir} and {et al.}(1990)]{1990IRASF.C......0M}
M.~{Moshir} and {et al.}
\newblock {IRAS Faint Source Catalogue, version 2.0.}
\newblock In \emph{IRAS Faint Source Catalogue, version 2.0 (1990)}, page~0,
  1990.

\bibitem[{Joint Iras Science}(1994)]{1994yCat.2125....0J}
W.~G. {Joint Iras Science}.
\newblock {VizieR Online Data Catalog: IRAS catalogue of Point Sources, Version
  2.0 (IPAC 1986)}.
\newblock \emph{VizieR Online Data Catalog}, 2125:\penalty0 0, January 1994.

\bibitem[{Planck Collaboration} et~al.(2011){Planck Collaboration}, {Ade},
  {Aghanim}, {Arnaud}, {Ashdown}, {Aumont}, {Baccigalupi}, {Balbi}, {Banday},
  {Barreiro}, and et~al.]{2011A&A...536A...7P}
{Planck Collaboration}, P.~A.~R. {Ade}, N.~{Aghanim}, M.~{Arnaud},
  M.~{Ashdown}, J.~{Aumont}, C.~{Baccigalupi}, A.~{Balbi}, A.~J. {Banday},
  R.~B. {Barreiro}, and et~al.
\newblock {Planck early results. VII. The Early Release Compact Source
  Catalogue}.
\newblock \emph{\aap}, 536:\penalty0 A7, December 2011.
\newblock \doi{10.1051/0004-6361/201116474}.

\bibitem[{Gregory} et~al.(1996){Gregory}, {Scott}, {Douglas}, and
  {Condon}]{1996ApJS..103..427G}
P.~C. {Gregory}, W.~K. {Scott}, K.~{Douglas}, and J.~J. {Condon}.
\newblock {The GB6 Catalog of Radio Sources}.
\newblock \emph{\apjs}, 103:\penalty0 427, April 1996.
\newblock \doi{10.1086/192282}.

\bibitem[{White} and {Becker}(1992)]{1992ApJS...79..331W}
R.~L. {White} and R.~H. {Becker}.
\newblock {A new catalog of 30,239 1.4 GHz sources}.
\newblock \emph{\apjs}, 79:\penalty0 331--467, April 1992.
\newblock \doi{10.1086/191656}.

\bibitem[{Planck Collaboration} et~al.(2014){Planck Collaboration}, {Ade},
  {Aghanim}, {Arg{\"u}eso}, {Armitage-Caplan}, {Arnaud}, {Ashdown},
  {Atrio-Barandela}, {Aumont}, {Baccigalupi}, and et~al.]{2014A&A...571A..28P}
{Planck Collaboration}, P.~A.~R. {Ade}, N.~{Aghanim}, F.~{Arg{\"u}eso},
  C.~{Armitage-Caplan}, M.~{Arnaud}, M.~{Ashdown}, F.~{Atrio-Barandela},
  J.~{Aumont}, C.~{Baccigalupi}, and et~al.
\newblock {Planck 2013 results. XXVIII. The Planck Catalogue of Compact
  Sources}.
\newblock \emph{\aap}, 571:\penalty0 A28, November 2014.
\newblock \doi{10.1051/0004-6361/201321524}.

\bibitem[{Planck Collaboration} et~al.(2015){Planck Collaboration}, {Ade},
  {Aghanim}, {Arg{\"u}eso}, {Arnaud}, {Ashdown}, {Aumont}, {Baccigalupi},
  {Banday}, {Barreiro}, and et~al.]{2015arXiv150702058P}
{Planck Collaboration}, P.~A.~R. {Ade}, N.~{Aghanim}, F.~{Arg{\"u}eso},
  M.~{Arnaud}, M.~{Ashdown}, J.~{Aumont}, C.~{Baccigalupi}, A.~J. {Banday},
  R.~B. {Barreiro}, and et~al.
\newblock {Planck 2015 results. XXVI. The Second Planck Catalogue of Compact
  Sources}.
\newblock \emph{ArXiv e-prints}, July 2015.

\bibitem[{Wright} et~al.(2009){Wright}, {Chen}, {Odegard}, {Bennett}, {Hill},
  {Hinshaw}, {Jarosik}, {Komatsu}, {Nolta}, {Page}, {Spergel}, {Weiland},
  {Wollack}, {Dunkley}, {Gold}, {Halpern}, {Kogut}, {Larson}, {Limon}, {Meyer},
  and {Tucker}]{2009ApJS..180..283W}
E.~L. {Wright}, X.~{Chen}, N.~{Odegard}, C.~L. {Bennett}, R.~S. {Hill},
  G.~{Hinshaw}, N.~{Jarosik}, E.~{Komatsu}, M.~R. {Nolta}, L.~{Page}, D.~N.
  {Spergel}, J.~L. {Weiland}, E.~{Wollack}, J.~{Dunkley}, B.~{Gold},
  M.~{Halpern}, A.~{Kogut}, D.~{Larson}, M.~{Limon}, S.~S. {Meyer}, and G.~S.
  {Tucker}.
\newblock {Five-Year Wilkinson Microwave Anisotropy Probe Observations: Source
  Catalog}.
\newblock \emph{\apjs}, 180:\penalty0 283--295, February 2009.
\newblock \doi{10.1088/0067-0049/180/2/283}.

\bibitem[{Wright} et~al.(2010){Wright}, {Eisenhardt}, {Mainzer}, {Ressler},
  {Cutri}, {Jarrett}, {Kirkpatrick}, {Padgett}, {McMillan}, {Skrutskie},
  {Stanford}, {Cohen}, {Walker}, {Mather}, {Leisawitz}, {Gautier}, {McLean},
  {Benford}, {Lonsdale}, {Blain}, {Mendez}, {Irace}, {Duval}, {Liu}, {Royer},
  {Heinrichsen}, {Howard}, {Shannon}, {Kendall}, {Walsh}, {Larsen}, {Cardon},
  {Schick}, {Schwalm}, {Abid}, {Fabinsky}, {Naes}, and
  {Tsai}]{2010AJ....140.1868W}
E.~L. {Wright}, P.~R.~M. {Eisenhardt}, A.~K. {Mainzer}, M.~E. {Ressler}, R.~M.
  {Cutri}, T.~{Jarrett}, J.~D. {Kirkpatrick}, D.~{Padgett}, R.~S. {McMillan},
  M.~{Skrutskie}, S.~A. {Stanford}, M.~{Cohen}, R.~G. {Walker}, J.~C. {Mather},
  D.~{Leisawitz}, T.~N. {Gautier}, III, I.~{McLean}, D.~{Benford}, C.~J.
  {Lonsdale}, A.~{Blain}, B.~{Mendez}, W.~R. {Irace}, V.~{Duval}, F.~{Liu},
  D.~{Royer}, I.~{Heinrichsen}, J.~{Howard}, M.~{Shannon}, M.~{Kendall}, A.~L.
  {Walsh}, M.~{Larsen}, J.~G. {Cardon}, S.~{Schick}, M.~{Schwalm}, M.~{Abid},
  B.~{Fabinsky}, L.~{Naes}, and C.-W. {Tsai}.
\newblock {The Wide-field Infrared Survey Explorer (WISE): Mission Description
  and Initial On-orbit Performance}.
\newblock \emph{\aj}, 140:\penalty0 1868, December 2010.
\newblock \doi{10.1088/0004-6256/140/6/1868}.

\bibitem[{Bianchi} et~al.(2011){Bianchi}, {Efremova}, {Herald}, {Girardi},
  {Zabot}, {Marigo}, and {Martin}]{2011MNRAS.411.2770B}
L.~{Bianchi}, B.~{Efremova}, J.~{Herald}, L.~{Girardi}, A.~{Zabot},
  P.~{Marigo}, and C.~{Martin}.
\newblock {Catalogues of hot white dwarfs in the Milky Way from GALEX's
  ultraviolet sky surveys: constraining stellar evolution}.
\newblock \emph{\mnras}, 411:\penalty0 2770--2791, March 2011.
\newblock \doi{10.1111/j.1365-2966.2010.17890.x}.

\bibitem[{D'Elia} et~al.(2013){D'Elia}, {Perri}, {Puccetti}, {Capalbi},
  {Giommi}, {Burrows}, {Campana}, {Tagliaferri}, {Cusumano}, {Evans},
  {Gehrels}, {Kennea}, {Moretti}, {Nousek}, {Osborne}, {Romano}, and
  {Stratta}]{2013A&A...551A.142D}
V.~{D'Elia}, M.~{Perri}, S.~{Puccetti}, M.~{Capalbi}, P.~{Giommi}, D.~N.
  {Burrows}, S.~{Campana}, G.~{Tagliaferri}, G.~{Cusumano}, P.~{Evans},
  N.~{Gehrels}, J.~{Kennea}, A.~{Moretti}, J.~A. {Nousek}, J.~P. {Osborne},
  P.~{Romano}, and G.~{Stratta}.
\newblock {The seven year Swift-XRT point source catalog (1SWXRT)}.
\newblock \emph{\aap}, 551:\penalty0 A142, March 2013.
\newblock \doi{10.1051/0004-6361/201220863}.

\bibitem[Evans et~al.(2014)Evans, Osborne, Beardmore, Page, Willingale,
  Mountford, Pagani, Burrows, Kennea, Perri, Tagliaferri, and
  Gehrels]{0067-0049-210-1-8}
P.~A. Evans, J.~P. Osborne, A.~P. Beardmore, K.~L. Page, R.~Willingale, C.~J.
  Mountford, C.~Pagani, D.~N. Burrows, J.~A. Kennea, M.~Perri, G.~Tagliaferri,
  and N.~Gehrels.
\newblock 1sxps: A deep swift x-ray telescope point source catalog with light
  curves and spectra.
\newblock \emph{The Astrophysical Journal Supplement Series}, 210\penalty0
  (1):\penalty0 8, 2014.
\newblock URL \url{http://stacks.iop.org/0067-0049/210/i=1/a=8}.

\bibitem[{Elvis} et~al.(1992){Elvis}, {Plummer}, {Schachter}, and
  {Fabbiano}]{1992ApJS...80..257E}
M.~{Elvis}, D.~{Plummer}, J.~{Schachter}, and G.~{Fabbiano}.
\newblock {The Einstein Slew Survey}.
\newblock \emph{\apjs}, 80:\penalty0 257--303, May 1992.
\newblock \doi{10.1086/191665}.

\bibitem[{Voges} et~al.(1999){Voges}, {Aschenbach}, {Boller}, {Br{\"a}uninger},
  {Briel}, {Burkert}, {Dennerl}, {Englhauser}, {Gruber}, {Haberl}, {Hartner},
  {Hasinger}, {K{\"u}rster}, {Pfeffermann}, {Pietsch}, {Predehl}, {Rosso},
  {Schmitt}, {Tr{\"u}mper}, and {Zimmermann}]{1999A&A...349..389V}
W.~{Voges}, B.~{Aschenbach}, T.~{Boller}, H.~{Br{\"a}uninger}, U.~{Briel},
  W.~{Burkert}, K.~{Dennerl}, J.~{Englhauser}, R.~{Gruber}, F.~{Haberl},
  G.~{Hartner}, G.~{Hasinger}, M.~{K{\"u}rster}, E.~{Pfeffermann},
  W.~{Pietsch}, P.~{Predehl}, C.~{Rosso}, J.~H.~M.~M. {Schmitt},
  J.~{Tr{\"u}mper}, and H.~U. {Zimmermann}.
\newblock {The ROSAT all-sky survey bright source catalogue}.
\newblock \emph{\aap}, 349:\penalty0 389--405, September 1999.

\bibitem[{Boller} et~al.(2016){Boller}, {Freyberg}, {Tr{""u}mper}, {Haberl},
  {Voges}, and {Nandra}]{2016A&A...588A.103B}
T.~{Boller}, M.~J. {Freyberg}, J.~{Tr{""u}mper}, F.~{Haberl}, W.~{Voges}, and
  K.~{Nandra}.
\newblock {Second ROSAT all-sky survey (2RXS) source catalogue}.
\newblock \emph{aap}, 588:\penalty0 A103, April 2016.
\newblock \doi{10.1051/0004-6361/201525648}.

\bibitem[{Saxton} et~al.(2008){Saxton}, {Read}, {Esquej}, {Freyberg},
  {Altieri}, and {Bermejo}]{2008A&A...480..611S}
R.~D. {Saxton}, A.~M. {Read}, P.~{Esquej}, M.~J. {Freyberg}, B.~{Altieri}, and
  D.~{Bermejo}.
\newblock {The first XMM-Newton slew survey catalogue: XMMSL1}.
\newblock \emph{\aap}, 480:\penalty0 611--622, March 2008.
\newblock \doi{10.1051/0004-6361:20079193}.

\bibitem[{Abdo} et~al.(2010){Abdo}, {Ackermann}, {Ajello}, {Allafort},
  {Antolini}, {Atwood}, {Axelsson}, {Baldini}, {Ballet}, {Barbiellini}, and
  et~al.]{2010ApJS..188..405A}
A.~A. {Abdo}, M.~{Ackermann}, M.~{Ajello}, A.~{Allafort}, E.~{Antolini}, W.~B.
  {Atwood}, M.~{Axelsson}, L.~{Baldini}, J.~{Ballet}, G.~{Barbiellini}, and
  et~al.
\newblock {Fermi Large Area Telescope First Source Catalog}.
\newblock \emph{\apjs}, 188:\penalty0 405--436, June 2010.
\newblock \doi{10.1088/0067-0049/188/2/405}.

\bibitem[{Nolan} et~al.(2012){Nolan}, {Abdo}, {Ackermann}, {Ajello},
  {Allafort}, {Antolini}, {Atwood}, {Axelsson}, {Baldini}, {Ballet}, and
  et~al.]{2012ApJS..199...31N}
P.~L. {Nolan}, A.~A. {Abdo}, M.~{Ackermann}, M.~{Ajello}, A.~{Allafort},
  E.~{Antolini}, W.~B. {Atwood}, M.~{Axelsson}, L.~{Baldini}, J.~{Ballet}, and
  et~al.
\newblock {Fermi Large Area Telescope Second Source Catalog}.
\newblock \emph{\apjs}, 199:\penalty0 31, April 2012.
\newblock \doi{10.1088/0067-0049/199/2/31}.

\bibitem[{Acero} et~al.(2015){Acero}, {Ackermann}, {Ajello}, {Albert},
  {Atwood}, {Axelsson}, {Baldini}, {Ballet}, {Barbiellini}, {Bastieri},
  {Belfiore}, {Bellazzini}, {Bissaldi}, {Blandford}, {Bloom}, {Bogart},
  {Bonino}, {Bottacini}, {Bregeon}, {Britto}, {Bruel}, {Buehler}, {Burnett},
  {Buson}, {Caliandro}, {Cameron}, {Caputo}, {Caragiulo}, {Caraveo},
  {Casandjian}, {Cavazzuti}, {Charles}, {Chaves}, {Chekhtman}, {Cheung},
  {Chiang}, {Chiaro}, {Ciprini}, {Claus}, {Cohen-Tanugi}, {Cominsky}, {Conrad},
  {Cutini}, {D'Ammando}, {de Angelis}, {DeKlotz}, {de Palma}, {Desiante},
  {Digel}, {Di Venere}, {Drell}, {Dubois}, {Dumora}, {Favuzzi}, {Fegan},
  {Ferrara}, {Finke}, {Franckowiak}, {Fukazawa}, {Funk}, {Fusco}, {Gargano},
  {Gasparrini}, {Giebels}, {Giglietto}, {Giommi}, {Giordano}, {Giroletti},
  {Glanzman}, {Godfrey}, {Grenier}, {Grondin}, {Grove}, {Guillemot}, {Guiriec},
  {Hadasch}, {Harding}, {Hays}, {Hewitt}, {Hill}, {Horan}, {Iafrate}, {Jogler},
  {J{\'o}hannesson}, {Johnson}, {Johnson}, {Johnson}, {Johnson}, {Kamae},
  {Kataoka}, {Katsuta}, {Kuss}, {La Mura}, {Landriu}, {Larsson}, {Latronico},
  {Lemoine-Goumard}, {Li}, {Li}, {Longo}, {Loparco}, {Lott}, {Lovellette},
  {Lubrano}, {Madejski}, {Massaro}, {Mayer}, {Mazziotta}, {McEnery},
  {Michelson}, {Mirabal}, {Mizuno}, {Moiseev}, {Mongelli}, {Monzani},
  {Morselli}, {Moskalenko}, {Murgia}, {Nuss}, {Ohno}, {Ohsugi}, {Omodei},
  {Orienti}, {Orlando}, {Ormes}, {Paneque}, {Panetta}, {Perkins},
  {Pesce-Rollins}, {Piron}, {Pivato}, {Porter}, {Racusin}, {Rando}, {Razzano},
  {Razzaque}, {Reimer}, {Reimer}, {Reposeur}, {Rochester}, {Romani},
  {Salvetti}, {S{\'a}nchez-Conde}, {Saz Parkinson}, {Schulz}, {Siskind},
  {Smith}, {Spada}, {Spandre}, {Spinelli}, {Stephens}, {Strong}, {Suson},
  {Takahashi}, {Takahashi}, {Tanaka}, {Thayer}, {Thayer}, {Thompson},
  {Tibaldo}, {Tibolla}, {Torres}, {Torresi}, {Tosti}, {Troja}, {Van Klaveren},
  {Vianello}, {Winer}, {Wood}, {Wood}, {Zimmer}, and {Fermi-LAT
  Collaboration}]{2015ApJS..218...23A}
F.~{Acero}, M.~{Ackermann}, M.~{Ajello}, A.~{Albert}, W.~B. {Atwood},
  M.~{Axelsson}, L.~{Baldini}, J.~{Ballet}, G.~{Barbiellini}, D.~{Bastieri},
  A.~{Belfiore}, R.~{Bellazzini}, E.~{Bissaldi}, R.~D. {Blandford}, E.~D.
  {Bloom}, J.~R. {Bogart}, R.~{Bonino}, E.~{Bottacini}, J.~{Bregeon}, R.~J.
  {Britto}, P.~{Bruel}, R.~{Buehler}, T.~H. {Burnett}, S.~{Buson}, G.~A.
  {Caliandro}, R.~A. {Cameron}, R.~{Caputo}, M.~{Caragiulo}, P.~A. {Caraveo},
  J.~M. {Casandjian}, E.~{Cavazzuti}, E.~{Charles}, R.~C.~G. {Chaves},
  A.~{Chekhtman}, C.~C. {Cheung}, J.~{Chiang}, G.~{Chiaro}, S.~{Ciprini},
  R.~{Claus}, J.~{Cohen-Tanugi}, L.~R. {Cominsky}, J.~{Conrad}, S.~{Cutini},
  F.~{D'Ammando}, A.~{de Angelis}, M.~{DeKlotz}, F.~{de Palma}, R.~{Desiante},
  S.~W. {Digel}, L.~{Di Venere}, P.~S. {Drell}, R.~{Dubois}, D.~{Dumora},
  C.~{Favuzzi}, S.~J. {Fegan}, E.~C. {Ferrara}, J.~{Finke}, A.~{Franckowiak},
  Y.~{Fukazawa}, S.~{Funk}, P.~{Fusco}, F.~{Gargano}, D.~{Gasparrini},
  B.~{Giebels}, N.~{Giglietto}, P.~{Giommi}, F.~{Giordano}, M.~{Giroletti},
  T.~{Glanzman}, G.~{Godfrey}, I.~A. {Grenier}, M.-H. {Grondin}, J.~E. {Grove},
  L.~{Guillemot}, S.~{Guiriec}, D.~{Hadasch}, A.~K. {Harding}, E.~{Hays}, J.~W.
  {Hewitt}, A.~B. {Hill}, D.~{Horan}, G.~{Iafrate}, T.~{Jogler},
  G.~{J{\'o}hannesson}, R.~P. {Johnson}, A.~S. {Johnson}, T.~J. {Johnson},
  W.~N. {Johnson}, T.~{Kamae}, J.~{Kataoka}, J.~{Katsuta}, M.~{Kuss}, G.~{La
  Mura}, D.~{Landriu}, S.~{Larsson}, L.~{Latronico}, M.~{Lemoine-Goumard},
  J.~{Li}, L.~{Li}, F.~{Longo}, F.~{Loparco}, B.~{Lott}, M.~N. {Lovellette},
  P.~{Lubrano}, G.~M. {Madejski}, F.~{Massaro}, M.~{Mayer}, M.~N. {Mazziotta},
  J.~E. {McEnery}, P.~F. {Michelson}, N.~{Mirabal}, T.~{Mizuno}, A.~A.
  {Moiseev}, M.~{Mongelli}, M.~E. {Monzani}, A.~{Morselli}, I.~V. {Moskalenko},
  S.~{Murgia}, E.~{Nuss}, M.~{Ohno}, T.~{Ohsugi}, N.~{Omodei}, M.~{Orienti},
  E.~{Orlando}, J.~F. {Ormes}, D.~{Paneque}, J.~H. {Panetta}, J.~S. {Perkins},
  M.~{Pesce-Rollins}, F.~{Piron}, G.~{Pivato}, T.~A. {Porter}, J.~L. {Racusin},
  R.~{Rando}, M.~{Razzano}, S.~{Razzaque}, A.~{Reimer}, O.~{Reimer},
  T.~{Reposeur}, L.~S. {Rochester}, R.~W. {Romani}, D.~{Salvetti},
  M.~{S{\'a}nchez-Conde}, P.~M. {Saz Parkinson}, A.~{Schulz}, E.~J. {Siskind},
  D.~A. {Smith}, F.~{Spada}, G.~{Spandre}, P.~{Spinelli}, T.~E. {Stephens},
  A.~W. {Strong}, D.~J. {Suson}, H.~{Takahashi}, T.~{Takahashi}, Y.~{Tanaka},
  J.~G. {Thayer}, J.~B. {Thayer}, D.~J. {Thompson}, L.~{Tibaldo}, O.~{Tibolla},
  D.~F. {Torres}, E.~{Torresi}, G.~{Tosti}, E.~{Troja}, B.~{Van Klaveren},
  G.~{Vianello}, B.~L. {Winer}, K.~S. {Wood}, M.~{Wood}, S.~{Zimmer}, and
  {Fermi-LAT Collaboration}.
\newblock {Fermi Large Area Telescope Third Source Catalog}.
\newblock \emph{\apjs}, 218:\penalty0 23, June 2015.
\newblock \doi{10.1088/0067-0049/218/2/23}.

\bibitem[Bartoli et~al.(2013)Bartoli, Bernardini, Bi, Bolognino, Branchini,
  Budano, Melcarne, Camarri, Cao, Cardarelli, Catalanotti, Chen, Chen, Chen,
  Creti, Cui, Dai, D'Amone, Danzengluobu, Mitri, Piazzoli, Girolamo, Ding,
  Sciascio, Feng, Feng, Feng, Gou, Guo, He, Hu, Hu, Huang, Iacovacci, Iuppa,
  Jia, Labaciren, Li, Li, Li, Liguori, Liu, Liu, Liu, Liu, Lu, Ma, Ma,
  Mancarella, Mari, Marsella, Martello, Mastroianni, Montini, Ning, Panareo,
  Panico, Perrone, Pistilli, Ruggieri, Salvini, Santonico, Sbano, Shen, Sheng,
  Shi, Surdo, Tan, Vallania, Vernetto, Vigorito, Wang, Wang, Wu, Wu, Xu, Xue,
  Yang, Yang, Yao, Yuan, Zha, Zhang, Zhang, Zhang, Zhang, Zhang, Zhang, Zhang,
  Zhao, Zhaxiciren, Zhaxisangzhu, Zhou, Zhu, Zhu, Zizzi, and
  Collaboration]{0004-637X-779-1-27}
B.~Bartoli, P.~Bernardini, X.~J. Bi, I.~Bolognino, P.~Branchini, A.~Budano,
  A.~K.~Calabrese Melcarne, P.~Camarri, Z.~Cao, R.~Cardarelli, S.~Catalanotti,
  S.~Z. Chen, T.~L. Chen, Y.~Chen, P.~Creti, S.~W. Cui, B.~Z. Dai, A.~D'Amone,
  Danzengluobu, I.~De Mitri, B.~D'Ettorre Piazzoli, T.~Di Girolamo, X.~H. Ding,
  G.~Di Sciascio, C.~F. Feng, Zhaoyang Feng, Zhenyong Feng, Q.~B. Gou, Y.~Q.
  Guo, H.~H. He, Haibing Hu, Hongbo Hu, Q.~Huang, M.~Iacovacci, R.~Iuppa, H.~Y.
  Jia, Labaciren, H.~J. Li, J.~Y. Li, X.~X. Li, G.~Liguori, C.~Liu, C.~Q. Liu,
  J.~Liu, M.~Y. Liu, H.~Lu, L.~L. Ma, X.~H. Ma, G.~Mancarella, S.~M. Mari,
  G.~Marsella, D.~Martello, S.~Mastroianni, P.~Montini, C.~C. Ning, M.~Panareo,
  B.~Panico, L.~Perrone, P.~Pistilli, F.~Ruggieri, P.~Salvini, R.~Santonico,
  S.~N. Sbano, P.~R. Shen, X.~D. Sheng, F.~Shi, A.~Surdo, Y.~H. Tan,
  P.~Vallania, S.~Vernetto, C.~Vigorito, B.~Wang, H.~Wang, C.~Y. Wu, H.~R. Wu,
  B.~Xu, L.~Xue, Q.~Y. Yang, X.~C. Yang, Z.~G. Yao, A.~F. Yuan, M.~Zha, H.~M.
  Zhang, Jilong Zhang, Jianli Zhang, L.~Zhang, P.~Zhang, X.~Y. Zhang, Y.~Zhang,
  J.~Zhao, Zhaxiciren, Zhaxisangzhu, X.~X. Zhou, F.~R. Zhu, Q.~Q. Zhu,
  G.~Zizzi, and The ARGO-YBJ Collaboration.
\newblock Tev gamma-ray survey of the northern sky using the argo-ybj detector.
\newblock \emph{The Astrophysical Journal}, 779\penalty0 (1):\penalty0 27,
  2013.
\newblock URL \url{http://stacks.iop.org/0004-637X/779/i=1/a=27}.

\bibitem[sed()]{sedbuilder}
{SSDC SED Builder}.
\newblock \url{https://tools.ssdc.asi.it/}.
\newblock Version 1.21.

\bibitem[Waxman and Bahcall(1999)]{Waxman:1998yy}
Eli Waxman and John~N. Bahcall.
\newblock {High-energy neutrinos from astrophysical sources: An Upper bound}.
\newblock \emph{Phys. Rev.}, D59:\penalty0 023002, 1999.
\newblock \doi{10.1103/PhysRevD.59.023002}.

\bibitem[Yuksel et~al.(2008)Yuksel, Kistler, Beacom, and
  Hopkins]{Yuksel:2008cu}
Hasan Yuksel, Matthew~D. Kistler, John~F. Beacom, and Andrew~M. Hopkins.
\newblock {Revealing the High-Redshift Star Formation Rate with Gamma-Ray
  Bursts}.
\newblock \emph{Astrophys. J.}, 683:\penalty0 L5--L8, 2008.
\newblock \doi{10.1086/591449}.

\bibitem[Hopkins and Beacom(2006)]{Hopkins:2006bw}
Andrew~M. Hopkins and John~F. Beacom.
\newblock {On the normalisation of the cosmic star formation history}.
\newblock \emph{Astrophys. J.}, 651:\penalty0 142--154, 2006.
\newblock \doi{10.1086/506610}.

\bibitem[{Wilson} et~al.(2014){Wilson}, {Rangwala}, {Glenn}, {Maloney},
  {Spinoglio}, and {Pereira-Santaella}]{2014ApJ...789L..36W}
C.~D. {Wilson}, N.~{Rangwala}, J.~{Glenn}, P.~R. {Maloney}, L.~{Spinoglio}, and
  M.~{Pereira-Santaella}.
\newblock {Extreme Dust Disks in Arp 220 as Revealed by ALMA}.
\newblock \emph{\apj}, 789\penalty0 (2):\penalty0 L36, Jul 2014.
\newblock \doi{10.1088/2041-8205/789/2/L36}.

\bibitem[Lonsdale et~al.(2006)Lonsdale, Farrah, and Smith]{Lonsdale:2006sr}
Carol Lonsdale, Duncan Farrah, and Harding Smith.
\newblock {Ultraluminous infrared galaxies}.
\newblock 2006.
\newblock \doi{10.1007/3-540-30313-8_9}.

\bibitem[{Clements} et~al.(2010){Clements}, {Dunne}, and
  {Eales}]{2010MNRAS.403..274C}
D.~L. {Clements}, L.~{Dunne}, and S.~{Eales}.
\newblock {The submillimetre properties of ultraluminous infrared galaxies}.
\newblock \emph{\mnras}, 403:\penalty0 274--286, March 2010.
\newblock \doi{10.1111/j.1365-2966.2009.16064.x}.

\bibitem[Nagar et~al.(2003)Nagar, Wilson, Falcke, Veilleux, and
  Maiolino]{Nagar:2003an}
Neil~M. Nagar, Andrew~S. Wilson, Heino Falcke, Sylvain Veilleux, and Roberto
  Maiolino.
\newblock {The AGN content of ultraluminous IR galaxies: High resolution VLA
  imaging of the IRAS 1Jy ULIRG sample}.
\newblock \emph{Astron. Astrophys.}, 409:\penalty0 115--122, 2003.
\newblock \doi{10.1051/0004-6361:20031069}.

\bibitem[Farrah et~al.(2003)Farrah, Afonso, Efstathiou, Rowan-Robinson, Fox,
  and Clements]{Farrah:2003ka}
D.~Farrah, J.~Afonso, A.~Efstathiou, M.~Rowan-Robinson, M.~Fox, and
  D.~Clements.
\newblock {Starburst and agn activity in ultraluminous infrared galaxies}.
\newblock \emph{Mon. Not. Roy. Astron. Soc.}, 343:\penalty0 585, 2003.
\newblock \doi{10.1046/j.1365-8711.2003.06696.x}.

\bibitem[{Hou} et~al.(2009){Hou}, {Wu}, and {Han}]{2009ApJ...704..789H}
L.~G. {Hou}, X.-B. {Wu}, and J.~L. {Han}.
\newblock {Ultra-luminous Infrared Galaxies in Sloan Digital Sky Survey Data
  Release 6}.
\newblock \emph{\apj}, 704:\penalty0 789--802, October 2009.
\newblock \doi{10.1088/0004-637X/704/1/789}.

\bibitem[{Fadda} et~al.(2010){Fadda}, {Yan}, {Lagache}, {Sajina}, {Lutz},
  {Wuyts}, {Frayer}, {Marcillac}, {Le Floc'h}, {Caputi}, {Spoon}, {Veilleux},
  {Blain}, and {Helou}]{2010ApJ...719..425F}
D.~{Fadda}, L.~{Yan}, G.~{Lagache}, A.~{Sajina}, D.~{Lutz}, S.~{Wuyts}, D.~T.
  {Frayer}, D.~{Marcillac}, E.~{Le Floc'h}, K.~{Caputi}, H.~W.~W. {Spoon},
  S.~{Veilleux}, A.~{Blain}, and G.~{Helou}.
\newblock {Ultra-deep Mid-infrared Spectroscopy of Luminous Infrared Galaxies
  at z \~{} 1 and z \~{} 2}.
\newblock \emph{\apj}, 719:\penalty0 425--450, August 2010.
\newblock \doi{10.1088/0004-637X/719/1/425}.

\bibitem[Nardini et~al.(2009)Nardini, Risaliti, Salvati, Sani, Watabe, Marconi,
  and Maiolino]{doi:10.1111/j.1365-2966.2009.15357.x}
E.~Nardini, G.~Risaliti, M.~Salvati, E.~Sani, Y.~Watabe, A.~Marconi, and
  R.~Maiolino.
\newblock Exploring the active galactic nucleus and starburst content of local
  ultraluminous infrared galaxies through 5–8 $\mu$m spectroscopy.
\newblock \emph{Monthly Notices of the Royal Astronomical Society},
  399\penalty0 (3):\penalty0 1373--1402, 2009.
\newblock \doi{10.1111/j.1365-2966.2009.15357.x}.
\newblock URL \url{http://dx.doi.org/10.1111/j.1365-2966.2009.15357.x}.

\bibitem[{Saturni} et~al.(2018){Saturni}, {Mancini}, {Pezzulli}, and
  {Tombesi}]{2018A&A...617A.131S}
F.~G. {Saturni}, M.~{Mancini}, E.~{Pezzulli}, and F.~{Tombesi}.
\newblock {``Zombie'' or active? An alternative explanation to the properties
  of star-forming galaxies at high redshift}.
\newblock \emph{\aap}, 617:\penalty0 A131, October 2018.
\newblock \doi{10.1051/0004-6361/201833261}.

\bibitem[{Kauffmann} and {Haehnelt}(2002)]{2002MNRAS.332..529K}
G.~{Kauffmann} and M.~G. {Haehnelt}.
\newblock {The clustering of galaxies around quasars}.
\newblock \emph{\mnras}, 332:\penalty0 529--535, May 2002.
\newblock \doi{10.1046/j.1365-8711.2002.05278.x}.

\bibitem[Dasyra et~al.(2006{\natexlab{a}})Dasyra, Tacconi, Davies, Genzel,
  Lutz, Naab, Burkert, Veilleux, and Sanders]{0004-637X-638-2-745}
K.~M. Dasyra, L.~J. Tacconi, R.~I. Davies, R.~Genzel, D.~Lutz, T.~Naab,
  A.~Burkert, S.~Veilleux, and D.~B. Sanders.
\newblock Dynamical properties of ultraluminous infrared galaxies. i. mass
  ratio conditions for ulirg activity in interacting pairs.
\newblock \emph{The Astrophysical Journal}, 638\penalty0 (2):\penalty0 745,
  2006{\natexlab{a}}.
\newblock URL \url{http://stacks.iop.org/0004-637X/638/i=2/a=745}.

\bibitem[Dasyra et~al.(2006{\natexlab{b}})Dasyra, Tacconi, Davies, Naab,
  Genzel, Lutz, Sturm, Baker, Veilleux, Sanders, and
  Burkert]{0004-637X-651-2-835}
K.~M. Dasyra, L.~J. Tacconi, R.~I. Davies, T.~Naab, R.~Genzel, D.~Lutz,
  E.~Sturm, A.~J. Baker, S.~Veilleux, D.~B. Sanders, and A.~Burkert.
\newblock Dynamical properties of ultraluminous infrared galaxies. ii. traces
  of dynamical evolution and end products of local ultraluminous mergers.
\newblock \emph{The Astrophysical Journal}, 651\penalty0 (2):\penalty0 835,
  2006{\natexlab{b}}.
\newblock URL \url{http://stacks.iop.org/0004-637X/651/i=2/a=835}.

\bibitem[Hopkins et~al.(2005)Hopkins, Hernquist, Martini, Cox, Robertson,
  Matteo, and Springel]{1538-4357-625-2-L71}
Philip~F. Hopkins, Lars Hernquist, Paul Martini, Thomas~J. Cox, Brant
  Robertson, Tiziana~Di Matteo, and Volker Springel.
\newblock A physical model for the origin of quasar lifetimes.
\newblock \emph{The Astrophysical Journal Letters}, 625\penalty0 (2):\penalty0
  L71, 2005.
\newblock URL \url{http://stacks.iop.org/1538-4357/625/i=2/a=L71}.

\bibitem[Hopkins et~al.(2006)Hopkins, Hernquist, Cox, Di~Matteo, Robertson, and
  Springel]{Hopkins:2005fb}
Philip~F. Hopkins, Lars Hernquist, Thomas~J. Cox, Tiziana Di~Matteo, Brant
  Robertson, and Volker Springel.
\newblock {A Unified, merger-driven model for the origin of starbursts,
  quasars, the cosmic x-ray background, supermassive black holes and galaxy
  spheroids}.
\newblock \emph{Astrophys. J. Suppl.}, 163:\penalty0 1--49, 2006.
\newblock \doi{10.1086/499298}.

\bibitem[Chapman et~al.(2005)Chapman, Blain, Smail, and Ivison]{Chapman:2004fu}
Scott~C. Chapman, A.~W. Blain, Ian Smail, and R.~J. Ivison.
\newblock {A Redshift survey of the submillimeter galaxy population}.
\newblock \emph{Astrophys. J.}, 622:\penalty0 772--796, 2005.
\newblock \doi{10.1086/428082}.

\bibitem[{Cowie} et~al.(2004){Cowie}, {Barger}, {Fomalont}, and
  {Capak}]{2004ApJ...603L..69C}
L.~L. {Cowie}, A.~J. {Barger}, E.~B. {Fomalont}, and P.~{Capak}.
\newblock {The Evolution of the Ultraluminous Infrared Galaxy Population from
  Redshift 0 to 1.5}.
\newblock \emph{\apjl}, 603:\penalty0 L69--L72, March 2004.
\newblock \doi{10.1086/383198}.

\bibitem[He et~al.(2013)He, Wang, Fan, Liu, and Wei]{He:2013cqa}
Hao-Ning He, Tao Wang, Yi-Zhong Fan, Si-Ming Liu, and Da-Ming Wei.
\newblock {Diffuse PeV neutrino emission from ultraluminous infrared galaxies}.
\newblock \emph{Phys. Rev.}, D87\penalty0 (6):\penalty0 063011, 2013.
\newblock \doi{10.1103/PhysRevD.87.063011}.

\bibitem[{Schmidt}(1959)]{1959ApJ...129..243S}
M.~{Schmidt}.
\newblock {The Rate of Star Formation.}
\newblock \emph{\apj}, 129:\penalty0 243, March 1959.
\newblock \doi{10.1086/146614}.

\bibitem[{Kennicutt}(1989)]{1989ApJ...344..685K}
R.~C. {Kennicutt}, Jr.
\newblock {The star formation law in galactic disks}.
\newblock \emph{\apj}, 344:\penalty0 685--703, September 1989.
\newblock \doi{10.1086/167834}.

\bibitem[{Kennicutt}(1998)]{1998ApJ...498..541K}
R.~C. {Kennicutt}, Jr.
\newblock {The Global Schmidt Law in Star-forming Galaxies}.
\newblock \emph{\apj}, 498:\penalty0 541--552, May 1998.
\newblock \doi{10.1086/305588}.

\bibitem[{Sargent} et~al.(2010){Sargent}, {Schinnerer}, {Murphy}, {Carilli},
  {Helou}, {Aussel}, {Le Floc'h}, {Frayer}, {Ilbert}, {Oesch}, {Salvato},
  {Smol{\v c}i{\'c}}, {Kartaltepe}, and {Sanders}]{2010ApJ...714L.190S}
M.~T. {Sargent}, E.~{Schinnerer}, E.~{Murphy}, C.~L. {Carilli}, G.~{Helou},
  H.~{Aussel}, E.~{Le Floc'h}, D.~T. {Frayer}, O.~{Ilbert}, P.~{Oesch},
  M.~{Salvato}, V.~{Smol{\v c}i{\'c}}, J.~{Kartaltepe}, and D.~B. {Sanders}.
\newblock {No Evolution in the IR-Radio Relation for IR-luminous Galaxies at
  $z<2$ in the COSMOS Field}.
\newblock \emph{\apjl}, 714:\penalty0 L190--L195, May 2010.
\newblock \doi{10.1088/2041-8205/714/2/L190}.

\bibitem[Hogg(1999)]{Hogg:1999ad}
David~W. Hogg.
\newblock {Distance measures in cosmology}.
\newblock 1999.

\bibitem[Weinberg(2008)]{weinberg2008cosmology}
S.~Weinberg.
\newblock \emph{Cosmology}.
\newblock OUP Oxford, 2008.
\newblock ISBN 9780198526827.
\newblock URL \url{https://books.google.be/books?id=rswTDAAAQBAJ}.

\bibitem[Berezinsky and Smirnov(1975)]{Berezinsky:1975zz}
V.~S. Berezinsky and A.~{\relax Yu}. Smirnov.
\newblock {Cosmic neutrinos of ultra-high energies and detection possibility}.
\newblock \emph{Astrophys. Space Sci.}, 32:\penalty0 461--482, 1975.
\newblock \doi{10.1007/BF00643157}.

\bibitem[Coppi and Aharonian(1997)]{Coppi:1996ze}
Paolo~S. Coppi and Felix~A. Aharonian.
\newblock {Constraints on the VHE emissivity of the universe from the diffuse
  GeV gamma-ray background}.
\newblock \emph{Astrophys. J.}, 487:\penalty0 L9--L12, 1997.
\newblock \doi{10.1086/310883}.

\bibitem[Murase(2012)]{Murase:2011yw}
Kohta Murase.
\newblock {High-Energy Emission Induced by Ultra-high-Energy Photons as a Probe
  of Ultra-high-Energy Cosmic-Ray Accelerators Embedded in the Cosmic Web}.
\newblock \emph{Astrophys. J.}, 745:\penalty0 L16, 2012.
\newblock \doi{10.1088/2041-8205/745/2/L16}.

\bibitem[{Murase} et~al.(2012){Murase}, {Dermer}, {Takami}, and
  {Migliori}]{2012ApJ...749...63M}
K.~{Murase}, C.~D. {Dermer}, H.~{Takami}, and G.~{Migliori}.
\newblock {Blazars as Ultra-high-energy Cosmic-ray Sources: Implications for
  TeV Gamma-Ray Observations}.
\newblock \emph{\apj}, 749:\penalty0 63, April 2012.
\newblock \doi{10.1088/0004-637X/749/1/63}.

\bibitem[Murase and Beacom(2012)]{Murase:2012xs}
Kohta Murase and John~F. Beacom.
\newblock {Constraining Very Heavy Dark Matter Using Diffuse Backgrounds of
  Neutrinos and Cascaded Gamma Rays}.
\newblock \emph{JCAP}, 1210:\penalty0 043, 2012.
\newblock \doi{10.1088/1475-7516/2012/10/043}.

\bibitem[Alves~Batista et~al.(2016)Alves~Batista, Dundovic, Erdmann, Kampert,
  Kuempel, Müller, Sigl, van Vliet, Walz, and Winchen]{Batista:2016yrx}
Rafael Alves~Batista, Andrej Dundovic, Martin Erdmann, Karl-Heinz Kampert,
  Daniel Kuempel, Gero Müller, Guenter Sigl, Arjen van Vliet, David Walz, and
  Tobias Winchen.
\newblock {CRPropa 3 - a Public Astrophysical Simulation Framework for
  Propagating Extraterrestrial Ultra-High Energy Particles}.
\newblock \emph{JCAP}, 1605\penalty0 (05):\penalty0 038, 2016.
\newblock \doi{10.1088/1475-7516/2016/05/038}.

\bibitem[Lee(1998)]{Lee:1996fp}
Sangjin Lee.
\newblock {On the propagation of extragalactic high-energy cosmic and
  gamma-rays}.
\newblock \emph{Phys. Rev.}, D58:\penalty0 043004, 1998.
\newblock \doi{10.1103/PhysRevD.58.043004}.

\bibitem[Ackermann et~al.(2015{\natexlab{b}})]{Ackermann:2014usa}
M.~Ackermann et~al.
\newblock {The spectrum of isotropic diffuse gamma-ray emission between 100 MeV
  and 820 GeV}.
\newblock \emph{Astrophys. J.}, 799:\penalty0 86, 2015{\natexlab{b}}.
\newblock \doi{10.1088/0004-637X/799/1/86}.

\bibitem[{Jackson} et~al.(2007){Jackson}, {Battye}, {Browne}, {Joshi},
  {Muxlow}, and {Wilkinson}]{2007MNRAS.376..371J}
N.~{Jackson}, R.~A. {Battye}, I.~W.~A. {Browne}, S.~{Joshi}, T.~W.~B. {Muxlow},
  and P.~N. {Wilkinson}.
\newblock {A survey of polarization in the JVAS/CLASS flat-spectrum radio
  source surveys - I. The data and catalogue production}.
\newblock \emph{\mnras}, 376:\penalty0 371--377, March 2007.
\newblock \doi{10.1111/j.1365-2966.2007.11442.x}.

\bibitem[{Wright} and {Otrupcek}(1990)]{1990PKS...C......0W}
A.~{Wright} and R.~{Otrupcek}.
\newblock {Parkes Catalog, 1990, Australia telescope national facility.}
\newblock In \emph{PKS Catalog (1990)}, page~0, 1990.

\bibitem[{White} et~al.(1997){White}, {Becker}, {Helfand}, and
  {Gregg}]{1997ApJ...475..479W}
R.~L. {White}, R.~H. {Becker}, D.~J. {Helfand}, and M.~D. {Gregg}.
\newblock {A Catalog of 1.4 GHz Radio Sources from the FIRST Survey}.
\newblock \emph{\apj}, 475:\penalty0 479--493, February 1997.

\bibitem[{Rosen} et~al.(2015){Rosen}, {Webb}, {Watson}, {Ballet}, {Barret},
  {Braito}, {Carrera}, {Ceballos}, {Coriat}, {Della Ceca}, {Denkinson},
  {Esquej}, {Farrell}, {Freyberg}, {Gris{\'e}}, {Guillout}, {Heil},
  {Law-Green}, {Lamer}, {Lin}, {Martino}, {Michel}, {Motch}, {Nebot
  Gomez-Moran}, {Page}, {Page}, {Page}, {Pakull}, {Pye}, {Read}, {Rodriguez},
  {Sakano}, {Saxton}, {Schwope}, {Scott}, {Sturm}, {Traulsen}, {Yershov}, and
  {Zolotukhin}]{2015arXiv150407051R}
S.~R. {Rosen}, N.~A. {Webb}, M.~G. {Watson}, J.~{Ballet}, D.~{Barret},
  V.~{Braito}, F.~J. {Carrera}, M.~T. {Ceballos}, M.~{Coriat}, R.~{Della Ceca},
  G.~{Denkinson}, P.~{Esquej}, S.~A. {Farrell}, M.~{Freyberg}, F.~{Gris{\'e}},
  P.~{Guillout}, L.~{Heil}, D.~{Law-Green}, G.~{Lamer}, D.~{Lin}, R.~{Martino},
  L.~{Michel}, C.~{Motch}, A.~{Nebot Gomez-Moran}, C.~G. {Page}, K.~{Page},
  M.~{Page}, M.~W. {Pakull}, J.~{Pye}, A.~{Read}, P.~{Rodriguez}, M.~{Sakano},
  R.~{Saxton}, A.~{Schwope}, A.~E. {Scott}, R.~{Sturm}, I.~{Traulsen},
  V.~{Yershov}, and I.~{Zolotukhin}.
\newblock {The XMM-Newton serendipitous survey. VII. The third XMM-Newton
  serendipitous source catalogue}.
\newblock \emph{ArXiv e-prints}, April 2015.

\bibitem[{Forman} et~al.(1978){Forman}, {Jones}, {Cominsky}, {Julien},
  {Murray}, {Peters}, {Tananbaum}, and {Giacconi}]{1978ApJS...38..357F}
W.~{Forman}, C.~{Jones}, L.~{Cominsky}, P.~{Julien}, S.~{Murray}, G.~{Peters},
  H.~{Tananbaum}, and R.~{Giacconi}.
\newblock {The fourth Uhuru catalog of X-ray sources.}
\newblock \emph{\apjs}, 38:\penalty0 357--412, December 1978.
\newblock \doi{10.1086/190561}.

\bibitem[{Verrecchia} et~al.(2007){Verrecchia}, {in't Zand}, {Giommi},
  {Santolamazza}, {Granata}, {Schuurmans}, and
  {Antonelli}]{2007A&A...472..705V}
F.~{Verrecchia}, J.~J.~M. {in't Zand}, P.~{Giommi}, P.~{Santolamazza},
  S.~{Granata}, J.~J. {Schuurmans}, and L.~A. {Antonelli}.
\newblock {The BeppoSAX WFC X-ray source catalogue}.
\newblock \emph{\aap}, 472:\penalty0 705--713, September 2007.
\newblock \doi{10.1051/0004-6361:20067040}.

\end{thebibliography}

\end{document}